\documentclass[structabstract]{aa} %paper
\usepackage{booktabs}
\usepackage[pdftex]{graphicx}
\usepackage{txfonts}
\usepackage{natbib}
\usepackage{longtable}
\usepackage{lscape}
\usepackage{multirow}
\usepackage{pdfpages}
\usepackage{longtable}
\usepackage{sidecap}
\usepackage{subfigure}
\usepackage{ulem}
\usepackage[colorlinks=true, allcolors=blue]{hyperref}

\bibpunct{(}{)}{;}{a}{}{,}

\newcommand{\kms}{km\,s$^{-1}$}

\newcommand{\Msun}{M$_{\odot}$}
\newcommand{\Lsun}{L$_{\odot}$}

\begin{document}
\title{Redshifted methanol absorption tracing infall motions of high-mass star formation regions}

\author{W. J. Yang\inst{1}, K. M. Menten\inst{1}, A. Y. Yang\inst{1}, F. Wyrowski\inst{1}, Y. Gong\inst{1}, S. P. Ellingsen\inst{2}, C. Henkel\inst{1,3,4}, X. Chen\inst{5}, Y. Xu\inst{6}}
\offprints{W. Yang, \email{wjyang@mpifr-bonn.mpg.de}}

\institute{
Max-Planck-Institut f{\"u}r Radioastronomie, Auf dem H{\"u}gel 69, D-53121 Bonn, Germany
\and
School of Natural Sciences, University of Tasmania, Private Bag 37, Hobart, Tasmania 7001, Australia
\and
Astronomy Department, Faculty of Science, King Abdulaziz University, PO Box 80203, Jeddah, 21589, Saudi Arabia
\and
Xinjiang Astronomical Observatory, Chinese Academy of Sciences, Urumqi 830011, PR China
\and
Center for Astrophysics, GuangZhou University, Guangzhou 510006, PR China
\and
Purple Mountain Observatory, Chinese Academy of Science, Nanjing 210023, PR China
}

\date{Received date ; accepted date}

\abstract
{Gravitational collapse is one of the most important processes in high-mass star formation. Compared with the classic blue-skewed profiles, redshifted absorption against continuum emission is a more reliable method to detect inward motions within high-mass star formation regions.}
{We aim to test if methanol transitions can be used to trace infall motions within high-mass star formation regions.}
{Using the Effelsberg-100 m, IRAM-30 m, and APEX-12 m telescopes, we carried out observations of 37 and 16 methanol transitions towards two well-known collapsing dense clumps, W31C (G10.6$-$0.4) and W3(OH), to search for redshifted absorption features or inverse P-Cygni profiles.}
{Redshifted absorption is observed in 14 and 11 methanol transitions towards W31C and W3(OH), respectively. The infall velocities fitted from a simple two-layer model agree with previously reported values derived from other tracers, suggesting that redshifted methanol absorption is a reliable tracer of infall motions within high-mass star formation regions. Our observations indicate the presence of large-scale inward motions, and the mass infall rates are roughly estimated to be $\gtrsim$10$^{-3}$~$M_{\odot}$~yr$^{-1}$, which supports the global hierarchical collapse and clump-fed scenario.}
%{Our work demonstrates that methanol redshifted absorption can be used to trace infall motions of high-mass star formation regions.}
%{Because the overcooling of methanol transitions due to collisional pumping will greatly aid in the detection of methanol absorption, we suggest that redshifted methanol absorption can be regarded as an excellent tracer to search for infall motions of high-mass star formation regions, and observing multiple methanol transitions allows us to study the inward motions at different scales and environments. }
{With the aid of bright continuum sources and the overcooling of methanol transitions leading to enhanced absorption, redshifted methanol absorption can trace infall motions within high-mass star formation regions hosting bright H{\scriptsize II} regions.}%, and observing multiple methanol transitions allows us to study the inward motions at different scales and environments.}


\titlerunning{Redshifted methanol absorption}

\authorrunning{Yang et al.}

\maketitle

%________________________________________________________________

\section{Introduction} \label{sec:intro}
%\subsection{Infall in high mass star forming regions}
Infall motions provide direct evidence of mass accretion and their observation plays a crucial role in star formation research. The classic blue-skewed profile
%: optically thick molecular transitions show a central absorption dip and a stronger blueshifted emission than the redshifted one, while the systemic velocity traced by optically thin lines is align with the central dip, 
is commonly taken as evidence of infall motions \citep[e.g.][]{1977ApJ...214L..73L,1993ApJ...404..232Z}. However, the interpretation of the blue-skewed profile as an infall signature can be %obscured
impaired by kinematic peculiarities (e.g. outflow and rotation) and chemical abundance variations \citep{2003cdsf.conf..157E}. This situation becomes even more severe when analysing spectra obtained towards complex high-mass star formation regions by single-dish observations with a resolution $\gtrsim$10$\arcsec$ (corresponding to 0.24 pc at a distance of 5 kpc), because they incorporate emission which is statistically located at farther distances.

%However, other kinematics such as outflow and rotation can also produce similar line profiles, and chemical abundance variations in clumps perplex the line shape, thus leading to the interpretation of the blue-skewed profile contain many pitfalls \citep[see details in][]{2003cdsf.conf..157E}, especially when spectra obtained by single-dish observations with resolution $>$ 10\arcsec.

Detecting redshifted absorption or inverse P-Cygni profiles towards background continuum sources turns out to be a more straightforward and reliable method to identify infall motions associated with high-mass young stellar objects. Such redshifted absorption has already been detected in lines from a few molecules (e.g. NH$_{3}$, \citealt{1987ApJ...318..712K,1988ApJ...324..920K,1996ApJ...472..742H,1997ApJ...488..241Z}; HCO$^{+}$, \citealt{1987Sci...238.1550W}; CS, \citealt{1998ApJ...494..636Z}; H$_2$CO, \citealt{2001ApJ...562..770D}) against strong radio- and millimetre-wavelength continuum emission. However, to date, such absorption has only been found in a couple of sources with interferometric observations (with typical angular resolutions of $<$10$\arcsec$). Redshifted absorption in sub-millimetre-wavelength rotational lines from NH$_{3}$ and H$_{2}$O against dust continuum emission has been detected in a number of objects \citep{2012A&A...542L..15W,2016A&A...585A.149W,2019A&A...625A.103V}. However, these transitions are not easily observed using ground-based telescopes because their frequencies are largely blocked by the Earth’s atmospheric absorption. Furthermore, these high frequency measurements by Herschel and SOFIA are rather expensive and only a very limited number of sources have been studied so far, leaving extensive searches of these redshifted absorption features towards a larger sample of sources unfeasible. Therefore, salient tracers that are more accessible by ground-based telescopes can help to gauge reliable infall motions in a much larger sample of high-mass star formation regions. 

\begin{figure*}[!htbp]
\centering
\includegraphics[width=0.45\textwidth]{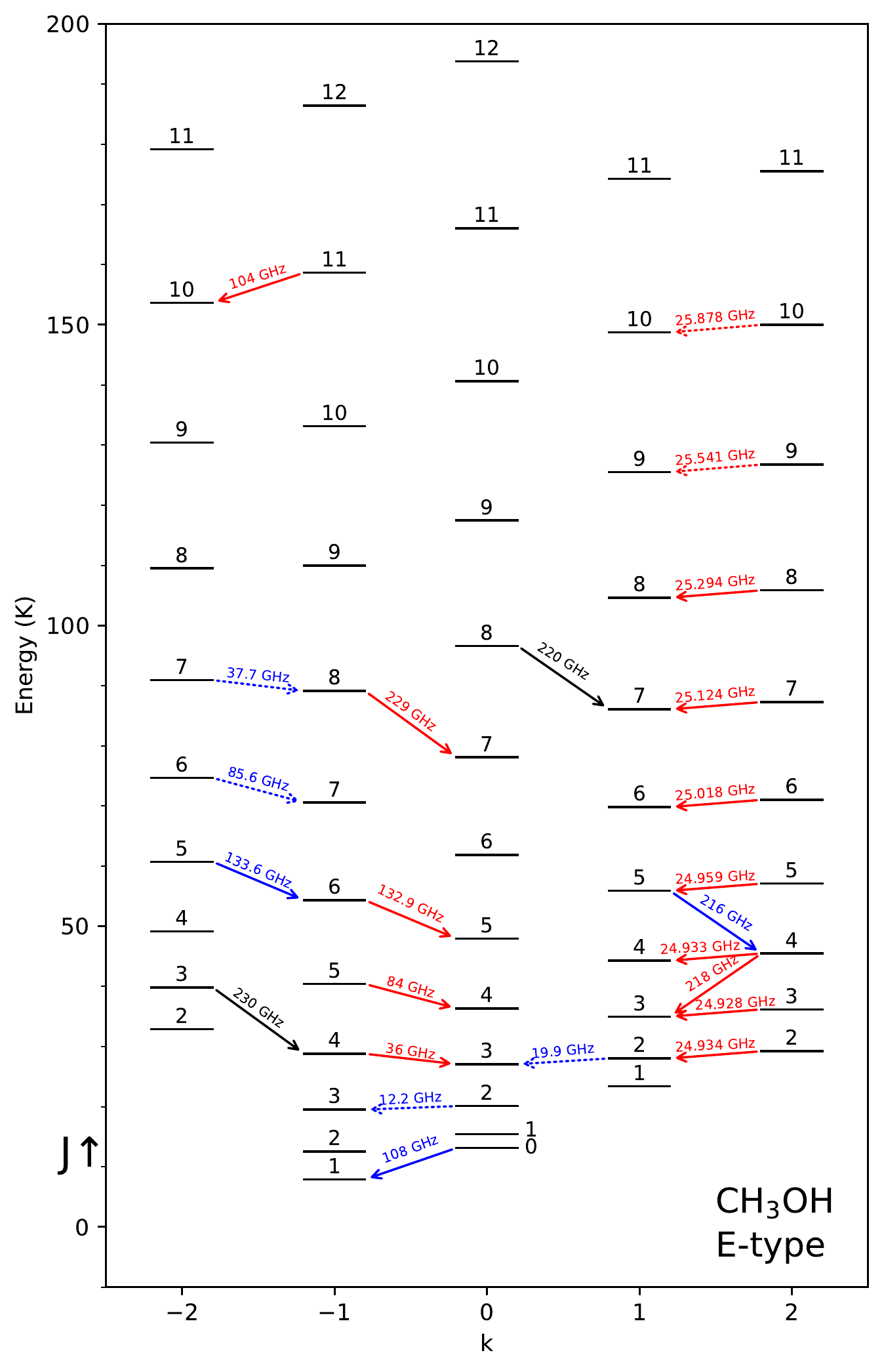}
\includegraphics[width=0.45\textwidth]{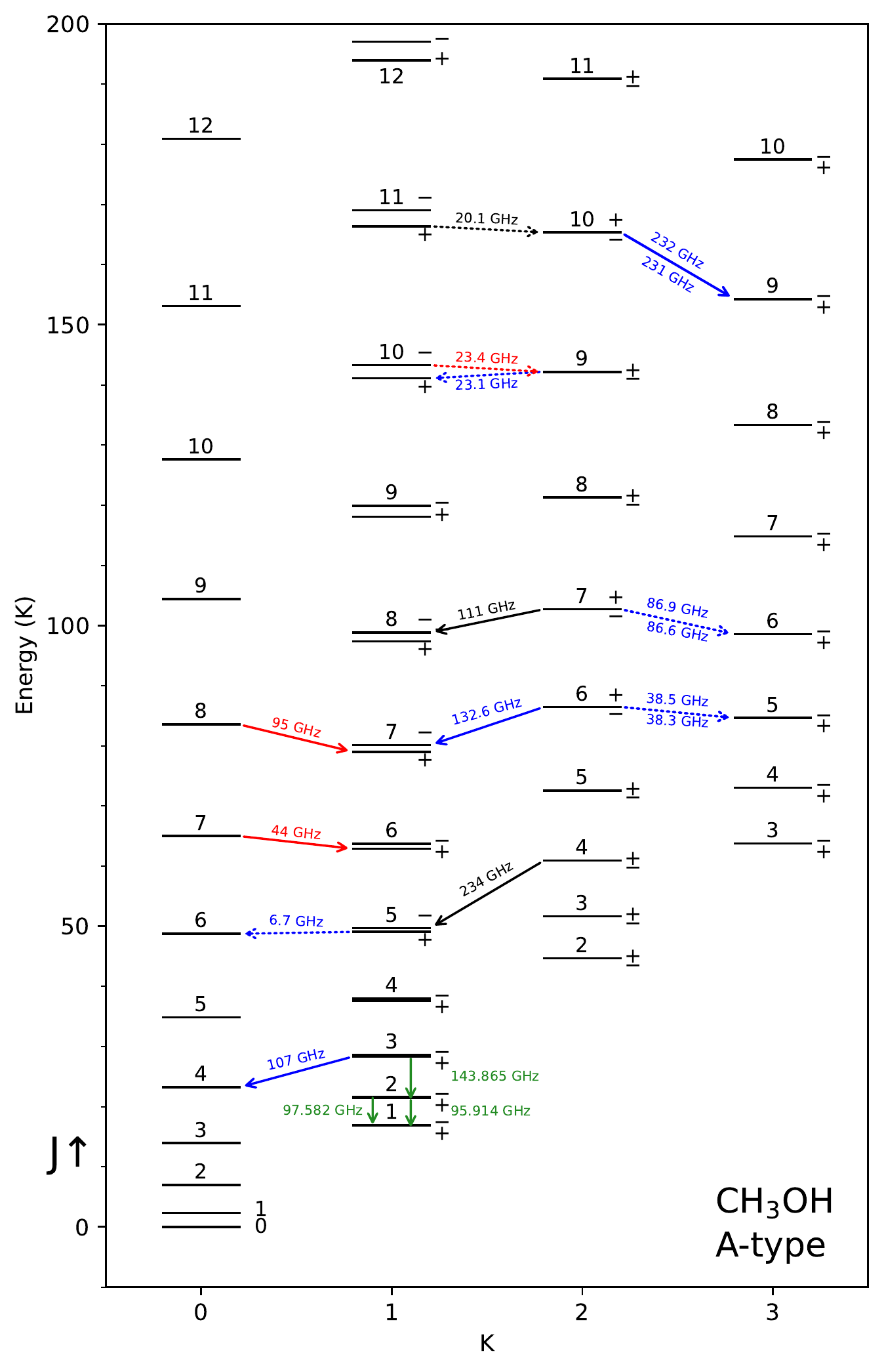}
\caption{Energy level diagram of E-type (left) and A-type  (right) methanol transitions. 
Transitions from \cite{1999ApJS..123..515S} belonging to the line series where absorption features are detected in W31C are also marked, at frequencies around 133~GHz. 
Known class I and II methanol masers transitions are marked with red and blue arrows, respectively. 
The black arrows represent the transitions that have no maser detection yet. The dotted arrows are transitions with detected absorption features in W31C. 
The green arrows indicate the thermal transitions of the $(J+1)_1-J_1A^{\pm}$ series observed in W3(OH). \label{fig:energy}}
\end{figure*}

%Methanol 
%\subsection{Interstellar methanol -- emission and absorption}\label{emiabs}
The methanol (CH$_3$OH) molecule has numerous transitions, including many maser lines, within the centimetre- and millimetre-wavelength ranges and hence provides a powerful tool to investigate star-forming regions. 
Early studies divided CH$_3$OH masers into two categories \citep{1987Natur.326...49B,1991ApJ...380L..75M} based on their different observational properties and pumping mechanisms.
Class~II CH$_3$OH masers are found in close proximity to infrared sources, OH masers, and ultracompact H{\sc ii} (UCH{\sc ii}) regions, and radiative pumping is believed to be responsible for their pumping \citep[e.g.][]{1991ASPC...16..119M,2005MNRAS.360..533C}.
In contrast, class~I CH$_3$OH masers %offset from those regions,
are thought to trace shocked regions and are produced by collisional pumping \citep[e.g.][]{2016A&A...592A..31L}. Experimentally 
determined rate coefficients for collisions between CH$_3$OH and H$_2$ or helium \citep{1974CaJPh..52.2250L,2010MNRAS.406...95R} have shown a tendency for the final state of a collisional excitation process of a CH$_3$OH molecule to have the same $k$-quantum number 
as the initial state, that is a propensity for $\Delta k$ = 0 collisions. This leads, for $E-$type CH$_3$OH\footnote{Methanol exists in two separate (non-interconverting) symmetry species: $E$-type and $A$-type CH$_3$OH. Its energy levels are described by the total angular momentum quantum number $J$ and its projection on the near symmetry (O--C) axis, which for the double degenerate $E-$type levels is lower case $k$ (with $-J\leq k \leq +J$). This leads to ladders of levels that have the same $k$, but (bottom to top) an increasing value of $J$. For $A-$type CH$_3$OH, upper case $K$ is used for the latter quantum number (with $0 \leq J$) and, except for the $K=0$ ladder,  the levels are split in pairs with opposite parity.}, to an overpopulation of the energy levels in the 
$k= -1$ ladder relative to the levels in the $k=0$ (or the $k= -2$) ladder that they can radiatively decay to (see Fig.~\ref{fig:energy}). 
For transitions originating in the  $k = -1$ ladder that have their upper energy 
levels in this ladder, that is, those with $J_{\rm upper} > 4$ (line numbers 3--8 in Table~\ref{Tab:freq}), 
this results in maser emission, while for $3, 2 $ and 1, it leads to enhanced 
absorption (anti-inversion or overcooling), in particular for the 12.2 GHz 
$2_{-1}-3_0$E line (line number 2). An analogous situation attains for $A-$type CH$_3$OH, for 
which the levels in the $K = 0$ ladder are overpopulated relative to those in the $K = 1$ ladder (see Fig.~\ref{fig:energy}), causing maser action for levels with  $J_{\rm upper} > 7$ (line numbers 27 and 28)  and enhanced absorption in the 6.7 GHz $5_1-6_0A^+$ (line number 26) transition. In the presence of a strong mid-infrared field, radiative pumping to torsionally excited states disturbs this simple picture and causes class II methanol maser action \citep{Sobolev1994, sobolev1997}. We note that the two strongest class II CH$_3$OH lines show, as explained above, absorption in regions that are conducive to class I maser excitation, which are often significantly removed from centres of activity marked by strong compact radio or IR sources \citep[see, e.g.][]{1986A&A...157..318M}.
The actual observational picture can be more complex, since in a single dish telescope beam, the signals from different regions can be merged, which is, among others, the case for the 12.2 GHz line in W31C, which shows absorption in addition to maser emission (see Sec.~\ref{Sec:thermal}).  These effects successfully explain the observed bright class I methanol masers \citep[e.g.][]{2016A&A...592A..31L} and overcooling, which enhances the absorption lines' detectability, in particular when strong background continuum emission is present, although even absorption against the cosmic microwave background radiation is observed in the 12.2 GHz line \citep{1988A&A...197..271W} and even the 6.7 GHz line, whose lower energy level is 49 K above the ground \citep{2008A&A...489.1175P}. % and being observed more easily. 
Absorption in the 6.7~GHz and 12.2~GHz CH$_3$OH transitions is indeed detected towards several sources, several of which host  I methanol masers \citep[e.g. NGC2264 and DR21-W;][]{1991ASPC...16..119M,2021A&A...651A..87O}, dark clouds \citep[TMC1 and L183;][]{1988A&A...197..271W}, hot corinos \citep[NGC1333;][]{2008A&A...489.1175P}, molecular clouds and H{\sc ii} region complexes \citep{1992MNRAS.254..301P}, and the Galactic Centre regions \citep[Sgr B2, Sgr A and G1.6$-$0.025;][]{1988MNRAS.235..655W,1989PASA....8..204P,1989MNRAS.239..677W,1995MNRAS.273.1033H}.
These facts suggest that methanol transitions that show absorption could have potential to aid in studying infall motions in high-mass star formation regions.

%W31C intro
%\subsection{The targets: high mass star forming regions: W31C and W3(OH)}\label{W3OH}
Two high-mass star formation regions, W31C and W3(OH), have been selected for this work. W31C, also known as G10.6$-$0.4, is an UCH{\sc ii} region located at a distance of 4.95$^{+0.51}_{-0.43}$~kpc \citep{2014ApJ...781..108S}. Based on previous dust spectral energy distribution (SED) fitting, % measurements, 
this region is found to have a mass of $\sim$ 2$\times 10^{4}$~\Msun\,and a luminosity of $\sim$ 3$\times 10^{6}$~\Lsun\, \citep[e.g.][]{2016ApJ...828...32L}.
W31C shows evidence for infall motions in both molecular gas (e.g. NH$_3$, \citealt{1986ApJ...304..501H,1987ApJ...318..712K,1988ApJ...324..920K,2005ApJ...630..987S}; HCO$^+$, \citealt{2008ApJ...684.1273K}; high-$J$ CO,  \citealt{2006A&A...454L..95W}) and ionised gas (in the H66$\alpha$ radio recombination line, \citealt{2002ApJ...568..754K}), which is well modelled by the gravitational collapse of a centrally condensed and centrally heated core \citep[e.g.][]{1987ApJ...318..712K,2005ApJ...624L..49S}.
A series of high resolution observations present the overall picture of a hierarchically collapsing system for W31C \citep{2010ApJ...725.2190L,2010ApJ...722..262L,2011ApJ...729..100L,2012ApJ...745...61L,2017A&A...597A..70L}.

The W3(OH) complex is located at a distance of 1.95$\pm$0.04~kpc \citep{2006Sci...311...54X}, and consists of two high-mass objects W3(OH) and W3(H$_2$O).
W3(OH) is associated with an UCH{\sc ii} region ionised by an O7 star \citep[e.g.][]{1981ApJ...245..857D}. In contrast, 
W3(H$_2$O), located 6$\arcsec$ (0.06 pc) east of W3(OH) \citep[e.g.][]{2011ApJ...740L..19Z,2015ApJ...803...39Q} is in an earlier stage of massive star formation and hosts a (double) hot molecular core  \citep{Wyrowski1999, Ahmadi2018}.
On larger ($\leq 1$~ pc size) scales, both single-pointing spectral line observations and mapping  reveal infall motions in the W3(OH) complex \citep[e.g.][]{2003ApJ...592L..79W,2009MNRAS.392..170S,2010ApJS..188..313W}. 
\cite{1987ApJ...323L.117K} reported that the velocity, density and temperature structure in the W3(OH) complex is similar to that in W31C.
That a blue-skewed profile is observed in the HCN (3$-$2) line over a region of 110$\arcsec \times 110\arcsec$ ($1.1 \times 1.1$ pc), \citet{2016MNRAS.456.2681Q} further supports the notion that the W3(OH) complex is overall undergoing large-scale collapse like W31C.
These properties make W31C and W3(OH) excellent test beds to examine the suitability of other spectral lines for tracing infall.%to examine whether methanol transitions can be good tracers for infall motions in high-mass star formation regions or not. 

In this work, we present observations of multiple methanol transitions towards W31C and W3(OH) to test whether methanol transitions can trace infall motions.  These data were collected using the Effelsberg-100 m, the IRAM-30 m, and the Atacama Pathfinder Experiment (APEX) 12 metre telescopes. 
We note that both W3(OH) and W3(H$_2$O) are covered within the same beam of single-dish observations. Here we use W3(OH) to refer to the W3(OH) complex.
Our observations and data reductions are described in Sec.~\ref{Sec:obs}. Sec.~\ref{Sec:result} presents the results which are discussed in Sec.~\ref{Sec:discuss}. A summary of this work is presented in Sec.~\ref{Sec:sum}

%The dust and gas kinetic temperature of W31C are about 34 K and 100 K \citep{2017A&A...599A.139K,2018A&A...611A...6T}. The dust and gas kinetic temperature of W3OH are about 25 K and 100 K \citep{1993ApJS...89..123M,2013ApJ...766...85R}.

\section{Observations and data reduction}\label{Sec:obs}
We observed 37 CH$_3$OH transitions towards W31C with rest frequencies ranging from 6 GHz to 236 GHz.  The observations were undertaken between 2011 and 2021 using the Effelsberg-100 m, IRAM-30 m, and APEX-12 m telescopes. Information on the observed transitions is given  in Table~\ref{Tab:freq}, and these transitions are denoted by arrows in the methanol energy level diagram (see Fig.~\ref{fig:energy}). 
We adopt the most accurate available rest frequencies, most of which are taken from  \citet{2004A&A...428.1019M}. We note that a given  error in frequency will lead to a larger error in velocity for transitions with a lower rest frequency. For all transitions in Table \ref{Tab:freq}, the corresponding errors in velocity range between 0.002~\kms\,(at 25.294 GHz) and 0.392~\kms\,(at 38.3 GHz).
%It is important to note that for the same frequency accuracy, the corresponding velocity uncertainty will be larger at lower frequency. The velocity uncertainties of all transitions in Table \ref{Tab:freq} range between 0.002~\kms\,(at 25.294 GHz) and 0.392~\kms\,(at 38.3 GHz).
%In this table, w
We adopt the prescription of \citet{2015PASP..127..299S} to estimate the optically thin critical density for each methanol transition. The spectroscopic data for the CH$_3$OH lines are taken from the Leiden Atomic and Molecular Database (LAMDA\footnote{\url{https://home.strw.leidenuniv.nl/~moldata/}}) with a data file version of 2021 April 12. 
Because the gas kinetic temperatures of W31C and W3(OH) are about 100 K \citep{1993ApJS...89..123M,2018A&A...611A...6T}, the optically thin critical density is calculated by assuming a gas kinetic temperature of 100 K for this work. The optically thin critical densities are determined to be 2$\times 10^{4}$--6$\times 10^{6}$~cm$^{-3}$, spanning two orders of magnitude. Observational parameters are given in Table~\ref{Tab:obs}. Four CH$_3$OH transitions, those at 84, 95, 108 and 111 GHz, were targeted towards an ATLASGAL source (AGAL010.624$-$00.384, at $\alpha_{\rm J2000}$=18$^{\rm h}$10$^{\rm m}$28$\rlap{.}^{\rm s}$6, $\delta_{\rm J2000}$=$-$19$^\circ$55$\arcmin$46$\arcsec$) near W31C, the other 33 transitions were observed towards the UCH{\sc ii} region G010.6234$-$0.3837 at $\alpha_{\rm J2000}$=18$^{\rm h}$10$^{\rm m}$28$\rlap{.}^{\rm s}$70, $\delta_{\rm J2000}$=$-$19$^\circ$55$\arcmin$49$\rlap{.}\arcsec$8 \citep{2019MNRAS.482.2681Y,2021A&A...645A.110Y}.
The angular offset between the two positions is 4$\arcsec$ which is less than one fifth of the IRAM-30 m beam size at 3 mm. 

We carried out K-band (18--26 GHz) observations towards W3(OH) in 2021 May 13 using the Effelsberg-100 m.
The telescope was pointed towards the position, $\alpha_{\rm J2000}$=02$^{\rm h}$27$^{\rm m}$04$\rlap{.}^{\rm s}$38, $\delta_{\rm J2000}$=61$^\circ$52$\arcmin$20$\rlap{.}\arcsec$5,
used by \citep{1986A&A...157..318M}, who reported absorption in the 25 GHz $J_2-J_1 E$ CH$_3$OH lines. The methanol transitions at 95.914, 97.582 and 143.865 GHz were observed in 2019 June 11 using the IRAM-30 m. 
The telescope was pointed towards the position of the UCH{\sc ii} region W3(OH), that is $\alpha_{\rm J2000}$=02$^{\rm h}$27$^{\rm m}$03$\rlap{.}^{\rm s}$90, $\delta_{\rm J2000}$=61$^\circ$52$\arcmin$24$\rlap{.}\arcsec$0, which is about 5$\arcsec$ offset (nearly one seventh of the Effelsberg-100 m beam size at 25~GHz) from the position that we used in the K-band observations.

%Fig. \ref{fig:bg} gives an overview of our targets with the observed positions indicated. 
Given that the offsets between different positions are much smaller than the beam sizes of our observations, the position offsets are neglected in the following analysis. Details of the observations and data reduction for each telescope are described below. 

%\begin{figure*}[!htbp]
%\centering
%\includegraphics[width=0.43\textwidth]{w31c_wise.eps}
%\includegraphics[width=0.43\textwidth]{w3oh_wise.eps}
%\caption{Three-color composite images of our observing targets W31C (left) and W3(OH) (right) where the red, green, and blue colors represent the WISE 22, 12, and 4.6~$\mu$m emission, respectively. In the left panel, the red %pentagram 
%star marks the position $\alpha_{\rm J2000}$=18$^{\rm h}$10$^{\rm m}$28$\rlap{.}^{\rm s}$70, $\delta_{\rm J2000}$=$-$19$^\circ$55$\arcmin$49$\rlap{.}\arcsec$8,towards which the data for a majority of the methanol transitions were obtained, and the big red circle indicates a typical beam size of 30$\arcsec$.
%The red hollow triangle represents the position of AGAL010.624$-$00.384 at $\alpha_{\rm J2000}$=18$^{\rm h}$10$^{\rm m}$28$\rlap{.}^{\rm s}$6, $\delta_{\rm J2000}$=$-$19$^\circ$55$\arcmin$46$\arcsec$. 
%In the right panel, the red star represents mark $\alpha_{\rm J2000}$=02$^{\rm h}$27$^{\rm m}$04$\rlap{.}^{\rm s}$38, $\delta_{\rm J2000}$=61$^\circ$52$\arcmin$20$\rlap{.}\arcsec$5, towards which our K-band (18--26 GHz) observations were conducted, and the big red circle indicates a typical beam size of 30$\arcsec$. The red hollow triangle marks $\alpha_{\rm J2000}$=02$^{\rm h}$27$^{\rm m}$03$\rlap{.}^{\rm s}$90, $\delta_{\rm J2000}$=61$^\circ$52$\arcmin$24$\rlap{.}\arcsec$0, towards which our 3 mm observations were pointed.
%\label{fig:bg}}
%\end{figure*}

\subsection{Effelsberg-100m observations}
Using the dual-polarisation S7mm Double Beam RX secondary receiver on the Effelsberg-100 m telescope\footnote{The 100-m telescope at Effelsberg is operated by the Max-Planck Institut f{\"u}r Radioastronomie (MPIfR) on behalf of the Max-Planck
Gesellschaft (MPG).}, we conducted Q band observations towards W31C ($\alpha_{\rm J2000}$=18$^{\rm h}$10$^{\rm m}$28$\rlap{.}^{\rm s}$70, $\delta_{\rm J2000}$=$-$19$^\circ$55$\arcmin$49$\rlap{.}\arcsec$8) to observe methanol transitions ranging from 34.5 to 39.5 GHz on 2020 September 14 and from 40 to 45 GHz on 2020 May 30 (project id: 77-19). Towards W31C, the 6.7 GHz and 12.2 GHz methanol transitions were observed on 2021 January 5 using the S45mm and S20mm secondary focus receivers, while K-band observations (18 -- 26 GHz) were performed towards both W3(OH) ($\alpha_{\rm J2000}$=02$^{\rm h}$27$^{\rm m}$04$\rlap{.}^{\rm s}$38, $\delta_{\rm J2000}$=61$^\circ$52$\arcmin$20$\rlap{.}\arcsec$5) and W31C on 2021 May 13 and 14 (project id: 17-21). All these observations were made in the position switching mode with an off-position offset of 10$\arcmin$ in right ascension. Fast Fourier Transform Spectrometers (FFTSs) were used as the backend to record signals. For the 6.7 GHz and 12.2 GHz methanol transitions, we used high spectral resolution FFTSs which provide a channel separation of 3.1~kHz and 4.6~kHz at 6.7~GHz and 12.2~GHz, corresponding to a velocity spacing of 0.14~km~s$^{-1}$ and 0.11~km~s$^{-1}$, respectively. For other methanol transitions, we adopted broad-bandwidth FFTSs which cover an instantaneous bandwidth of 2.5 GHz with 65536 channels, yielding a channel separation of 38.1 kHz and a corresponding velocity spacing of 0.33~\kms at 35 GHz (multiply by 1.16 to convert to velocity resolution, see \citealt{2012A&A...542L...3K}).

%The Full width at half maximum (FWHM) beam size of the Effelsberg 100m ranges from 25\arcsec\,to 20\arcsec.
NGC 7027 and 3C 286 were used for pointing, focus and flux density calibration, and the pointing uncertainty was better than 10$\arcsec$, 5$\arcsec$, 5$\arcsec$, and 10$\arcsec$ at 6.7 GHz, 12.2 GHz, 24 GHz, and 40 GHz, respectively. The observed methanol transitions, corresponding observation dates, full width at half maximum (FWHM) beam sizes, velocity resolutions, the 1$\sigma$ rms noise levels at the corresponding velocity resolutions, main beam efficiencies, $\eta_{\rm mb}$, and the scaling factors to convert antenna temperature, $T_{\rm A}^*$, to flux density, $S_{\nu}$, are listed in Table \ref{Tab:obs}.

%The data were processed using the GILDAS/CLASS package \citep{2005sf2a.conf..721P,2013ascl.soft05010G}.
%For the observations performed with the double beams receiver at 7~mm, we only analyzed the data for the beam that tracked the targeted position.
%Because radio frequency interferences caused a poor baseline for the 12.2 GHz spectrum, a fourth order polynomial baseline subtraction was performed on the 12.2 GHz spectrum within a velocity coverage of 60~km~s$^{-1}$ around the systemic velocity. A low-order ($<$4) polynomial baseline was removed from the other spectra.

\subsection{IRAM-30 m observations}
We observed multiple methanol transitions within the frequency ranges of 84.6--92.4~GHz and 100.3--108.1~GHz towards W31C ($\alpha_{\rm J2000}$=18$^{\rm h}$10$^{\rm m}$28$\rlap{.}^{\rm s}$70, $\delta_{\rm J2000}$=$-$19$^\circ$55$\arcmin$49$\rlap{.}\arcsec$8) using the Eight MIxer Receiver \citep[EMIR,][]{2012A&A...538A..89C} of the IRAM 30m telescope on 2019 June 29 (project id: 043-19). The methanol transitions at 95.914, 97.582 and 143.865 GHz were observed towards W3(OH) ($\alpha_{\rm J2000}$=02$^{\rm h}$27$^{\rm m}$03$\rlap{.}^{\rm s}$90, $\delta_{\rm J2000}$=61$^\circ$52$\arcmin$24$\rlap{.}\arcsec$0) on 2019 June 11 (project id: 045-19) using the EMIR receivers (E090 and E150). The dual-polarisation EMIR receivers provide nearly 8 GHz bandwidth per sideband, which is connected to two 4~GHz wide FFTSs. These FFTSs provide a channel width of $\sim$195 kHz, resulting in a velocity spacing of $\sim$0.6 km s$^{-1}$ at 95~GHz. 
The wobbler switching mode, through small movements of the sub-reflector to change the optical path from antenna to the receiver, was used for pointing towards the source with a throw of 120$\arcsec$ in azimuth. Additional methanol line data at 84, 95, 108 and 111 GHz towards W31C were taken from an unbiased line survey towards ATLASGAL sources at $\alpha_{\rm J2000}$=18$^{\rm h}$10$^{\rm m}$28$\rlap{.}^{\rm s}$6, $\delta_{\rm J2000}$=$-$19$^\circ$55$\arcmin$46$\arcsec$. The observations were carried out during 2011 April 8 to 11 (see \citealt{2016A&A...586A.149C} for details, project id: 181-10). The observational parameters including the main beam efficiency, the scaling factor, and the 1$\sigma$ rms noise level of each transition are listed in Table~\ref{Tab:obs}.

\subsection{APEX-12 m observations}
We observed methanol transitions ranging from 209.2 to 220.9 GHz and from 225 to 236.7 GHz towards W31C ($\alpha_{\rm J2000}$=18$^{\rm h}$10$^{\rm m}$28$\rlap{.}^{\rm s}$70, $\delta_{\rm J2000}$=$-$19$^\circ$55$\arcmin$49$\rlap{.}\arcsec$8) with the PI230 receiver at the APEX\footnote{This publication is based on data acquired with the Atacama Pathfinder Experiment 12 m sub-millimetre telescope (APEX). APEX is a collaboration between the Max-Planck-Institut f{\"u}r Radioastronomie, the European Southern Observatory, and the Onsala Space Observatory.} telescope \citep{2006A&A...454L..13G} on 2019 June 1 and June 9 (project id: 0103.F-9516A).
The PI230 receiver\footnote{\url{https://www.eso.org/public/teles-instr/apex/pi230/}} is a dual polarisation and sideband-separating heterodyne system that covers a frequency range of 200$-$270 GHz.
The FFTS backend provides 65536 channels over 4 GHz bandwidth, and an overlap of 0.2 GHz between the two backends results in a total 7.8 GHz coverage for each sideband and polarisation per tuning. 
The wobbler switching mode with a throw of 120$\arcsec$ in azimuth was used for our observations.
The pointing was accurate to $\lesssim$2$\arcsec$.
The main beam efficiency, the scaling factor, and the rms noise level after an on-source integration time of 5 minutes are listed in Table \ref{Tab:obs}.
%The APEX data were also processed with the GILDAS/CLASS package. A linear baseline subtraction and Hanning smoothing were performed on the spectra within a velocity range of 80~\kms\,around the systemic velocity of observed methanol transitions. Velocities are given with respect to the local standard of rest (LSR) throughout this work. 
%(the velocity resolution between 0.098 km/s to 0.090 km/s before smoothing)

\subsection{Data reduction}
The CLASS programme that is part of the GILDAS package \citep{2005sf2a.conf..721P,2013ascl.soft05010G} was used for processing all of our data.
For the observations performed with the 7~mm dual-beam receiver, we only analysed the data for the beam that tracked the selected targeted positions.
Because radio frequency interference caused a poor baseline for  12.2 GHz spectrum (for W31C), a fourth order polynomial baseline was subtracted % was performed on the 12.2 GHz spectrum 
within the baseline parameters determined over a velocity range of 60~\kms\, around the systemic velocity. A low-order ($<$4) polynomial baseline was subtracted from the other spectra.
To improve the signal-to-noise ratios (S/Ns), Hanning smoothing was performed on the spectra obtained with the APEX-12~m telescope. %, within a velocity range of 80~\kms\,around the systemic velocity of observed methanol transitions. 
Velocities are given with respect to the local standard of rest (LSR) throughout this work.

\section{Results}\label{Sec:result}
Our observations resulted in the detection of 37 methanol transitions in W31C and 16 methanol transitions in W3(OH). 
%\textbf{\sout{The spectroscopic information for the observed  transitions is summarized in Table~\ref{Tab:freq}, and these transitions are denoted by arrows in the methanol energy level diagram (see Fig.~\ref{fig:energy}).}} 
The observational results for the two sources are given separately in the following.  

%In both W31C and W3(OH), we mainly focused on the known CH$_3$OH maser transitions and the CH$_3$OH lines belong to the series where we detected absorption (see Figure~\ref{fig:energy} and Table~\ref{Tab:freq}). 

\subsection{Detection of 14 redshifted methanol absorption lines towards W31C} \label{Sec:thermal}

\begin{figure*}[!htbp]
\centering
\includegraphics[width=0.9\textwidth]{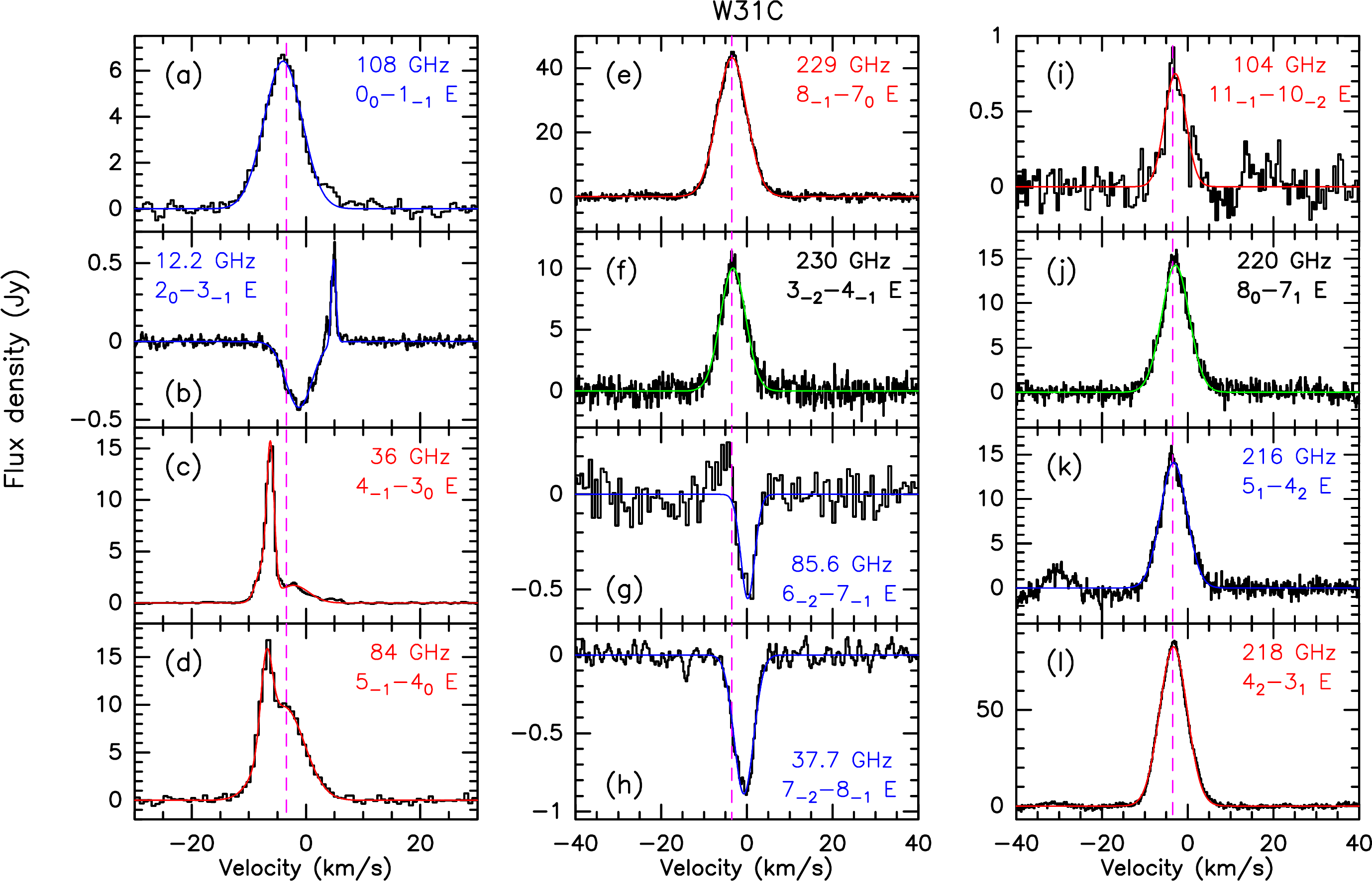}
\caption{Spectra of observed $E$-type CH$_{3}$OH transitions towards W31C, excluding those in K-band (18 -- 26 GHz), which are collected in Fig.~\ref{fig:spec-k}. 
The panels from (a) to (l) are arranged following the order in Table~\ref{Tab:freq}.
The quantum numbers and rest frequencies of these transitions are labelled in their respective panels. 
Class I and II CH$_3$OH maser transitions are labelled in red and blue, respectively, and their Gaussian fitting results are plotted in the corresponding colour.
The transitions that have no maser detection yet are labelled in black, and the Gaussian fitting results are plotted in green. In panel (k), the $-$30~\kms~component is caused by the blend of four CH$_{3}$OCHO transitions (i.e. $20_{0,20}-19_{0,19}\,E$, $20_{0,20}-19_{0,19}\,A$, $20_{1,20}-19_{1,19}\,E$, and $20_{1,20}-19_{1,19}\,A$).
The vertical magenta dashed lines represent the systemic velocity of $V_{\rm LSR}$=$-$3.43~\kms. \label{fig:spec-e}}
\end{figure*}

\begin{figure*}[!htbp]
\centering
\includegraphics[width=0.9\textwidth]{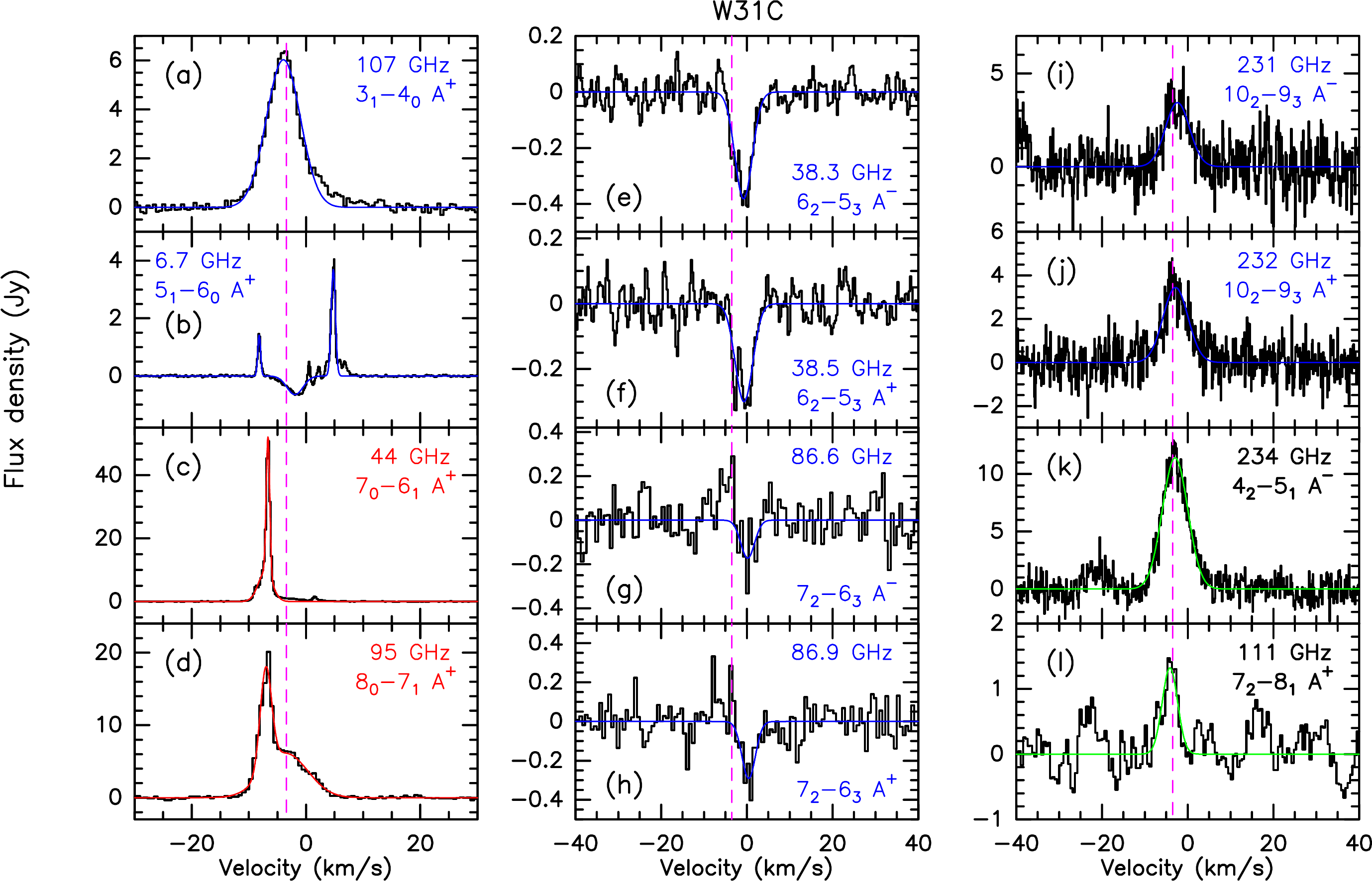}
\caption{Spectra of observed $A$-type CH$_{3}$OH transitions towards W31C, excluding those from K-band (18--26 GHz), which are collected in Fig.~\ref{fig:spec-k}.
The panels from (a) to (l) are arranged following the order in Table~\ref{Tab:freq}.
The quantum numbers and rest frequencies of transitions are labelled in their respective panels. 
Class I and II CH$_3$OH maser transitions are labelled in red and blue, respectively, and their Gaussian fitting results are plotted in the corresponding colour.
The transitions that have no maser detection yet are labelled in black, and the Gaussian fitting results are plotted in green. In panel (k), the $-$18~\kms~component arises from CH$_{3}$OH ($5_{-4} - 6_{-3}\,E$).
The vertical magenta dashed lines represent the systemic velocity of $V_{\rm LSR}$=$-$3.43~\kms. \label{fig:spec-a}}
\end{figure*}

\begin{figure*}[!htbp]
\centering
\includegraphics[width=0.9\textwidth]{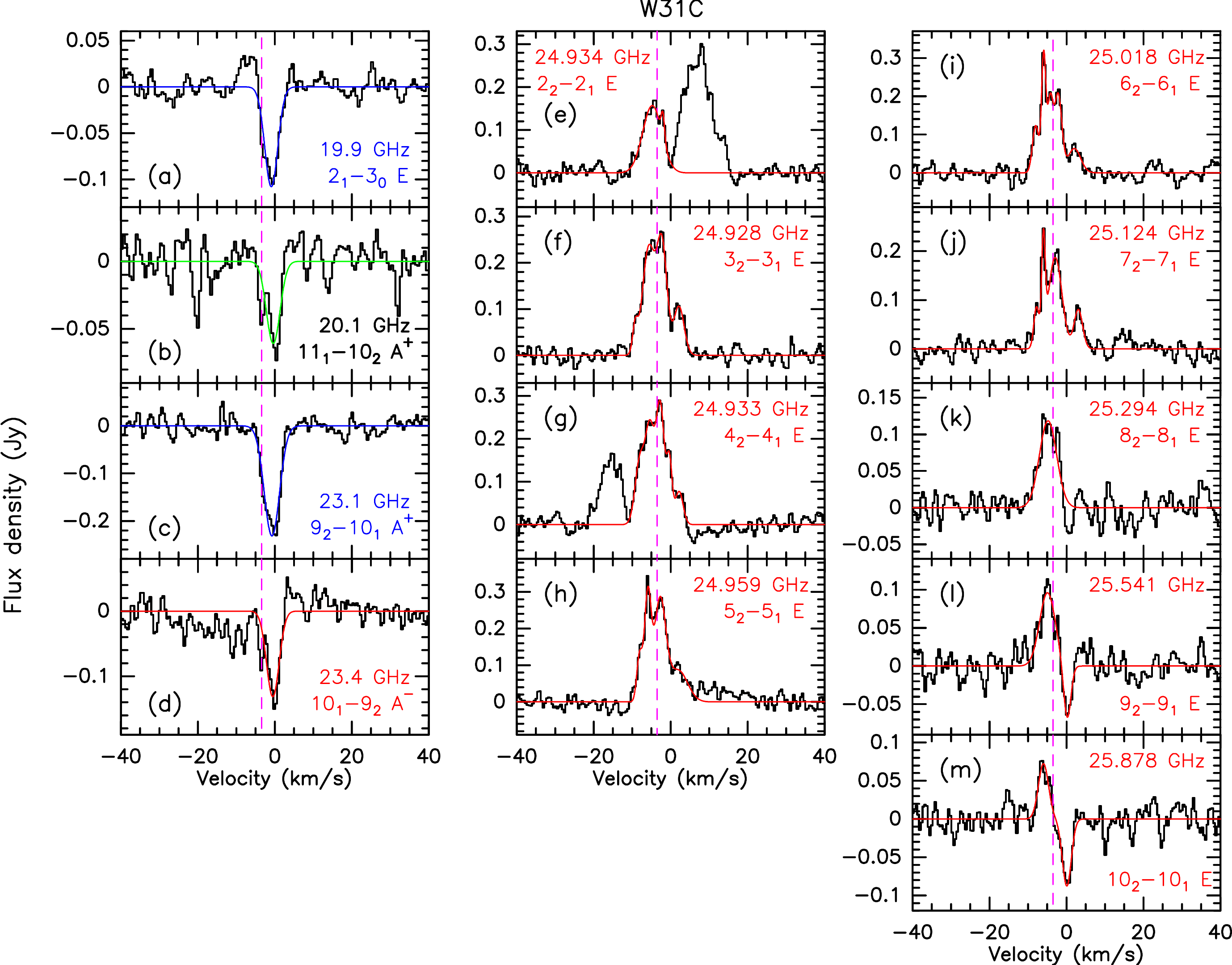}
\caption{Observed spectra of CH$_{3}$OH transitions in K-band (18 -- 26 GHz) towards W31C. The panels from (a) to (m) are arranged based on increasing frequency or increasing $J$ for the $J_2-J_1E$ line series. The quantum numbers and rest frequencies of transitions are labelled in their respective panels.
Class I and II CH$_3$OH maser transitions are labelled in red and blue, respectively, and their Gaussian fitting results are plotted in the corresponding colour.
The transitions that have no maser detection yet are labelled in black, and the Gaussian fitting results are plotted in green.
The vertical magenta dashed lines represent the systemic velocity of $V_{\rm LSR}$=$-$3.43~\kms.
\label{fig:spec-k}}
\end{figure*}

Figures~\ref{fig:spec-e}--\ref{fig:spec-k} present the observed spectra of 37 CH$_{3}$OH transitions observed towards W31C, showing diverse line shapes. Gaussian fitting was performed to characterise the spectral features of each transition. Single-peaked spectra observed in emission are fitted with a single Gaussian component. 

We %also 
used a single Gaussian component to fit the absorption signal %component 
for the transitions at 19.9, 20.1, 23.4, 37.7, 38.3, 38.5, 85.6, 86.6 and 86.9 GHz, in order to %better depict 
clearly characterise the absorption feature. For other spectra with multiple emission peaks and absorption dips, we employed multiple Gaussian components for the fitting. 
The results of the fitting are listed in Table~\ref{Tab:result}.

\begin{figure*}[!htbp]
%\centering
\begin{minipage}[c]{0.65\textwidth}
\includegraphics[width=1\textwidth]{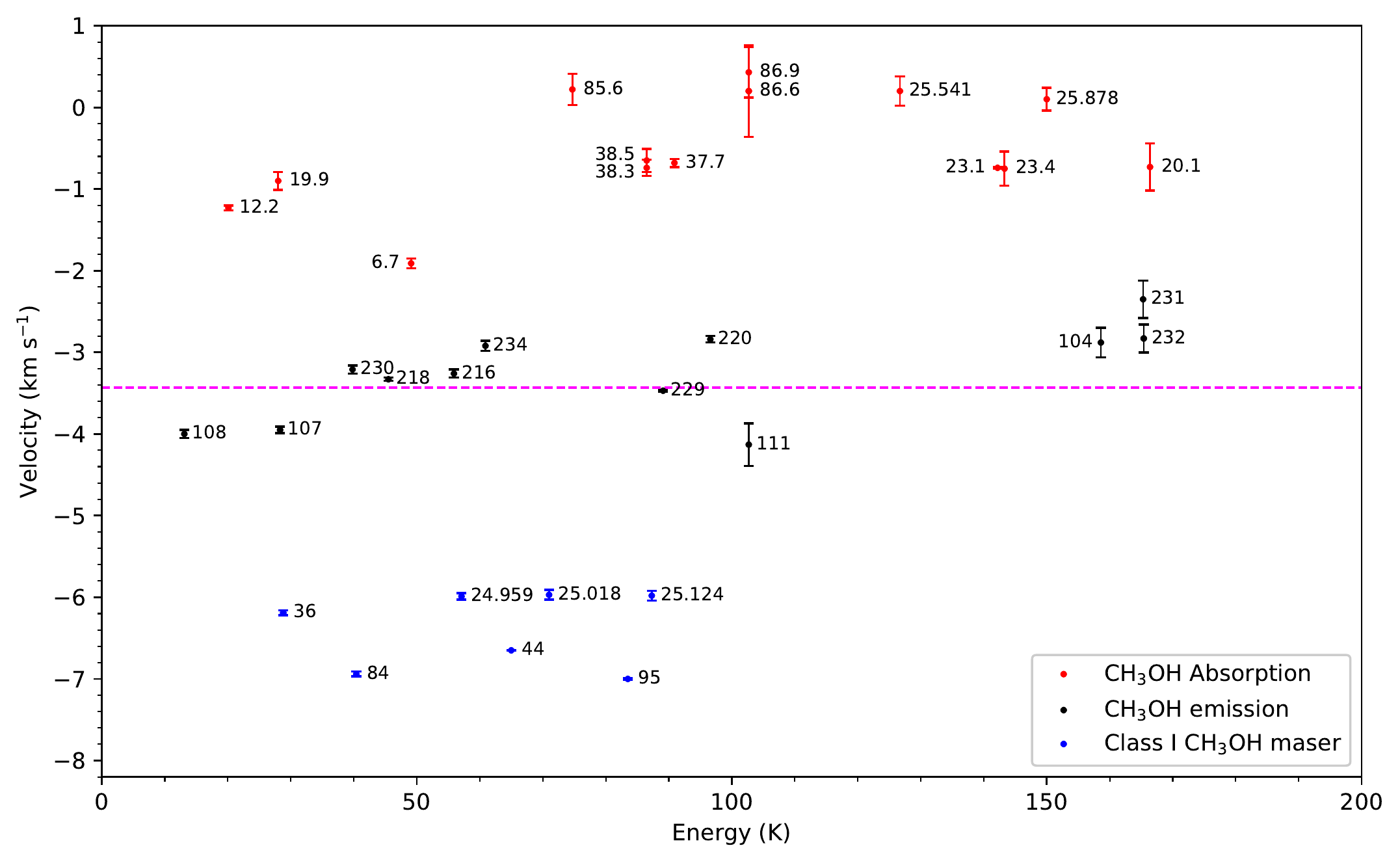}
\includegraphics[width=1\textwidth]{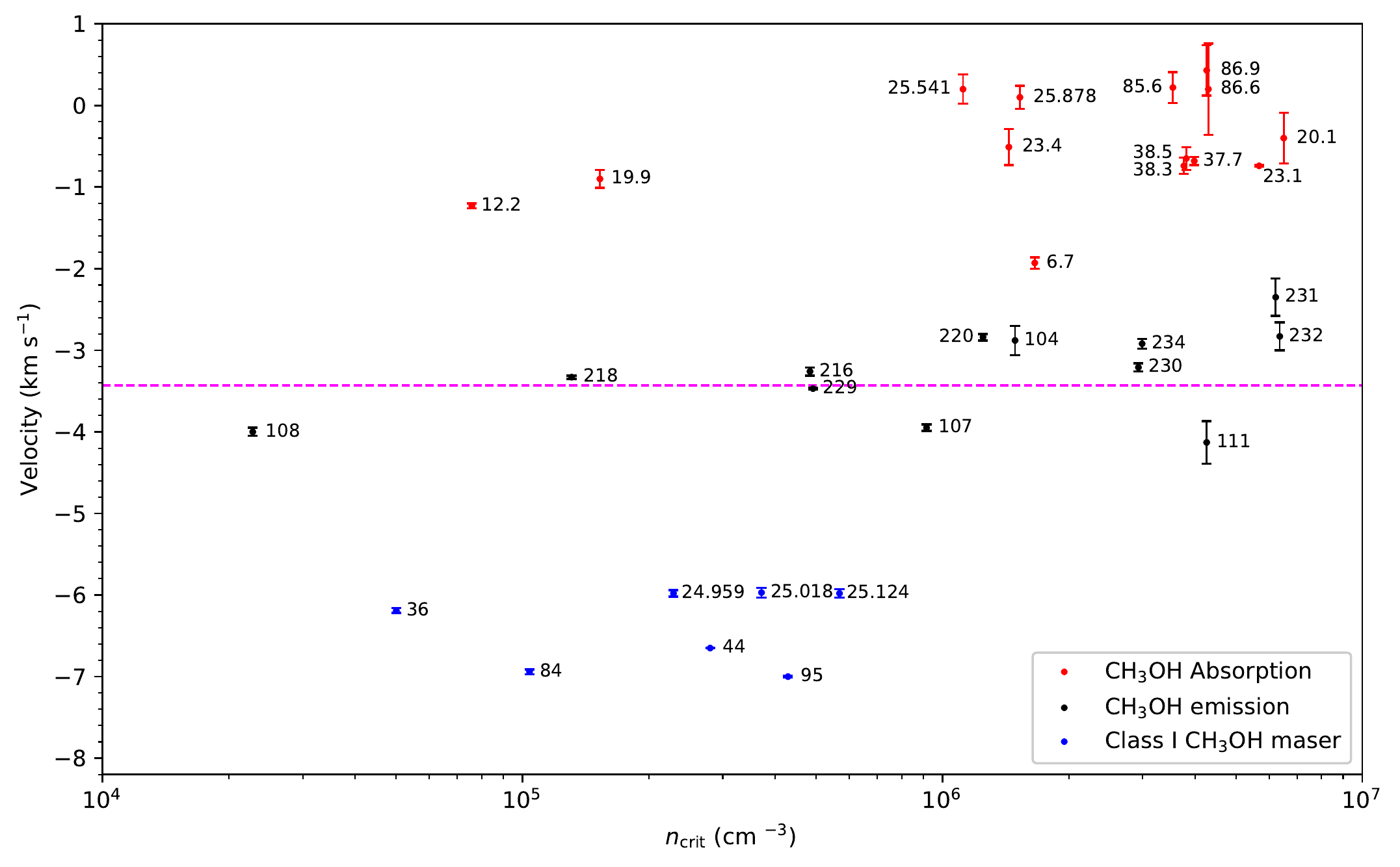}
\end{minipage}\hfill
\begin{minipage}[c]{0.3\textwidth}
\caption{LSR velocity as a function of the upper energy level (upper panel) and critical density (lower panel) for W31C.
%Upper panel: Line velocity as a function of the upper energy level for W31C. Lower panel: Line velocity as a function of the critical density for W31C. 
The red and blue dots indicate the methanol lines with absorption features and maser features, respectively.
The black dots indicate the methanol emission without a narrow and strong maser emission feature.
We only show the $J_2-J_1E$ lines with absorption features or maser features in the panels.
Their velocities are derived from Gaussian fitting to the emission peaks or absorption dips.  
The error bars indicate Gaussian fitting errors.
The horizontal magenta dashed line represents the systemic velocity of $V_{\rm LSR}$=$-$3.43~\kms.
\label{fig:Vel-E}}
\end{minipage}
\end{figure*}

Among the observed spectra, we find 12 methanol transitions showing single-peaked emission profiles. These are the 104, 107, 108, 111, 216, 218, 220, 229, 230, 231, 232, and 234 GHz lines. Although the 104, 107, 108, 216, 218, 229, 231 and 232~GHz lines have been found to show maser action towards other sources \citep[e.g.][]{1999MNRAS.310.1077V,2002ARep...46...49S,2006ARep...50..965S,2006MNRAS.373..411V,2012ApJ...759L...5E,2014ApJ...788..187H}, the observed broad widths of the lines are similar to the widths of lines from other species, which indicates that they are dominated by thermally excited emission towards W31C. Therefore, we use the fitted velocities of these 12 transitions to derive the systemic velocity of W31C. Adopting inverse-variance weighting, we use the weighted mean velocity as the systemic velocity, and it is determined to be $V_{\rm LSR}$=$-$3.43$\pm$0.44~\kms. The systemic velocity we adopt is consistent with the value of $\sim -$3~\kms\,used in previous studies \citep[e.g.][]{1990ApJ...355..190K,2010ApJ...725.2190L,2020A&A...644A.160K}. In Figure~\ref{fig:Vel-E}, it can be seen that the velocity centroids of these 12 methanol transitions become more redshifted with increasing energy level and critical density, which indicates that there is a velocity gradient from the outer envelope to the centre.  

Absorption features are detected in 14 transitions, the 6.7, 12.2, 19.9, 20.1, 23.1, 23.4, 25.541, 25.878, 37.7, 38.3, 38.5, 85.6, 86.6 and 86.9 GHz lines (see Column~3 in Table~\ref{Tab:freq}). 
Absorption in five of these transitions (6.7, 12.2, 19.9, 23.1 and 23.4~GHz) towards W31C has been discussed in previous studies \citep{1984A&A...134L...7W,1985A&A...147L..19W,1991ApJ...380L..75M,1992MNRAS.254..301P,1995MNRAS.272...96C,2010MNRAS.401.2219B,2014MNRAS.438.3368B}.
The remaining nine absorption profiles at 20.1, 25.541, 25.878, 37.7, 38.3, 38.5, 85.6, 86.6 and 86.9~GHz are reported here for the first time. Furthermore, the CH$_{3}$OH transitions at 19.9, 25.541, 25.878 and 85.6 GHz show a typical inverse P-Cygni profile.
The absorption features have a large scatter in their LSR velocities, ranging from $-$1.91~\kms\,to 0.43~\kms\,(see Table~\ref{Tab:result} and Fig.~\ref{fig:Vel-E}). Nevertheless, all 14 absorption features are redshifted with respect to the systemic velocity ($V_{\rm LSR}$=$-$3.43$\pm$0.44~\kms, see also Fig.~\ref{fig:Vel-E}). 

%while the class I CH$_{3}$OH masers are blueshifted to the systemic velocity (see Figure~\ref{fig:Vel-E}).

Both the $5_1-6_0A^+$ and $2_0-3_{-1}E$ lines at 6.7 (Fig.~\ref{fig:spec-a}b) and 12.2~GHz (Fig.~\ref{fig:spec-e}b) show broad absorption and maser features. 
The absorption features in these two transitions lie in the velocity range $-$2 to $-$1 \kms\, and are therefore redshifted with respect to the systemic velocity of the source.
The line width of the absorption feature at 12.2~GHz is the broadest we have detected for absorption towards W31C. The 6.7 GHz spectral profile shows absorption sandwiched between two maser features, which is consistent with previous observations \citep{1991ApJ...380L..75M,1995MNRAS.272...96C}.
\cite{2010MNRAS.401.2219B,2014MNRAS.438.3368B} reported the detection of a 12.2~GHz maser. In their spectra obtained in 2008, the intensity of the absorption feature is about $-$0.5~Jy, similar to our result. On the other hand, the peak flux density of the maser feature at 4.6~\kms\/ is 1.4~Jy, which is nearly three times our value. This indicates that the 12.2~GHz maser flux density has declined over the past 13 years.

The 19.9~GHz line (Fig.~\ref{fig:spec-k}a) shows an inverse P-Cygni profile with an emission feature at about $-$6.6~\kms\,and an absorption feature at about $-$0.9~\kms. 
The 23.1~GHz line (Fig.~\ref{fig:spec-k}c) exhibits one prominent absorption feature at about $-$0.74~\kms.
These two spectra are consistent with previous studies \citep{1984A&A...134L...7W,1985A&A...147L..19W}. 
We adopted a rest frequency of 23444.778~MHz \citep{1985ZNatA..40..683M,2011MNRAS.413.2339V} for the $10_1-9_2A^{-}$ line, which is 42~kHz lower than the value used by \cite{1986A&A...169..271M}. Thus, to compare with the velocity of the absorption feature observed by \cite{1986A&A...169..271M}, a value of   $-$0.54~\kms should be subtracted from the velocity they reported, which shifts the peak absorption to 0.26~\kms and reduces the difference with our result to 0.77~\kms.
The remaining difference may be due to the noisy spectra and slightly different fitting methods. Despite the low S/N ratio in the 20.1~GHz line (Fig.~\ref{fig:spec-k}b), its absorption feature has a similar velocity as those of the 19.9, 23.1, 23.4 GHz lines, indicating that it is a reliable detection.

A series of nine $J_2-J_1$ lines ($J$ from 2 to 10) of $E$-type methanol near 25 GHz are detected by our observations. Both the $9_2-9_1E$ line at 25.541 GHz (Fig.~\ref{fig:spec-k}l) and the $10_2-10_1E$ line at 25.878 GHz (Fig.~\ref{fig:spec-k}m) exhibit typical inverse P-Cygni profiles with the absorption dips at about 0.1~\kms. We note that such absorption dips are also present in other $J_2-J_1$ lines at a nearly identical velocity but blended with emission, which make them deviate from the typical inverse P-Cygni profile. The dips appear to become more prominent with increasing $J$. The presence of narrow 25~GHz Class I methanol maser emission  also makes the profiles more complex for the $5_2-5_1E$ line at 24.959~GHz, the $6_2-6_1E$ line at 25.018~GHz and the $7_2-7_1E$ line at 25.124~GHz (see Figs.~\ref{fig:spec-k}h--\ref{fig:spec-k}j).
\citet{1986A&A...157..318M} reported the detection of five $J_2-J_1E$ lines ($J$ = 2, 3, 4, 5, 6) towards W31C, but the maser emission peaks at a location 9$\arcsec$ offset from our pointing centre. Nevertheless, the masers should be covered by our Effelsberg beam of $\sim$34$\arcsec$, and they give rise to the observed narrow spikes between $-$6.0~\kms\,and $-$8.0~\kms\,in our spectra of the $5_2-5_1E$, $6_2-6_1E$ and $7_2-7_1E$ transitions (see Figs.~\ref{fig:spec-k}h--\ref{fig:spec-k}j). On the other hand, likely influenced by our offset position, the flux densities are much lower than the values reported in \citet{1986A&A...157..318M}. In addition to the position difference, our low spectral resolution may be further reducing the measured peak flux densities.  

An absorption feature with a velocity of $-$0.68 \kms\/ was detected in the $7_{-2}-8_{-1}E$ line at 37.7 GHz (see Fig.~\ref{fig:spec-e}g), and, with a flux density of $-$0.89~Jy.  This is the strongest absorption feature among our detected absorption lines. Both  $6_{2}-5_{1}A^{\mp}$ lines (from levels of different parity) at 38.3 and 38.5 GHz (see Figs.~\ref{fig:spec-a}e--\ref{fig:spec-a}f) have a narrow emission spike with a velocity of $-$1.9 \kms\/ that is superimposed on a broad absorption feature. Similar to the $9_2-9_1 E$ and the $6_2-6_1 E$ methanol transitions near 25 GHz in W3(OH) \citep{1986A&A...157..318M}, the narrow spike may arise from a weak maser. The velocities of the absorption features in the 38.3 and 38.5 GHz transitions match that of the 37.7 GHz line. The line widths of the absorption features in the 38.3 and 38.5 GHz transitions are $\sim$4.8~\kms, which is slightly narrower than the line width ($\sim$5.35~\kms) of the 37.7 GHz line. \cite{2011ApJ...742..109E} also searched for the 37.7, 38.3 and 38.5 GHz CH$_3$OH lines towards W31C but nothing was detected, owing to their rms noise of $\sim$1.2 Jy with a channel width of 0.27~\kms. On the other hand, \citet{2018MNRAS.480.4851E} discovered an absorption feature in the 37.7, 38.3 and 38.5 GHz transitions towards G337.705$-$0.053, but their velocities are not redshifted with respect to its systemic velocity traced by NH$_{3}$ (1,1) \citep{2012MNRAS.426.1972P}. Our observations are therefore the first detection of redshifted absorption in these three lines.

Absorption features of methanol transitions at 85.6, 86.6 and 86.9 GHz in this work are detected for the first time. The CH$_3$OH transition at 85.6 GHz (Fig.~\ref{fig:spec-e}g) clearly shows an inverse P-Cygni profile. The emission feature peaks at about $-$4~\kms, and the absorption feature lies at a velocity of about 0.22~\kms.
Despite the low S/N ratios in the CH$_3$OH spectra at 86.6 and 86.9 GHz (Figs.~\ref{fig:spec-a}g--\ref{fig:spec-a}h), their shapes are indicative of an inverse P-Cygni profile. The velocities and line widths of the absorption features of the 86.6 and 86.9 GHz lines are similar to that of the 85.6 GHz line, but their absorption line strengths are much weaker than that of the 85.6 GHz line. This could be due to a lower excitation temperature of the 85.6 GHz line.

The most common class I CH$_3$OH transitions at 36, 44, 84 and 95 GHz show strong and narrow maser features, and the velocities of these maser features range from $-$7 to $-$6~\kms\,(see Table~\ref{Tab:result}). Therefore, all the class I methanol masers are significantly blueshifted with respect to the systemic velocity (see Fig.~\ref{fig:Vel-E}). 
These masers have been already reported by previous studies. The peak velocity and flux density of the 84 and 95 GHz CH$_{3}$OH masers from our observations are consistent with previous observations \citep{2012ApJS..200....5C,2019MNRAS.484.5072B}. However, the flux densities of the 36 and 44 GHz CH$_{3}$OH masers of our Effelsberg-100 m observations are about 2--3 times lower than previous single-dish measurements \citep{1989ApJ...339..949H,2016ApJS..227...17K,2019MNRAS.484.5072B,2019ApJS..244....2K}. The discrepancy might arise from the different beam sizes, spectral resolutions, and pointing positions. On the other hand, maser variability may also contribute to the observed difference. 
To our knowledge, there are few published variability studies of class I methanol masers in the Galaxy. Despite previous claims to the contrary, \citet{Menten1988}, using the Effelsberg 100 m telescope, found no significant variability in the 25 GHz  $J_2-J_1E$ class I CH$_3$OH maser lines ($J$ = 2 -- 8) in the Orion Kleinmann-Low Nebula, neither on a time baseline of 2 years nor over 13 years, comparing their data with those that \cite{Hills1975} obtained with the same telescope.  Because class I methanol masers are often distributed over larger scales than the class II masers, it is more difficult to infer variability from observations using different telescopes with different spatial and spectral resolutions. For example, in Orion KL the 25 GHz masers arise from different location spread over an area that is comparable to the Effelsberg beam ($34$\arcsec).

%\clearpage

\subsection{Detection of 11 redshifted methanol absorption lines towards W3(OH)} \label{Sec:w3oh}

\begin{figure*}[!htbp]
%\centering
\begin{minipage}[c]{0.65\textwidth}
\includegraphics[width=1\textwidth]{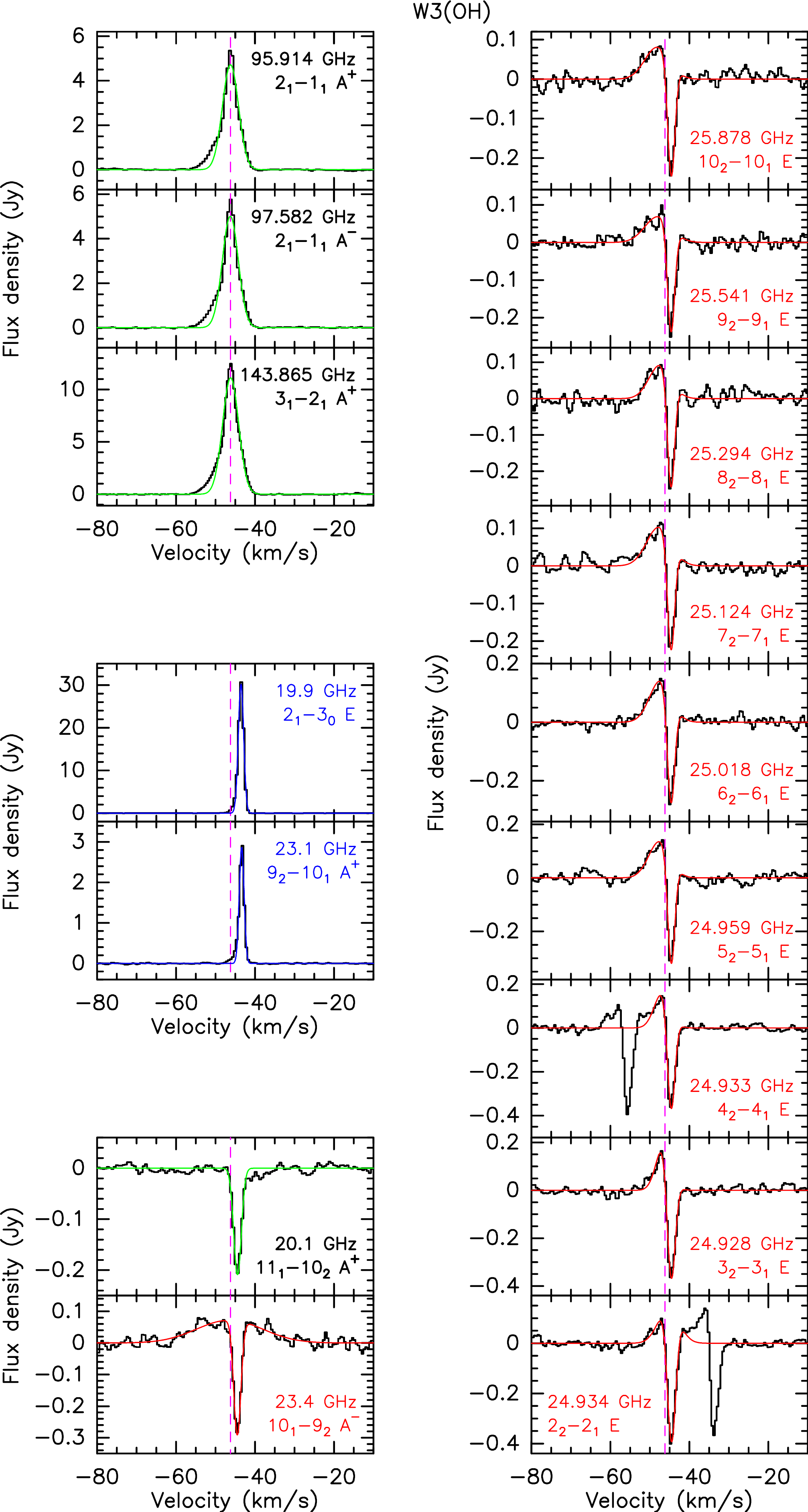}
\end{minipage}\hfill
\begin{minipage}[c]{0.3\textwidth}
\caption{Observed spectra of 16 CH$_{3}$OH transitions towards W3(OH). The quantum numbers and rest frequencies of transitions are labelled in their respective panels. 
Class I and II CH$_3$OH maser transitions are labelled in red and blue, respectively. Their Gaussian fitting results are plotted in the corresponding colour.
The transitions that have no maser detection yet are labelled in black, and the Gaussian fitting results are plotted in green.
The vertical magenta dashed line represents the systemic velocity of $V_{\rm LSR}$=$-$46.19~\kms.
\label{fig:spec-w3oh}}
\end{minipage}
\end{figure*}

Figure~\ref{fig:spec-w3oh} shows the spectra of 16 methanol transitions observed towards W3(OH). The 19.9, 23.1, 95.914, 97.582, 143.865~GHz transitions appear to be single-peaked. The 20.1 and 23.4 GHz lines show absorption features, while the nine $J_2-J_1$ lines ($J$ from 2 to 10) of $E$-type methanol near 25 GHz show inverse P-Cygni profiles. We fitted the emission and absorption features using two Gaussian components for the 23.4 GHz and the $J_2-J_1E$ line series near 25 GHz. One Gaussian component is assumed for the other transitions. The fitted line parameters are given in Table~\ref{Tab:result-w3oh}. The systemic velocity of the W3(OH) complex is derived from three thermal methanol transitions at 95.914, 97.582 and 143.865~GHz. Similar to W31C, the weighted mean velocity is $V_{\rm LSR}$= $-$46.19$\pm$0.02~\kms.
This velocity is consistent with previous high angular resolution observations of the W3(OH) region whose systemic velocity is determined to range from $-$46.5~\kms\,to~$-$46.0~\kms\,\citep{1995ApJ...444..765K,2011ApJ...740L..19Z}. In this work, we simply take $V_{\rm LSR}$=$-$46.19$\pm$0.02~\kms\,from our single-dish observations as the systemic velocity for the subsequent analysis.

%Since the UCH{\sc ii} region in W3(OH) is a bright background continuum source, the absorption feature could be more relevant to W3(OH) rather than W3(H$_2$O), however, molecular and ionised gases may trace different parts of UCH{\sc ii} region resulting in the velocity difference. On the other side, interferometric observations reveal gas distributions and kinematics in small scales, while single-dish observations focus on large scales. 

The nine $J_2-J_1$ lines ($J$ from 2 to 10) of $E$-type methanol near 25 GHz show inverse P-Cygni profiles. The emission features peak at about $-$47.1~\kms\/ for all $J_2-J_1E$ lines, while the absorption features lie at about $-$44.6~\kms. In the nine transitions, the systemic velocity is nearly at the point where the emission feature transits to the absorption feature. The intensities of the absorption features show a descending trend with increasing $J$ (see Fig.~\ref{fig:spec-w3oh}), which appears to be opposite to the case in W31C (see Fig.~\ref{fig:spec-k}). Because the difference between the rest frequencies of these lines is small, the continuum levels should be almost identical. Hence, the trend strongly suggests that the excitation temperatures of these transitions decreases with increasing $J$ in W3(OH). Our results are generally consistent with the line profiles of the $J_2-J_1$ ($J$=2, 3, 4, 5, 6, 9) lines in \citet{1986A&A...157..318M} except that different line profiles are found in the $6_2-6_1E$ and $9_2-9_1E$ lines. Their observations revealed double absorption features in these two transitions, but only one single absorption component is seen in our observations. The double absorption features were interpreted as a weak maser at a velocity of $-$44~\kms\/ that partially covers the absorption trough \citep{1986A&A...157..318M}. The non-detection of the double absorption features by our observations may indicate temporal variation of this weak maser.

The absorption feature in both the 20.1 and 23.4 GHz lines has an LSR velocity of about $-$44.4~\kms\, with a line width of $\sim$2~\kms, which is in good agreement with the results reported by \citet{1986A&A...169..271M}.
Despite the different intensities that might be caused by different spectral resolutions and calibrations, the line profiles of the observed masers at 19.9 and 23.1 GHz are also consistent with the spectra obtained more than 30 years ago \citep{1984A&A...134L...7W,1985A&A...147L..19W}. 
The $4_0-3_1E$ line at 28.3~GHz and the $2_1-3_0E$ maser line at 19.9~GHz belong to the same series, but absorption features were detected in the former line by \cite{1992ApJ...397L..43S} and \cite{1993A&A...268..249W}, and the absorption features appear to be nearly at the same velocity as those at 20.1 and 23.4 GHz.

\begin{figure*}[!htbp]
%\centering
\begin{minipage}[c]{0.7\textwidth}
\includegraphics[width=1\textwidth]{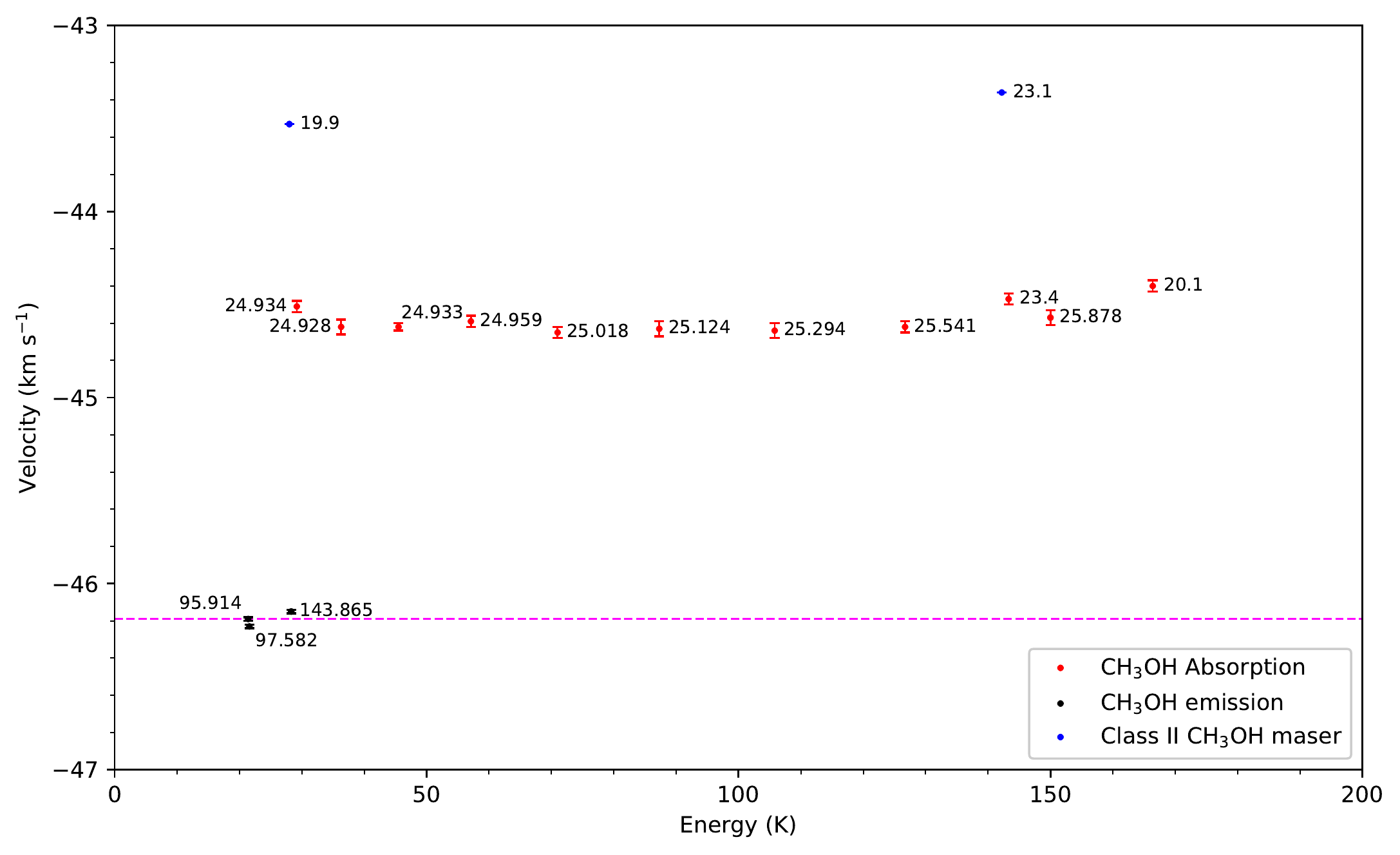}
\includegraphics[width=1\textwidth]{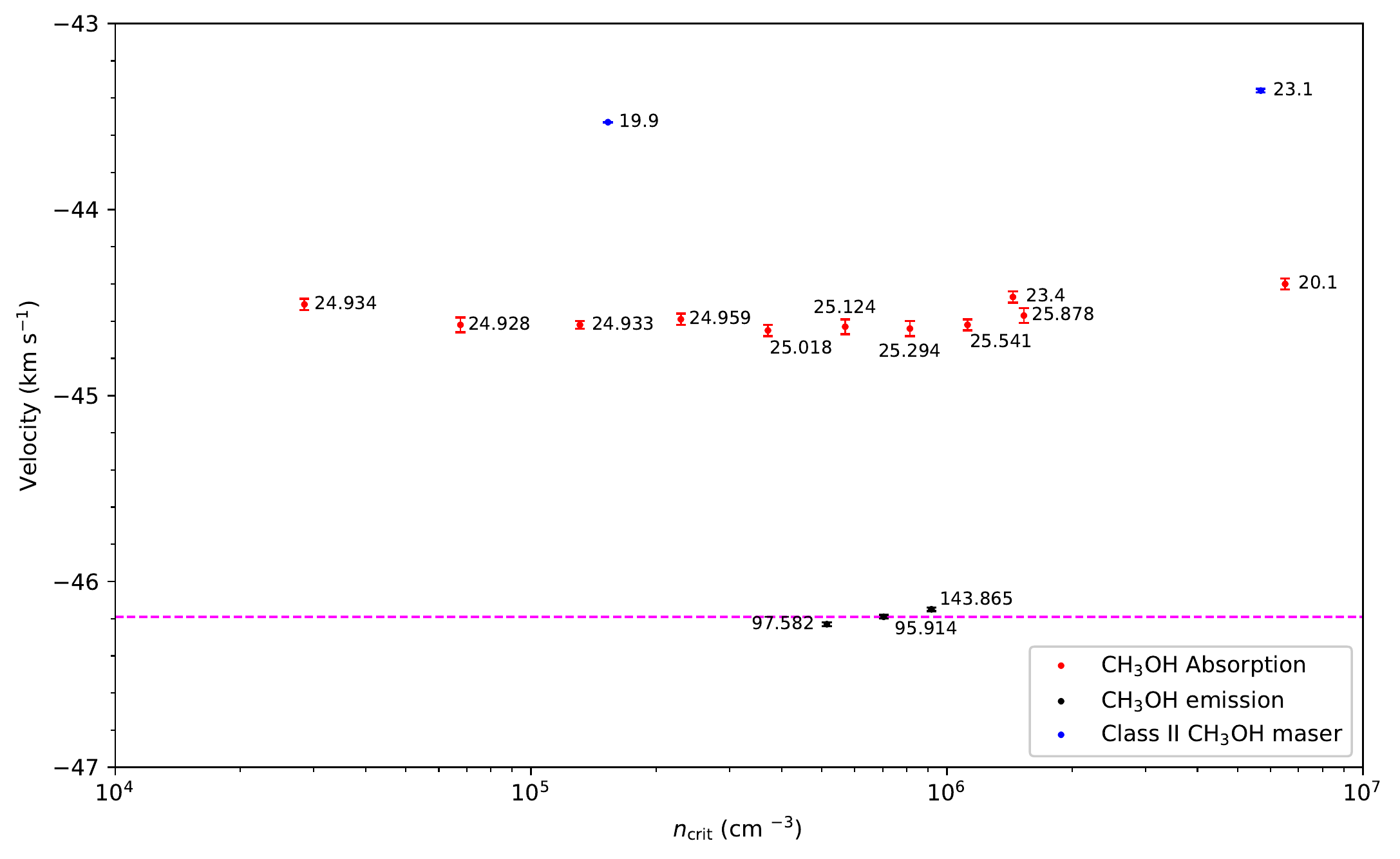}
\end{minipage}\hfill
\begin{minipage}[c]{0.25\textwidth}
\caption{
LSR velocity as a function of the upper energy level (upper panel) and critical density (lower panel) for W3(OH).
%Upper panel: LSR velocity as a function of the upper energy level for CH$_3$OH line observed towards W3(OH). Lower panel: LSR velocity as a function of the critical density for W3(OH).
The red and blue dots indicate the methanol lines with absorption and maser features, respectively.
The black dots indicate the methanol emission without a narrow and strong maser emission feature.
Their velocities are derived from Gaussian fitting to the emission peaks or absorption dips.  
The error bars indicate Gaussian fitting errors.
The horizontal magenta dashed line represents the systemic velocity of $V_{\rm LSR}$=$-$46.19~\kms.
\label{fig:Vel-E-w3oh}}
\end{minipage}
\end{figure*}

Figure~\ref{fig:Vel-E-w3oh} shows the line velocity as a function of the upper energy level and critical density of each transition for W3(OH). We find that all detected absorption features have  a velocity of about $-$44.6~\kms, which is redshifted with respect to the systemic velocity. The detected class II methanol masers at 19.9 and 23.1 GHz are even more redshifted than the absorption features. The three types of methanol emission indicate the presence of three distinct groups of lines in Fig.~\ref{fig:Vel-E-w3oh}, and the velocities of the three groups seem to be independent of the energy level of the respective transitions and the critical density. 

\section{Discussion}\label{Sec:discuss}

\subsection{Modelling redshifted methanol absorption}\label{sec:model}
Our observations have shown that redshifted methanol absorption is more prominent in the lower frequency (radio wavelength) transitions than in those at higher frequencies. This is expected since the two sources host UCH{\scriptsize II} regions, whose free-free continuum emission becomes optically thin regime at higher frequencies, which results in a reduction of the continuum brightness temperature \citep[e.g.][]{2005IAUS..227..111K,2021A&A...645A.110Y}. Furthermore, the observed absorption is mainly caused by cool gas that lies in the foreground of the free-free continuum emission. Therefore,  low frequency lines that are absorbed from levels with low energy above the ground state tend to be more prominent. We also note that the absorption features of methanol lines from transitions at nearby frequencies can exhibit quite different intensities, indicating different excitation temperatures. This suggests that their populations deviate from local thermodynamic equilibrium (LTE). As discussed in Sec.~\ref{sec:intro}, such non-LTE effects can be easily achieved for various methanol lines by collisional pumping \citep[e.g.][]{2016A&A...592A..31L}, and such effects could lead to very low excitation temperatures (even $<$2.73~K) in methanol transitions. 

Our single-dish spectra are likely tracing large-scale inward motions, but the large-scale density, temperature, and velocity structures might not be well described by simple analytical solutions. Hence, we use a simple two-layer model for this study instead of more sophisticated radiative transfer models. %like LIME \citep{2010A&A...523A..25B}. 
In order to demonstrate the feasibility of using redshifted methanol absorption to quantify infall motions, we make use of a modified two-layer model as discussed by \citet{1996ApJ...465L.133M,2001ApJ...562..770D} to fit the observed redshifted methanol absorption profiles.
In this model, the two layers lie along the line of sight. The ``front" layer lies between the observer and the continuum source with an excitation temperature $T_{\rm f}$, while the ``rear" layer lies behind the source with an excitation temperature $T_{\rm r}$. The continuum source is assumed to be optically thick with a blackbody temperature $T_{\rm c}$, and the cosmic background radiation $T_{\rm b}$ is also included. 
Each layer has a peak optical depth, $\tau_0$, velocity dispersion, $\sigma$, and infall velocity, $V_{\rm in}$, towards the source.
The Planck-corrected brightness temperature at the rest frequency $\nu$ is defined as $J~\equiv~T_0/{\rm exp}[(T_0/T)-1]$, where $T_0~\equiv~h\nu/k$, $h$ is Planck's constant, and $k$ is Boltzmann's constant. $J_{\rm f}$, $J_{\rm r}$, $J_{\rm c}$, and $J_{\rm b}$ are the Planck-corrected brightness temperature of the ``front" layer, ``rear" layer, central continuum source, and the cosmic background radiation, respectively. The central continuum source fills a fraction of $\Phi$ of the observing beam with and $V_{\rm LSR}$ is the systemic LSR velocity of the source. According to this model \citep{2001ApJ...562..770D}, the observed line brightness temperature can be expressed as,
\begin{equation}\label{eq:twolayer1}
    \Delta T_{\rm B} = (J_{\rm f}- J_{\rm cr})[1-{\rm exp}(-\tau_{\rm f})]+(1-\Phi)(J_{\rm r}- J_{\rm b})[1-{\rm exp}(-\tau_{\rm r}-\tau_{\rm f})] \,,
\end{equation}
where
\begin{equation}\label{eq:twolayer2}
    J_{\rm cr} = \Phi J_{\rm c}+ (1-\Phi)J_{\rm r} \,,
\end{equation}

\begin{equation}\label{eq:twolayer3}
    \tau_{\rm f} = \tau_{\rm 0}\,{\rm exp}[\frac{-(V-V_{\rm in}-V_{\rm LSR})^2}{2 \sigma^2}] \,,
\end{equation}

\begin{equation}\label{eq:twolayer4}
   \tau_{\rm r} = \tau_{\rm 0}\,{\rm exp}[\frac{-(V+V_{\rm in}-V_{\rm LSR})^2}{2 \sigma^2}] \,.
\end{equation}

In order to reduce the number of free parameters, we fixed $V_{\rm LSR}$=$-$3.43~\kms\,for W31C and $V_{\rm LSR}$=$-$46.19~\kms\,for W3(OH), while $T_{\rm b}$=2.73~K and $\Phi$=0.9 are adopted for both sources. 
The fixed parameter $\Phi$ is arbitrarily selected. However, we tested the model with $\Phi$=0.1--0.9, and the resulting $V_{\rm in}$, $\tau_0$ and $\sigma$ stayed nearly identical, while $T_{\rm f}$ and $T_{\rm r}$ changed dramatically.
For each transition, the model inputs also include the corresponding rest frequency and the radiation from the central continuum source ($T_{\rm c}$).
The continuum flux densities contributed by the central UCH{\sc ii} region were obtained from interferometric radio continuum measurements (see \citealt{2021A&A...645A.110Y} for W31C, see \citealt{1991A&A...251..220W} for W3(OH)), and corresponding $T_{\rm c}$ values are thus calculated by adopting different frequencies and beam sizes. After fixing these parameters, there are still five free parameters $V_{\rm in}$, $\tau_0$, $\sigma$, $T_{\rm f}$, $T_{\rm r}$ which will be determined by comparison with our spectra.

\begin{figure*}[!htbp]
\centering
\includegraphics[width=0.6\textwidth]{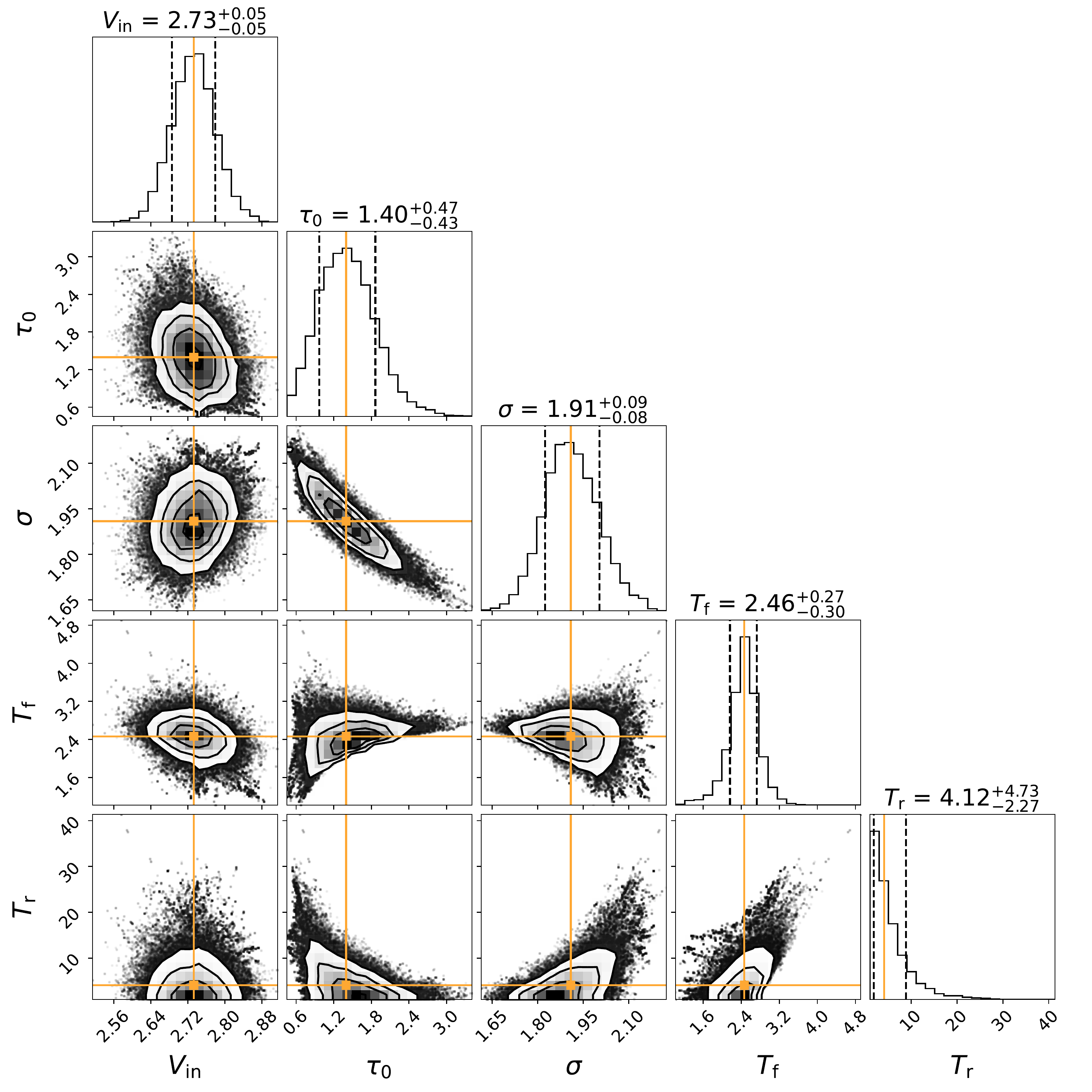}
\includegraphics[width=0.36\textwidth]{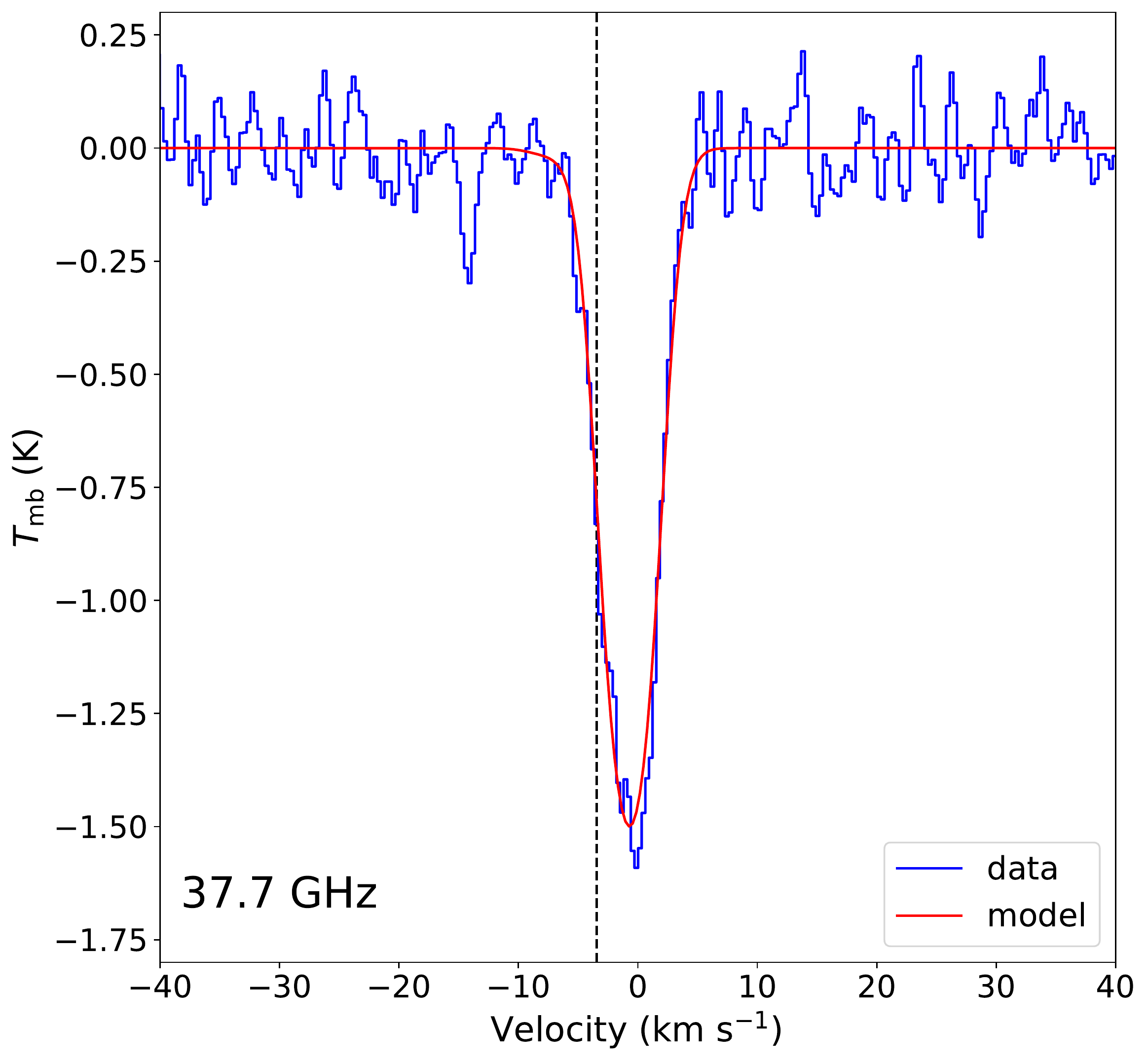}
\caption{Left: Posterior probability distributions of infall velocity, $V_{\rm in}$, peak optical depth, $\tau_{0}$, velocity dispersion, $\sigma$, excitation temperature of the ``front" layer, $T_{\rm f}$, and excitation temperature of the ``rear" layer, $T_{\rm r}$ for the 37.7 GHz CH$_{3}$OH transition towards W31C. The maximum posterior possibility point in the parameter space is shown in orange lines and points. The contours represent the 0.5, 1.0, 1.5, and 2.0$\sigma$ confidence intervals. Right: Observed and modelled spectra for the 37.7~GHz CH$_{3}$OH transition. The blue spectrum represents the observed data, and the red curve indicates the fitted line from the two-layer model (see Sec.~\ref{sec:model}). The vertical black dashed line represents the systemic velocity of $V_{\rm LSR}$=$-$3.43 \kms.
\label{fig:model-37}}
\end{figure*}

\begin{figure*}[!htbp]
%\centering
\includegraphics[width=0.24\textwidth]{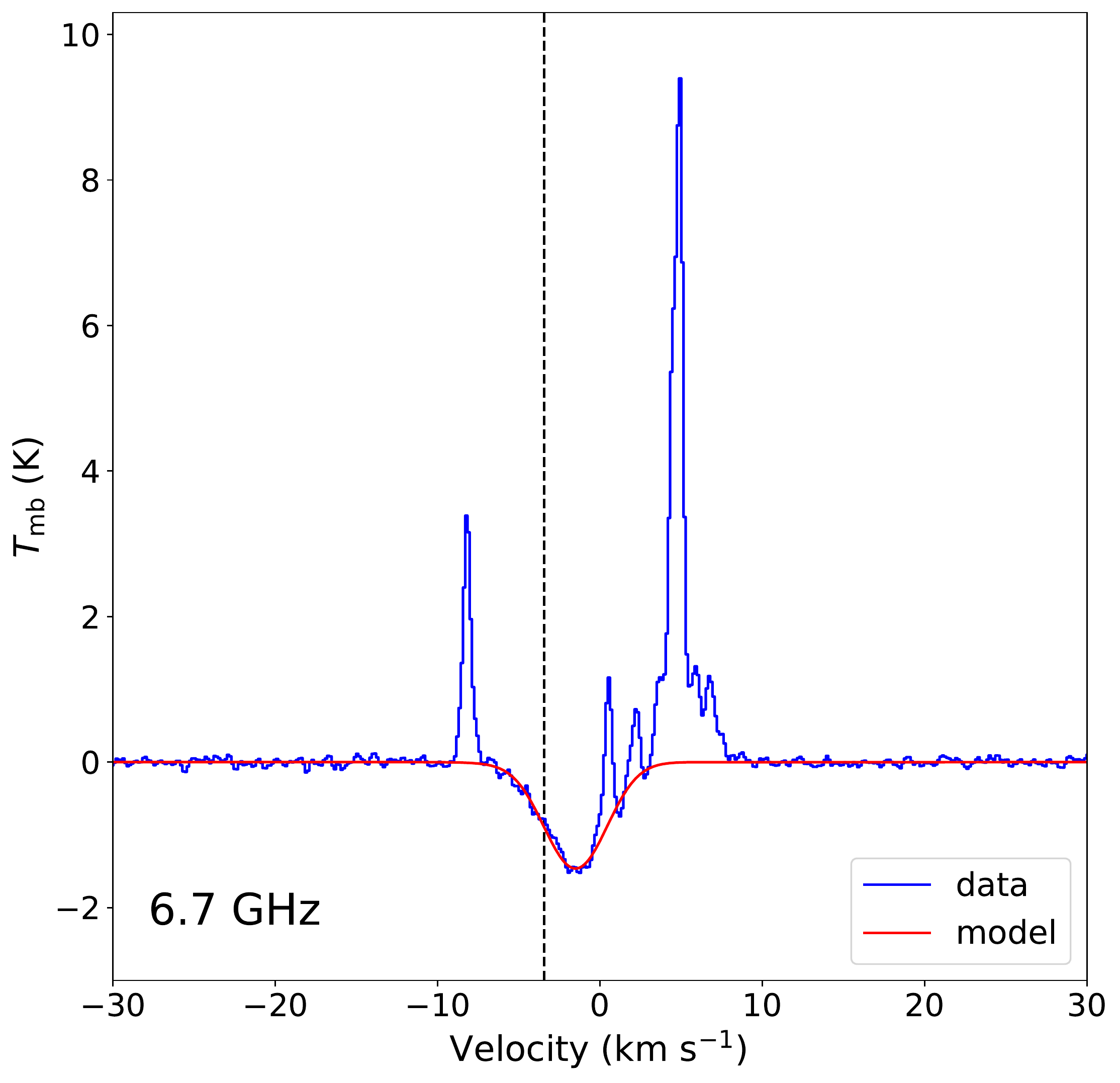}
\includegraphics[width=0.24\textwidth]{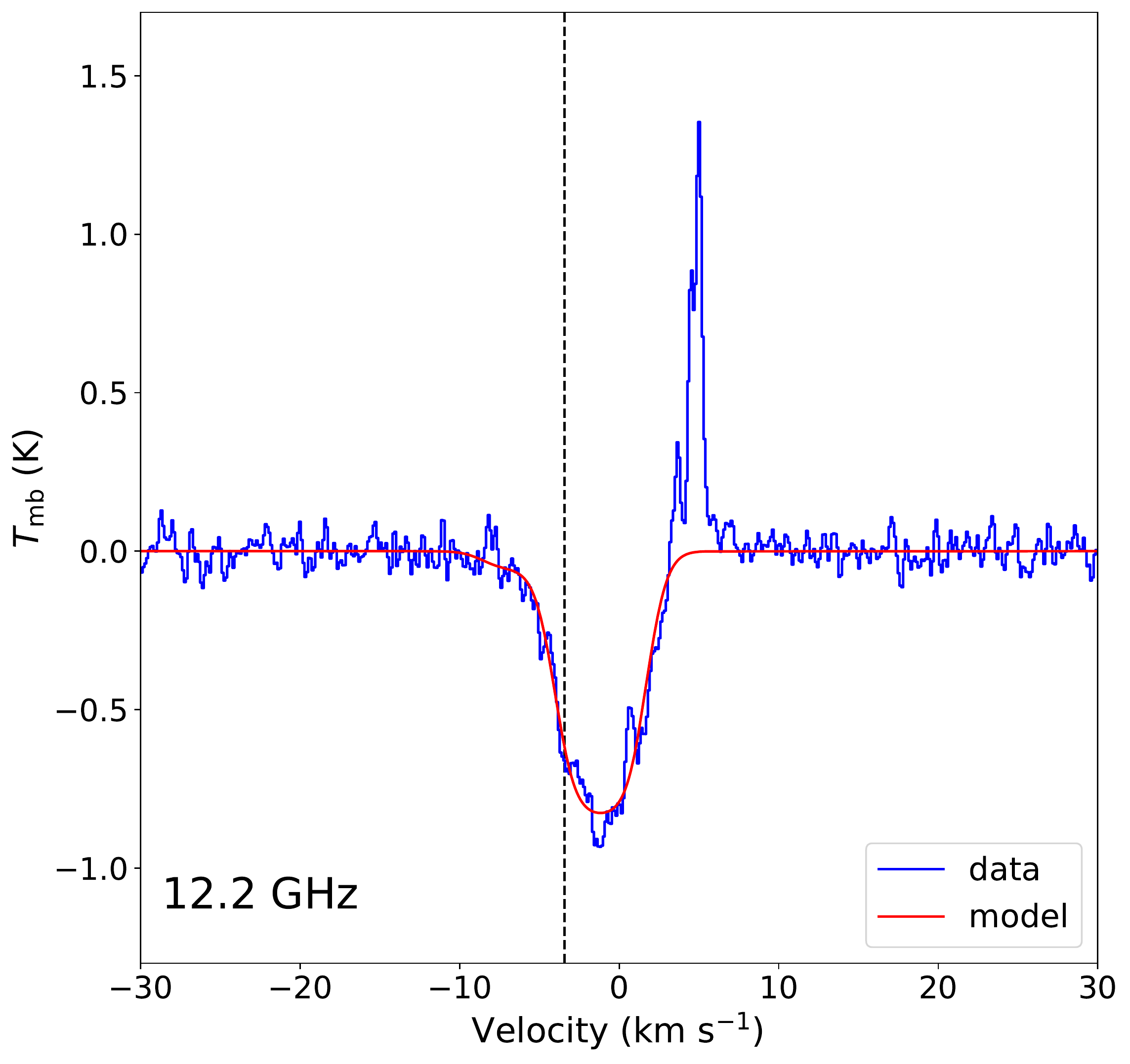}
\includegraphics[width=0.24\textwidth]{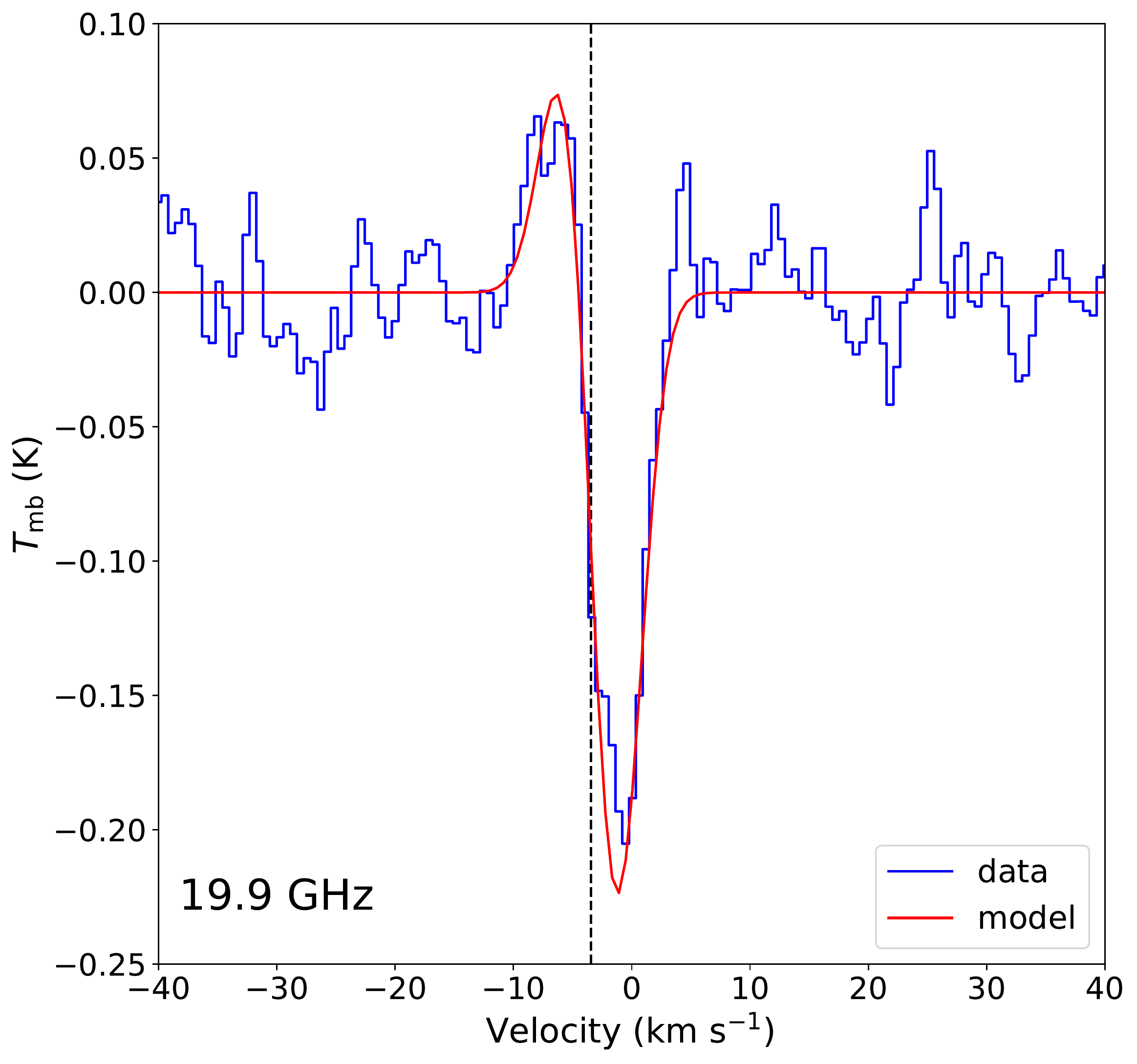}
\includegraphics[width=0.24\textwidth]{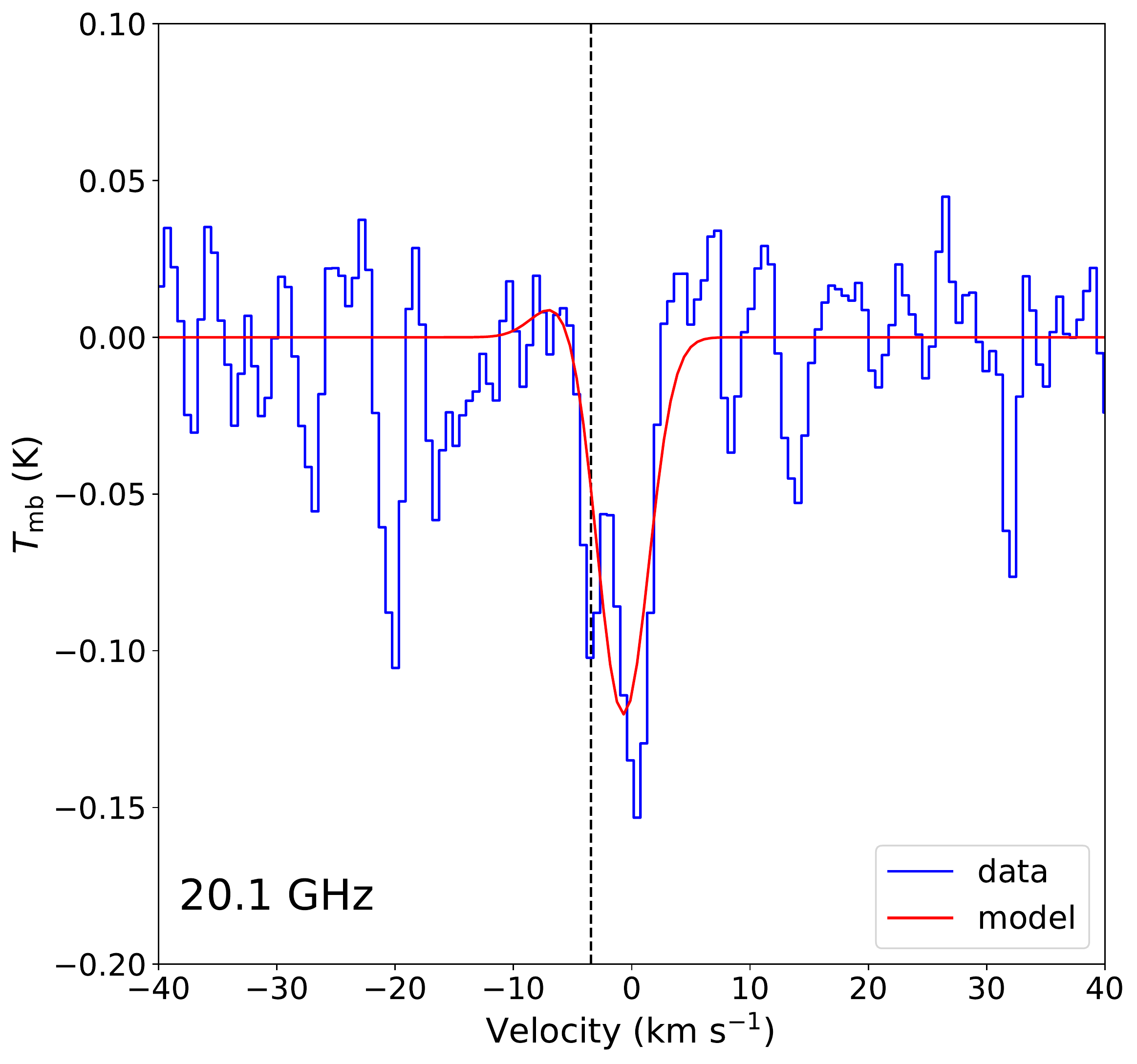}
\includegraphics[width=0.24\textwidth]{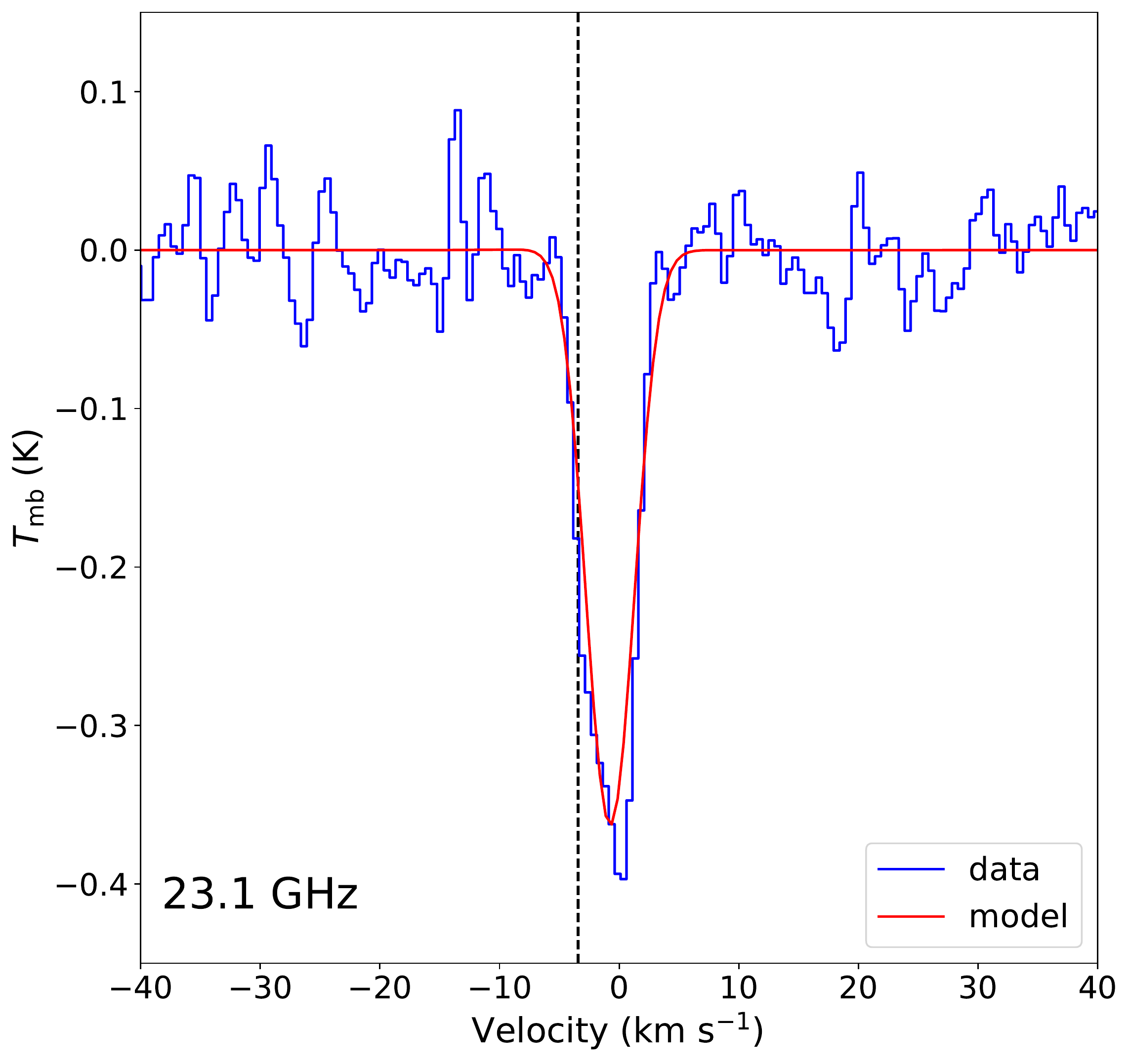}
\includegraphics[width=0.24\textwidth]{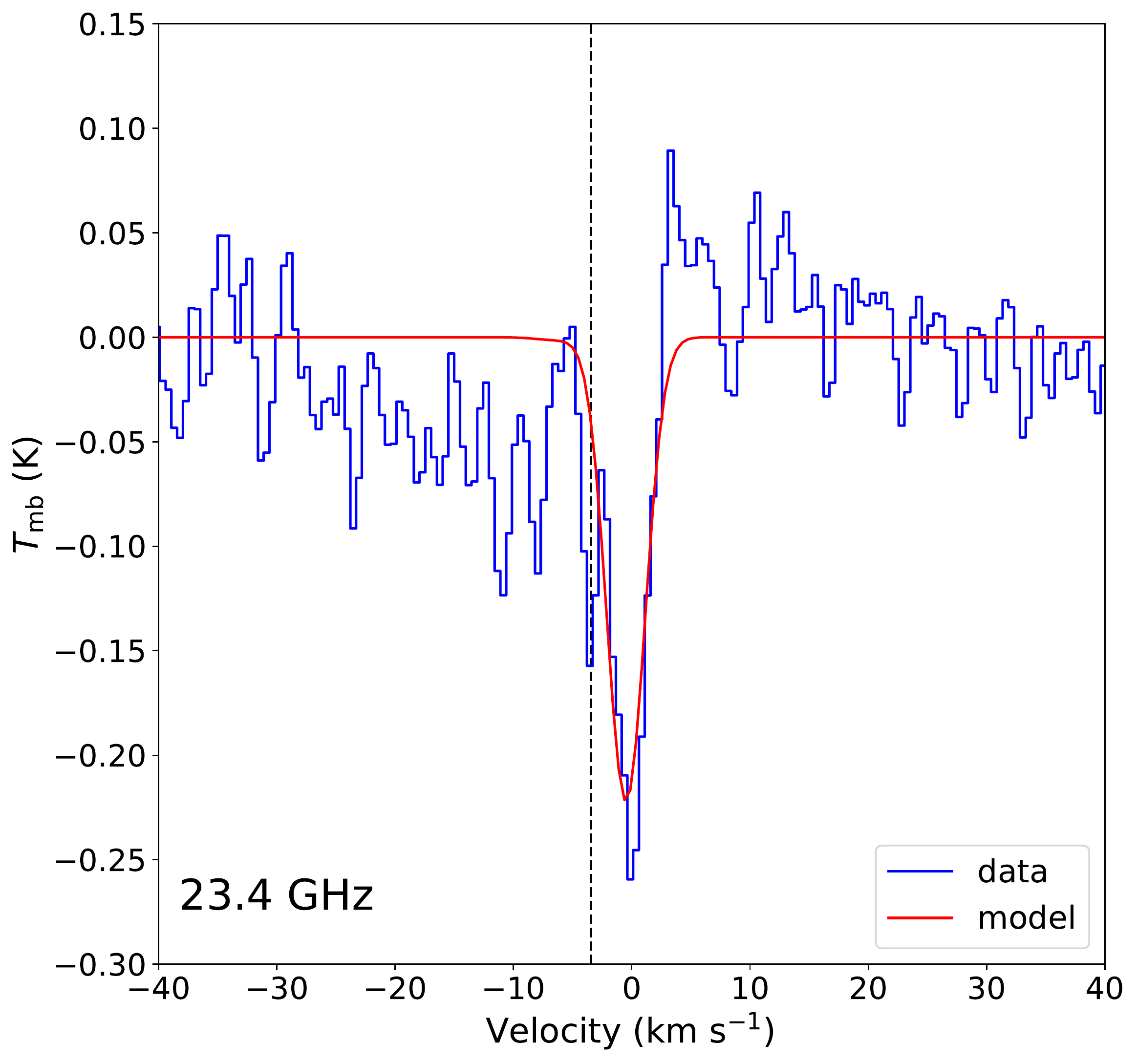}
\includegraphics[width=0.24\textwidth]{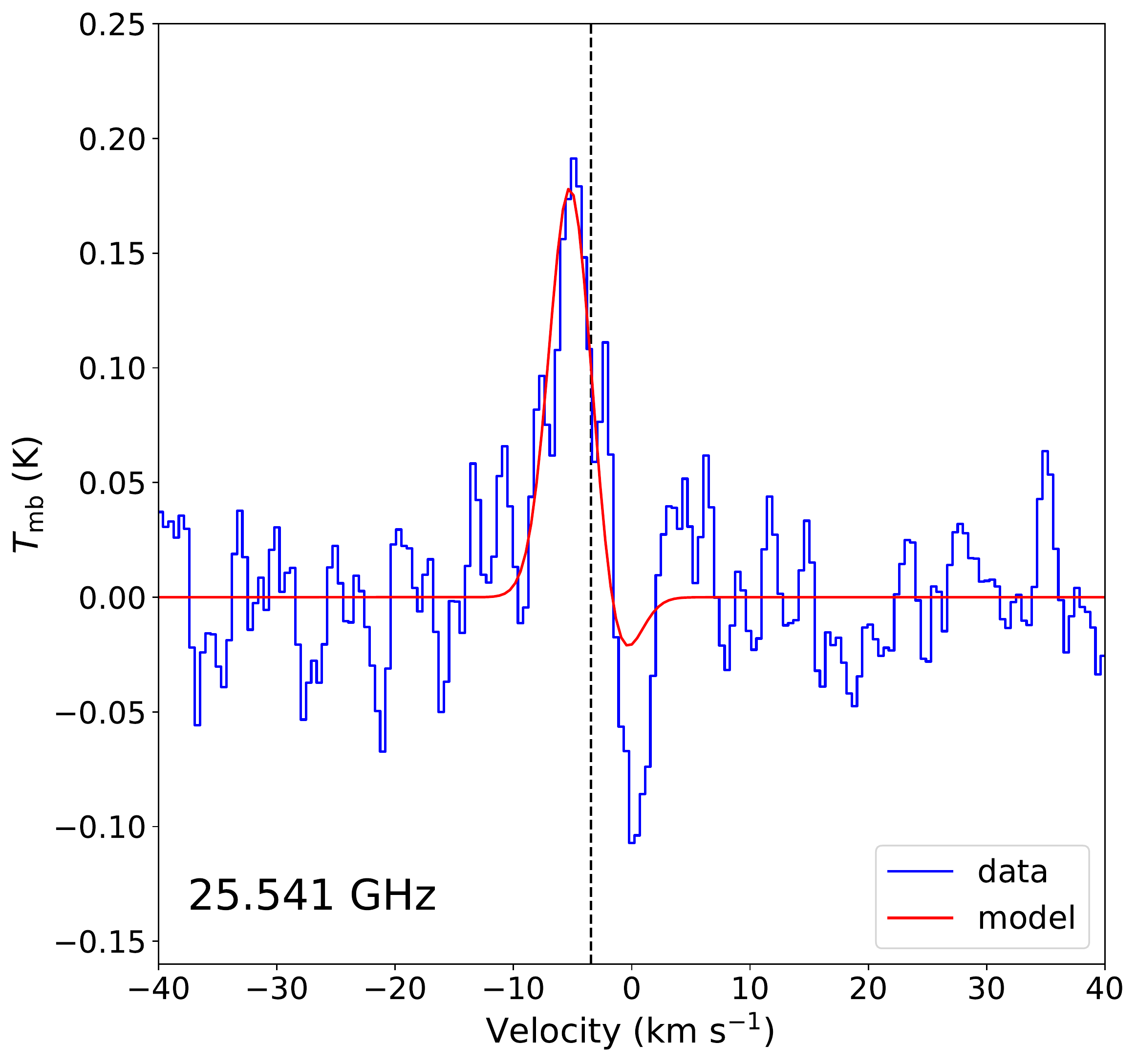}
\includegraphics[width=0.24\textwidth]{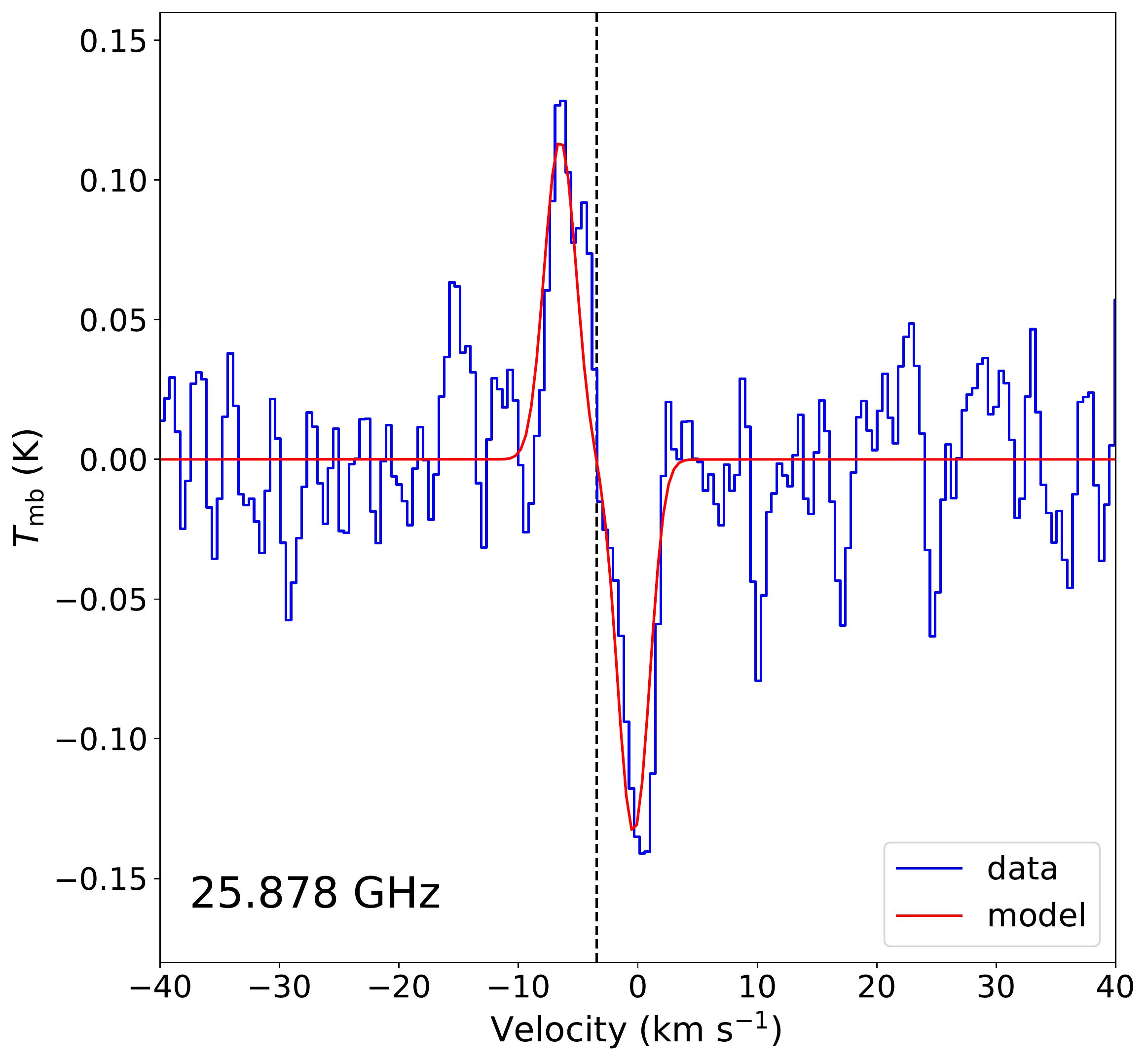}
\includegraphics[width=0.24\textwidth]{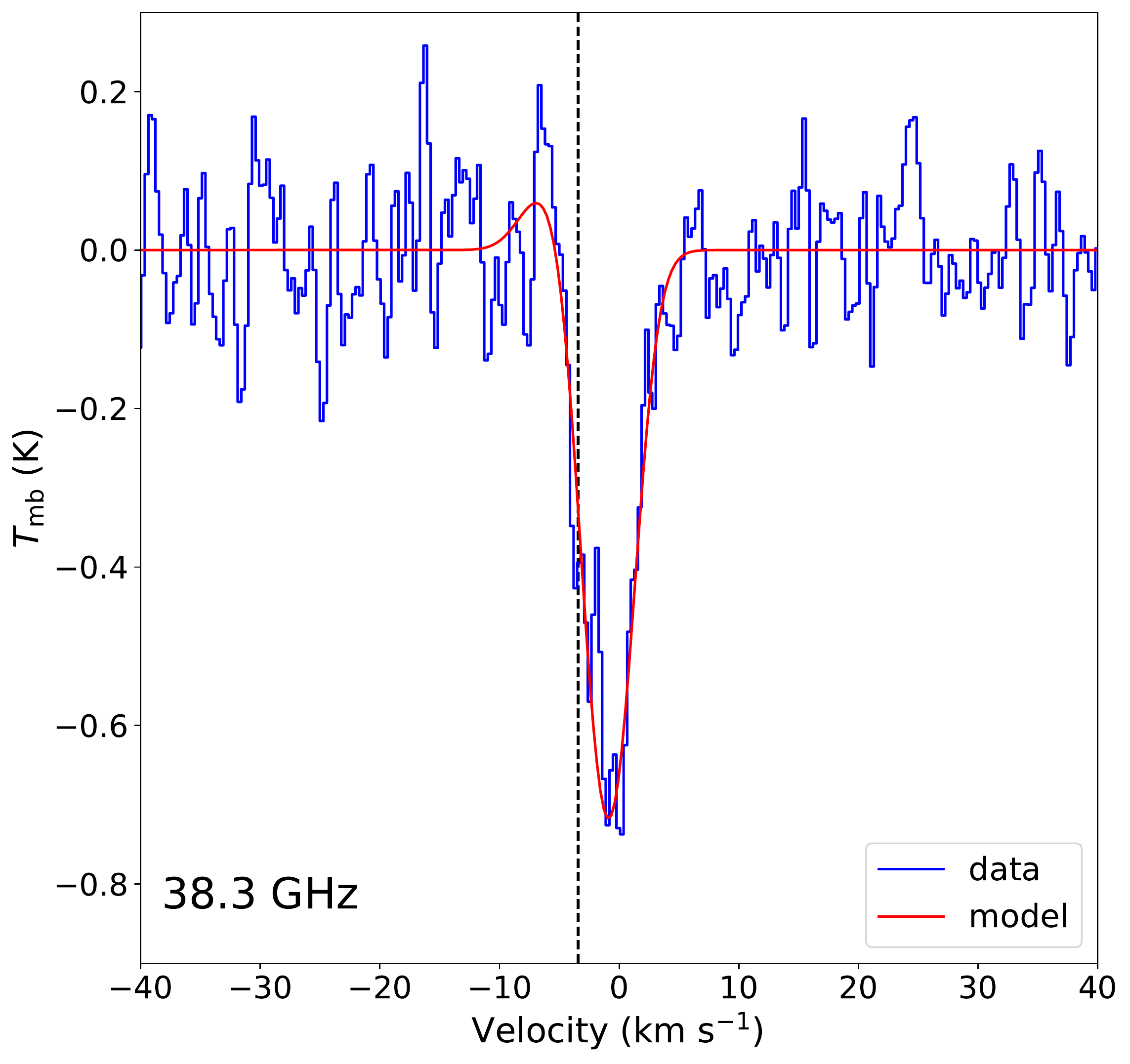}
\includegraphics[width=0.24\textwidth]{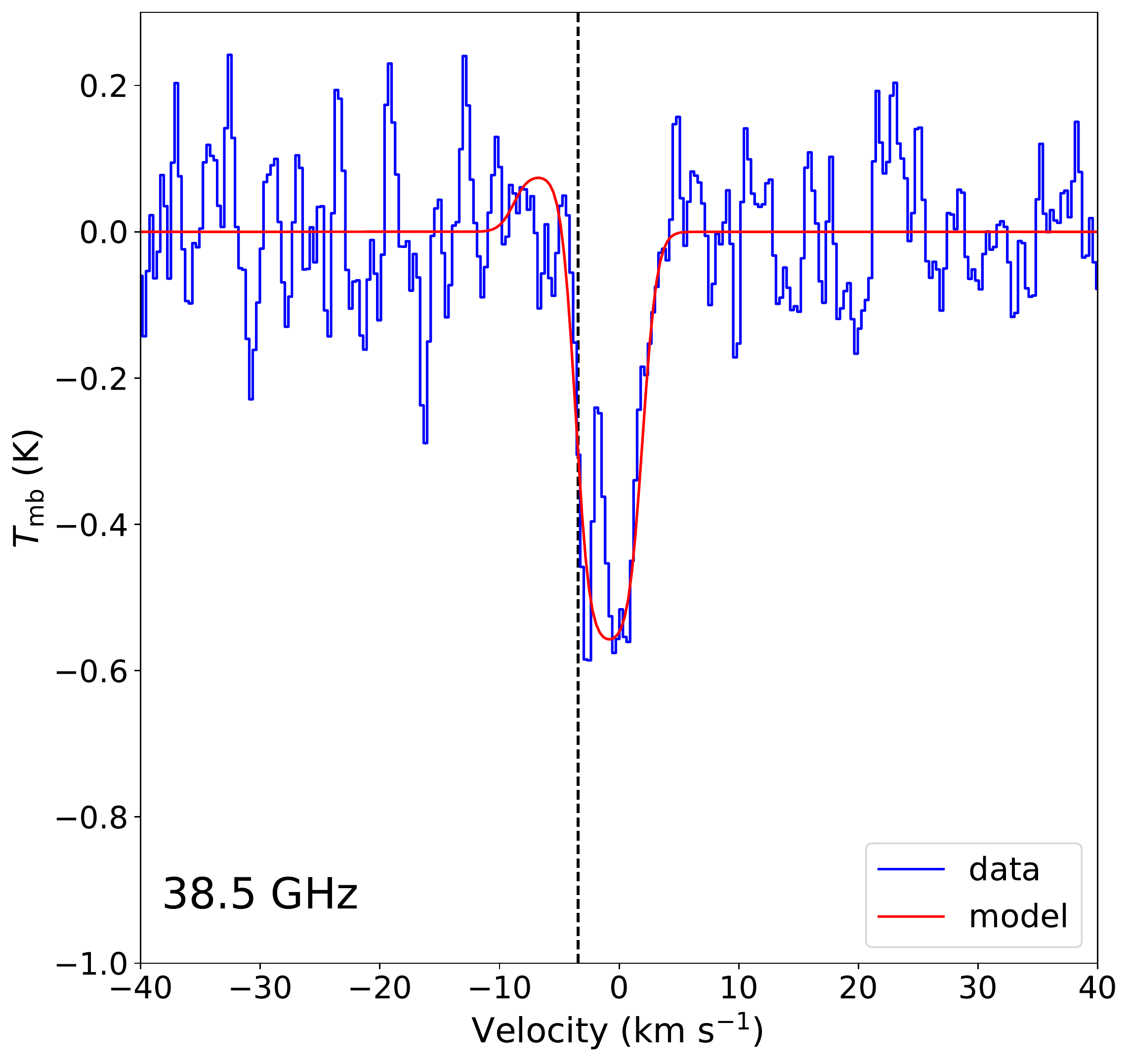}
\includegraphics[width=0.24\textwidth]{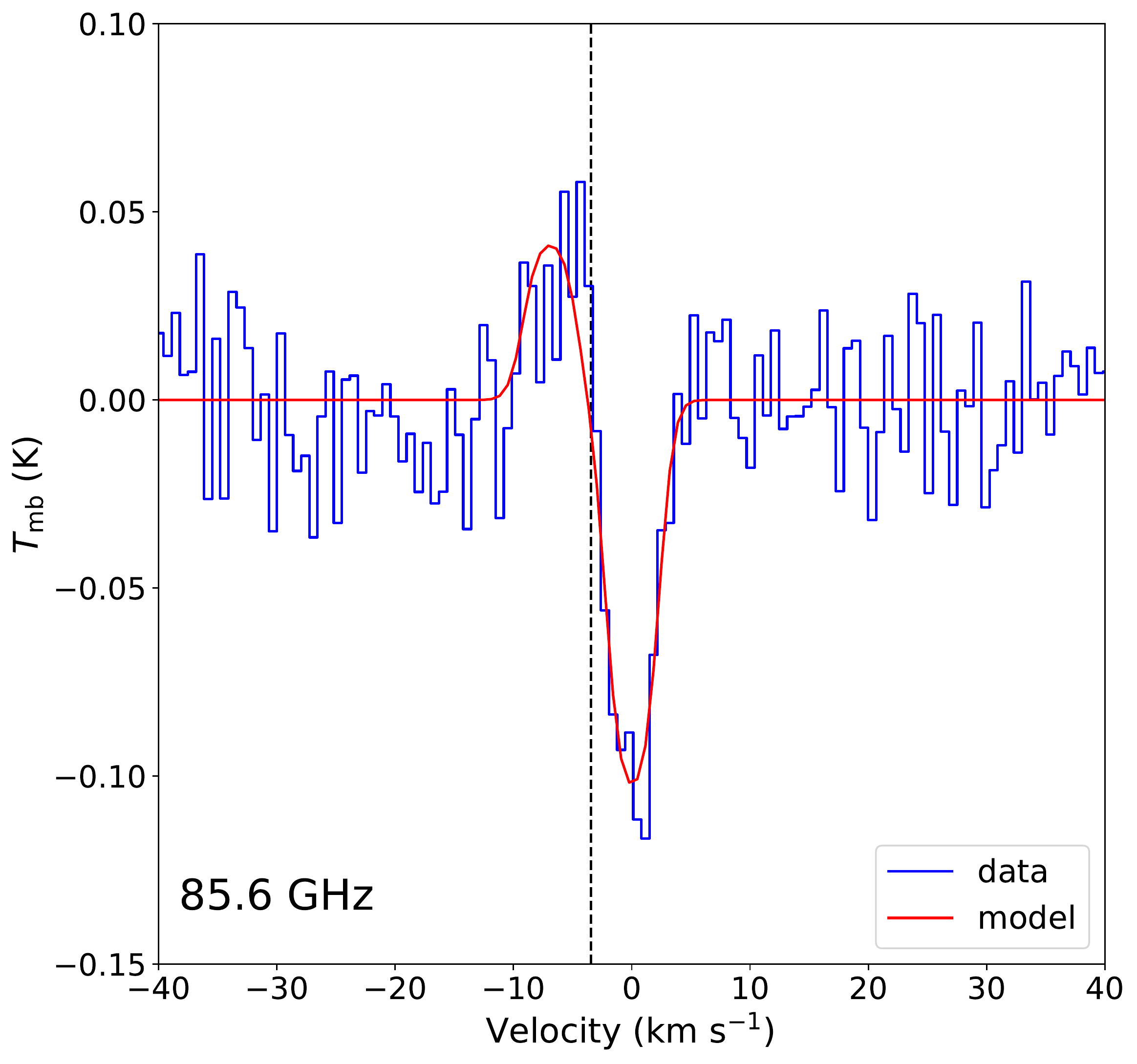}
\includegraphics[width=0.24\textwidth]{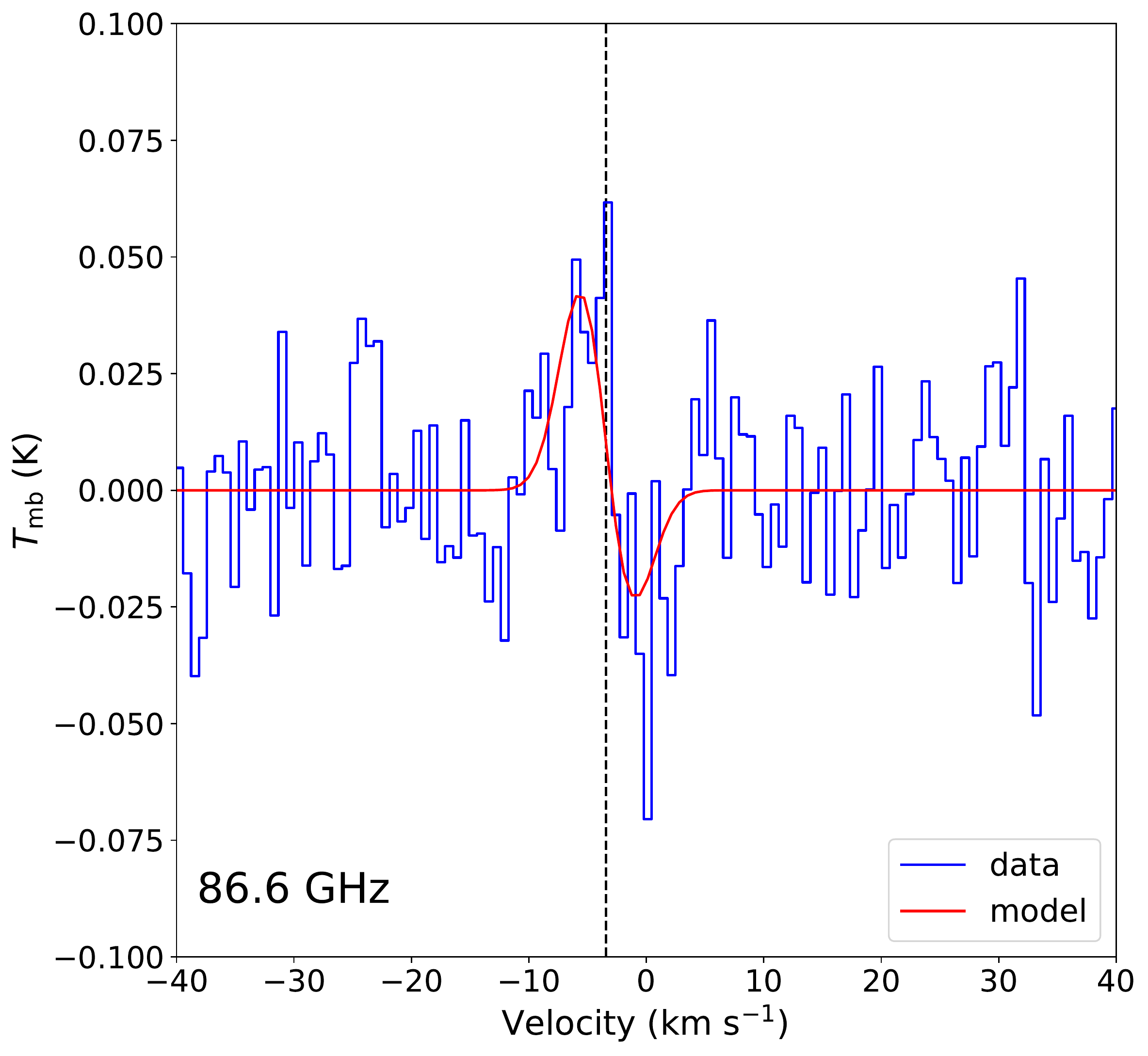}
\includegraphics[width=0.24\textwidth]{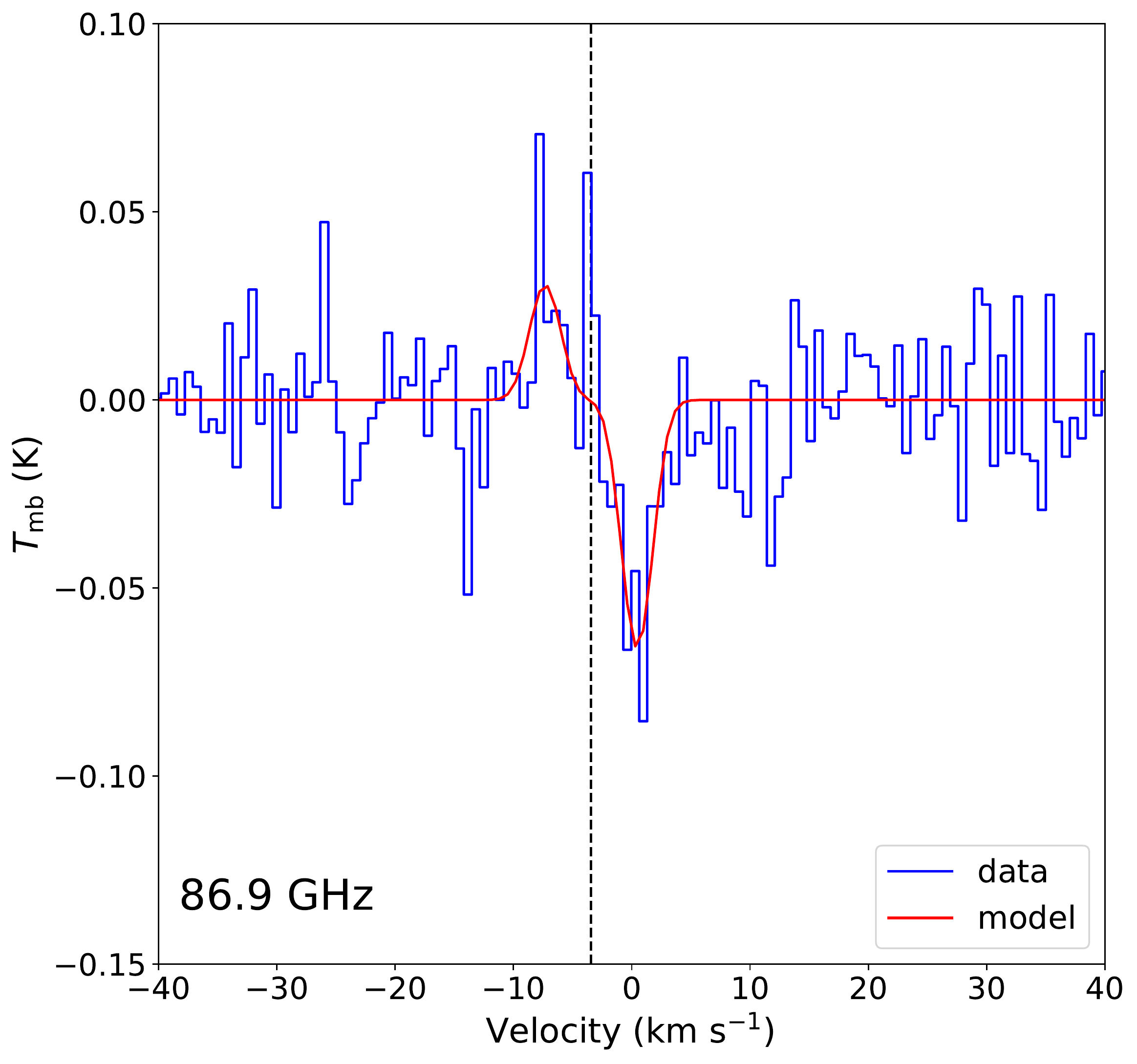}
\caption{Observed and modelled spectra for the methanol transitions with absorption features towards W31C. The short line name of each transition is denoted in the lower left corner of each panel. The blue spectra represent the observed data, and the red curves indicate the fitted line from the two-layer model with adopted parameters listed in Table~\ref{Tab:model}. The vertical black dashed lines represent the systemic velocity of $V_{\rm LSR}$=$-$3.43 \kms.
\label{fig:model}}
\end{figure*}

\begin{figure*}[!htbp]
%\centering
\includegraphics[width=0.24\textwidth]{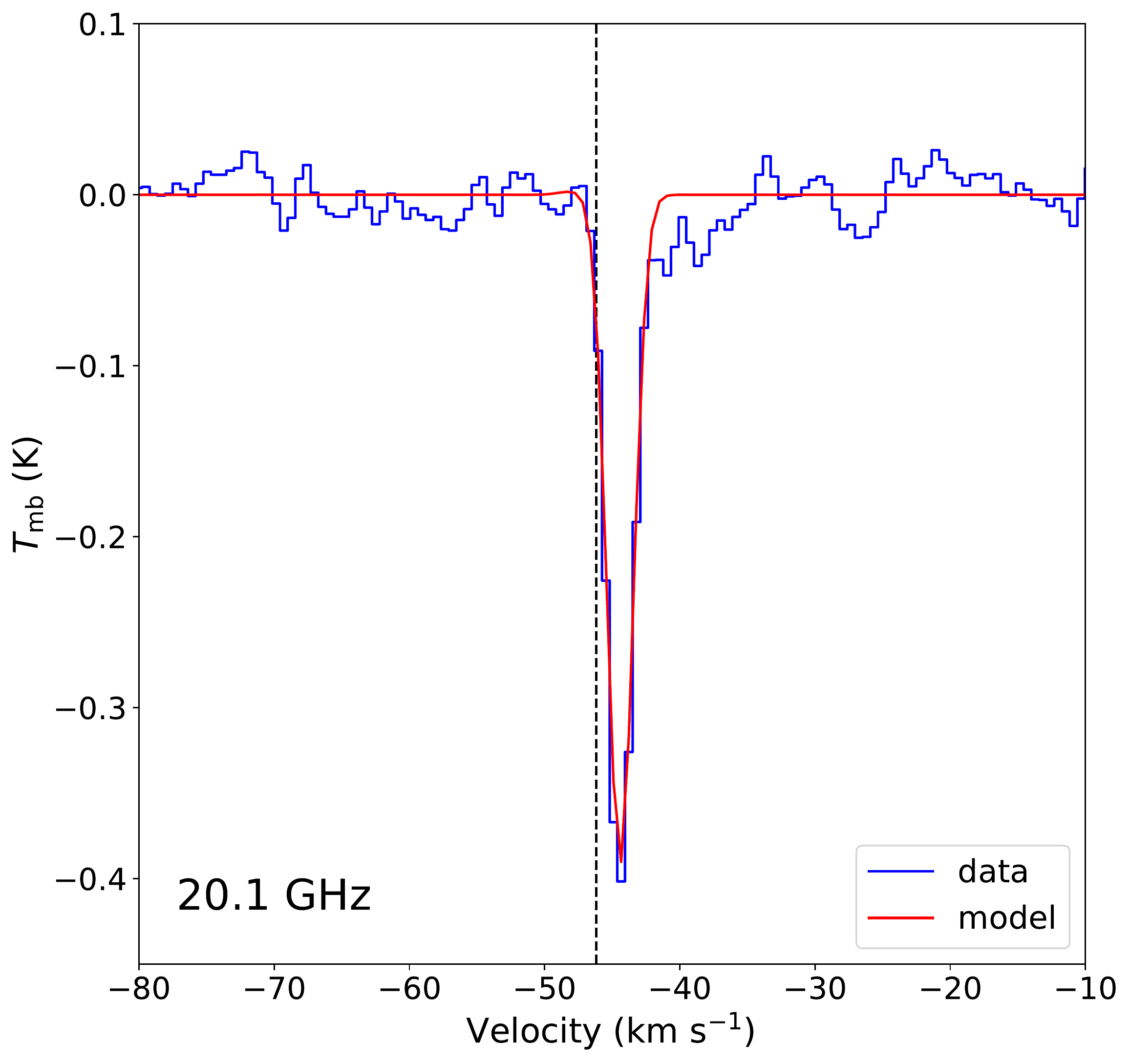}
\includegraphics[width=0.24\textwidth]{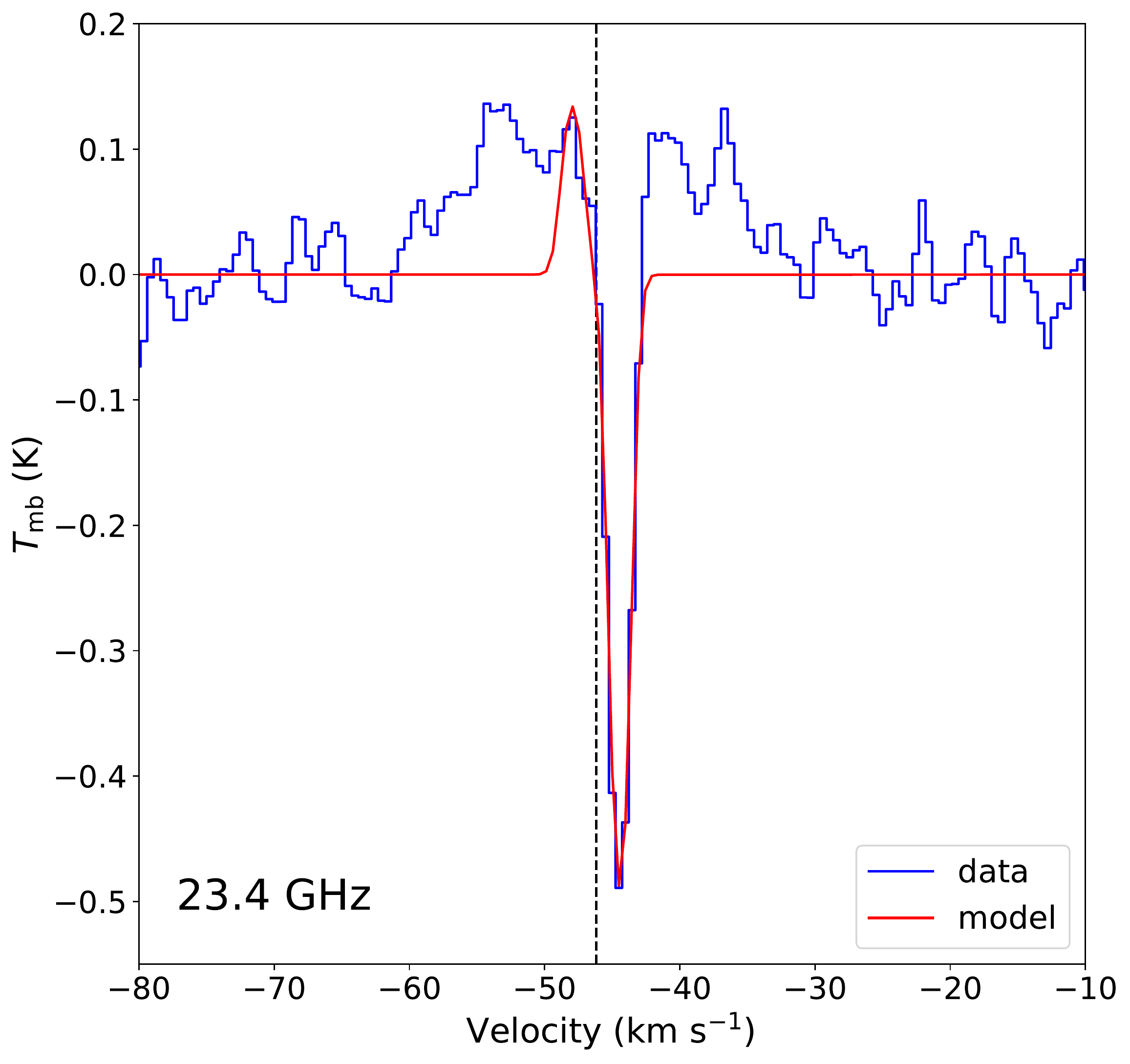}
\includegraphics[width=0.24\textwidth]{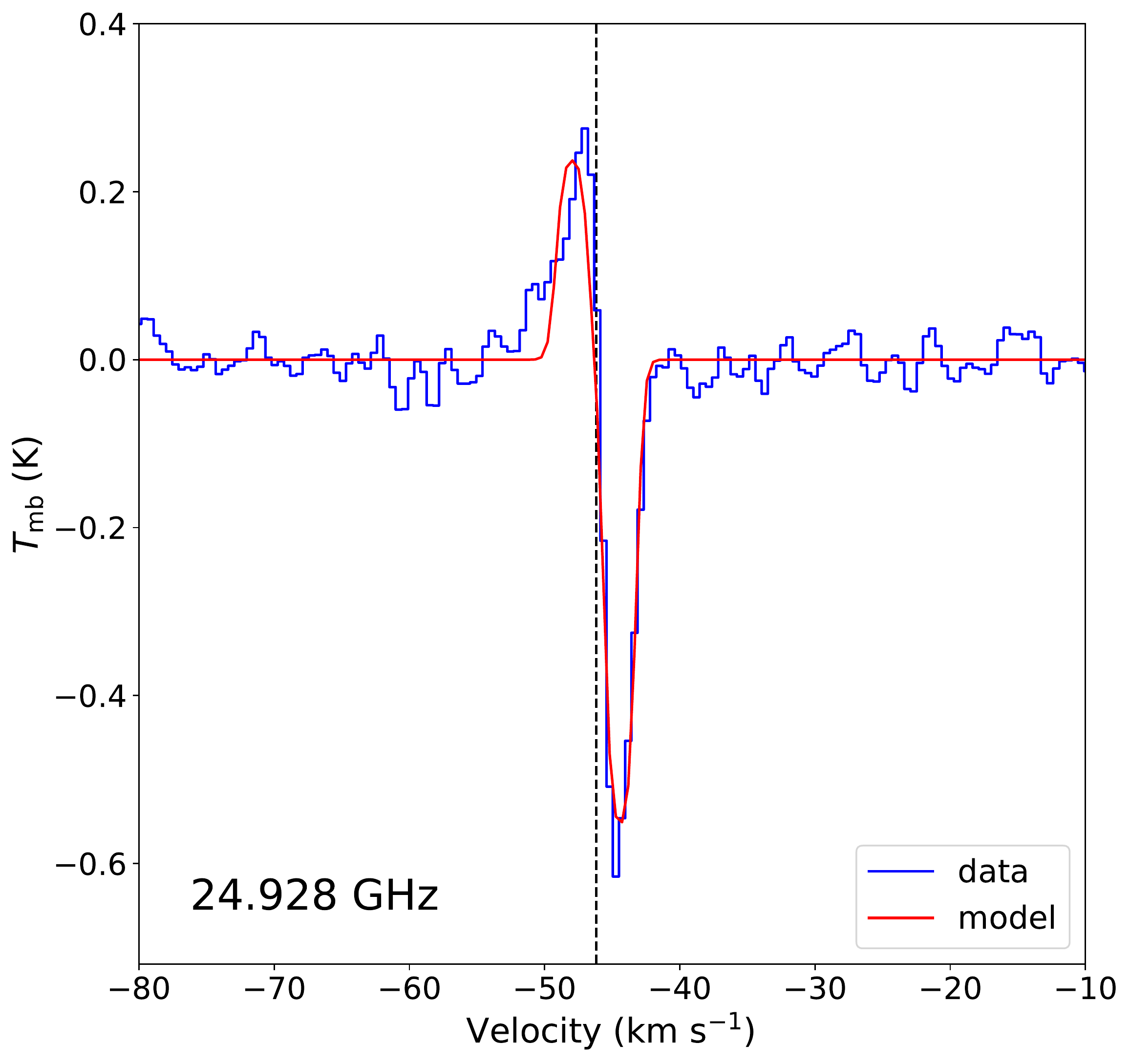}
\includegraphics[width=0.24\textwidth]{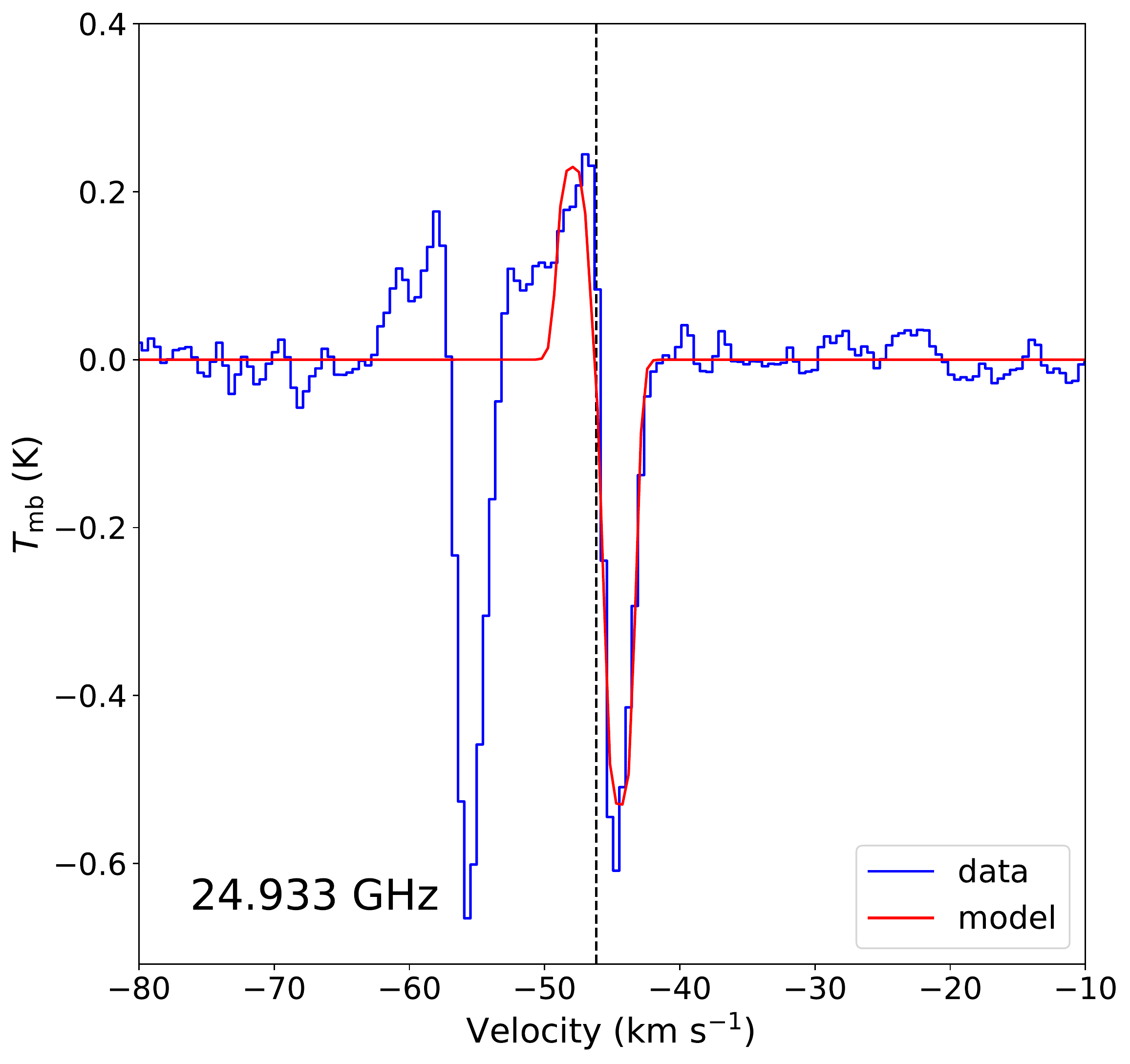}
\includegraphics[width=0.24\textwidth]{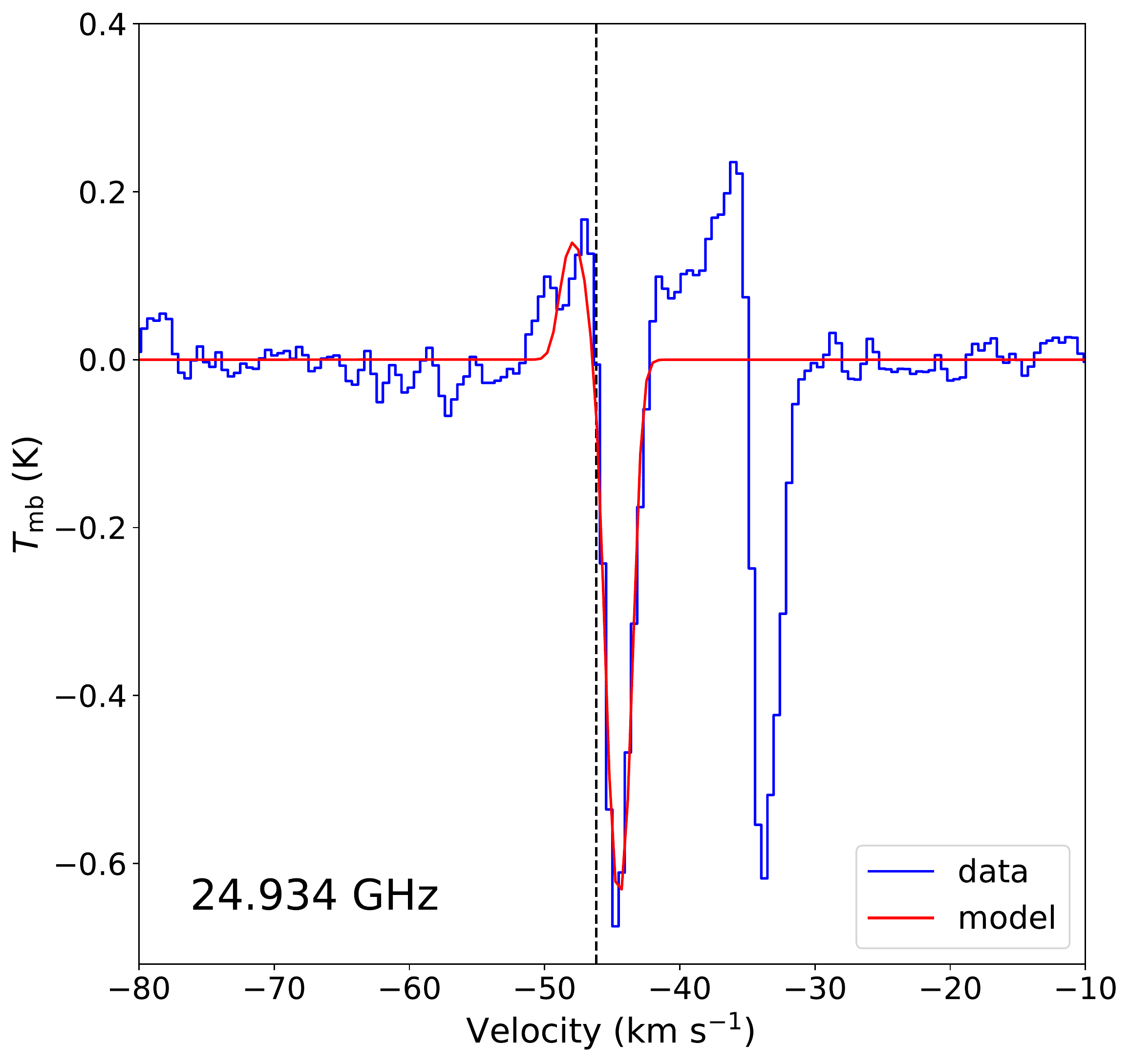}
\includegraphics[width=0.24\textwidth]{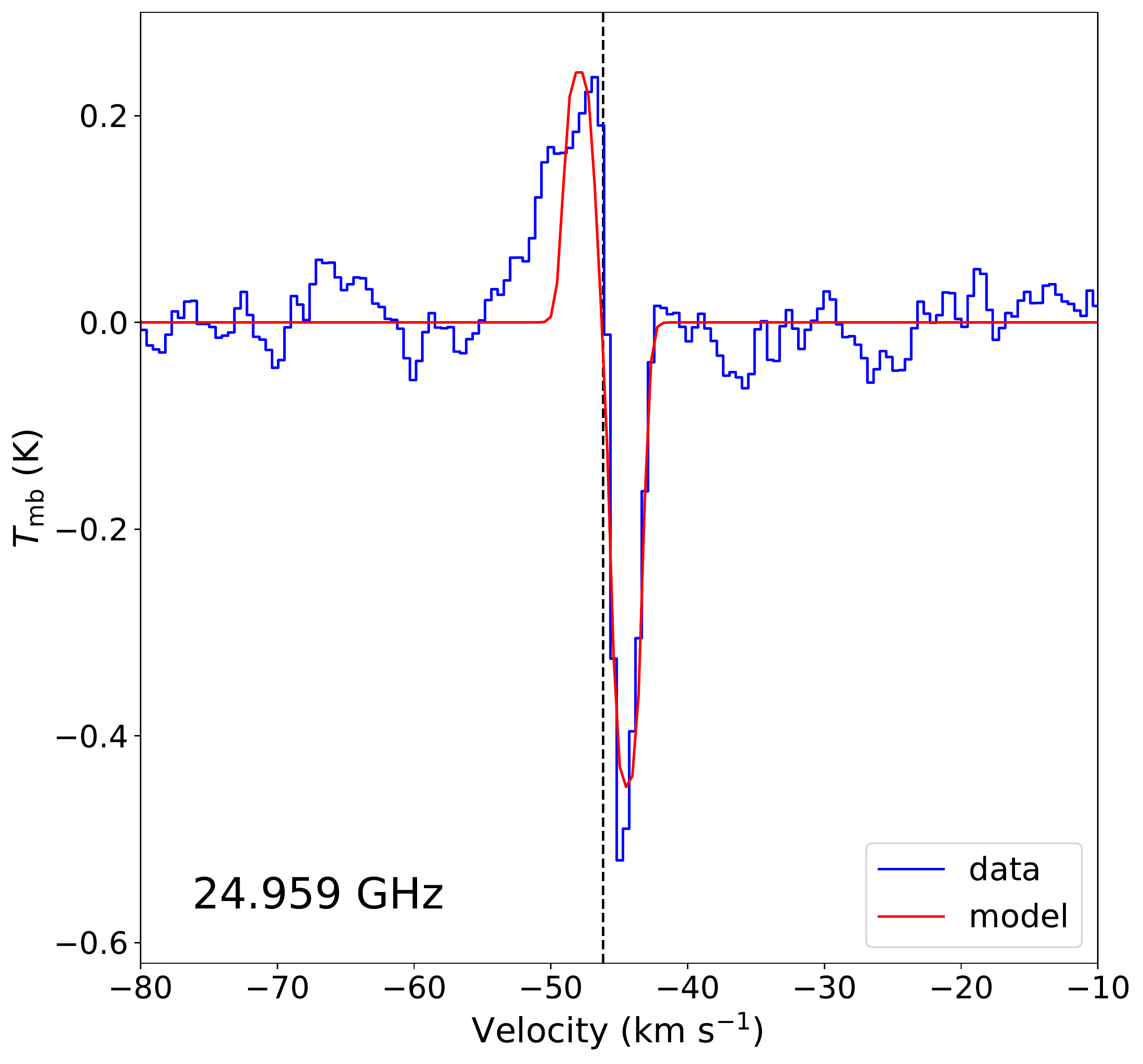}
\includegraphics[width=0.24\textwidth]{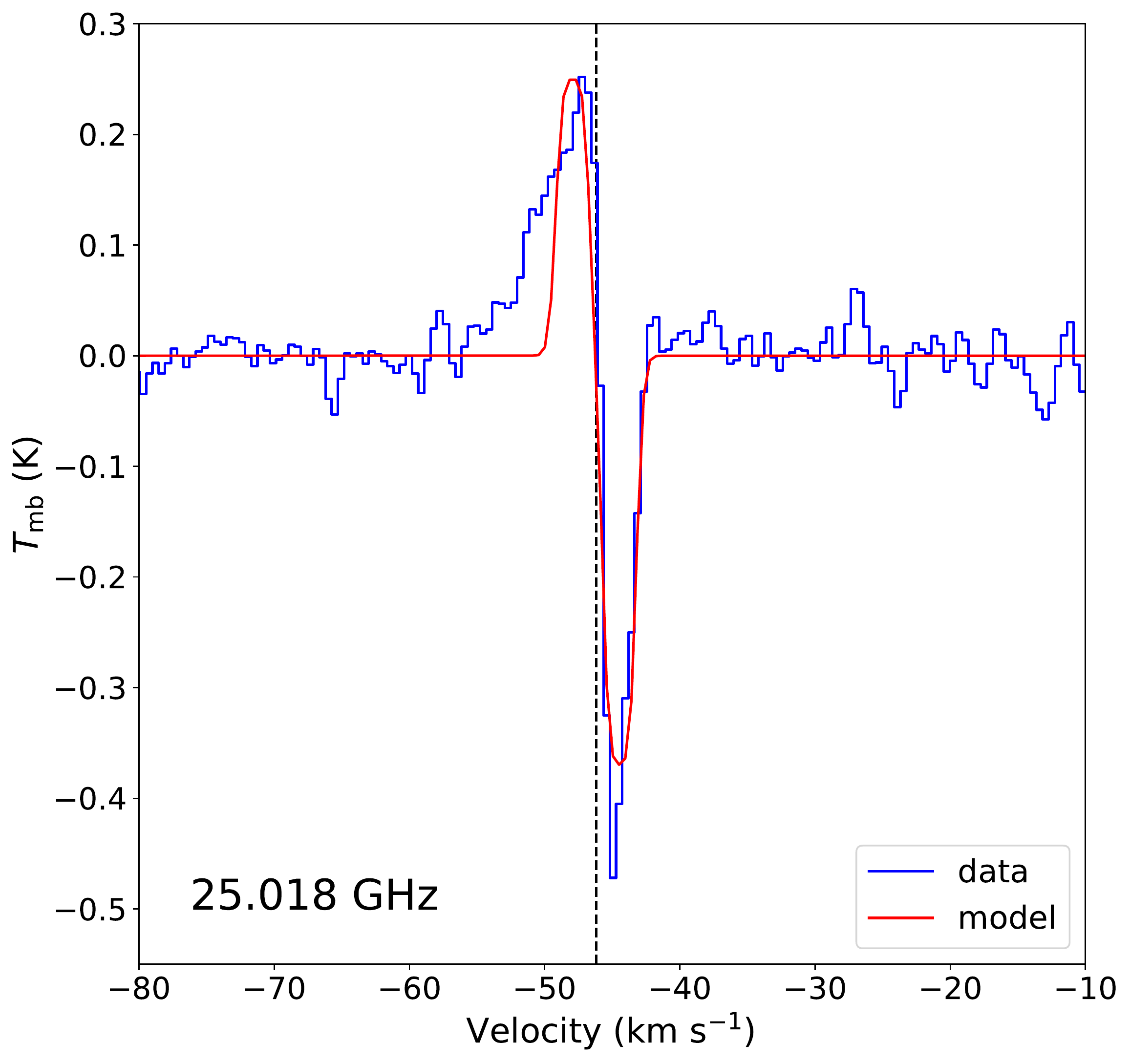}
\includegraphics[width=0.24\textwidth]{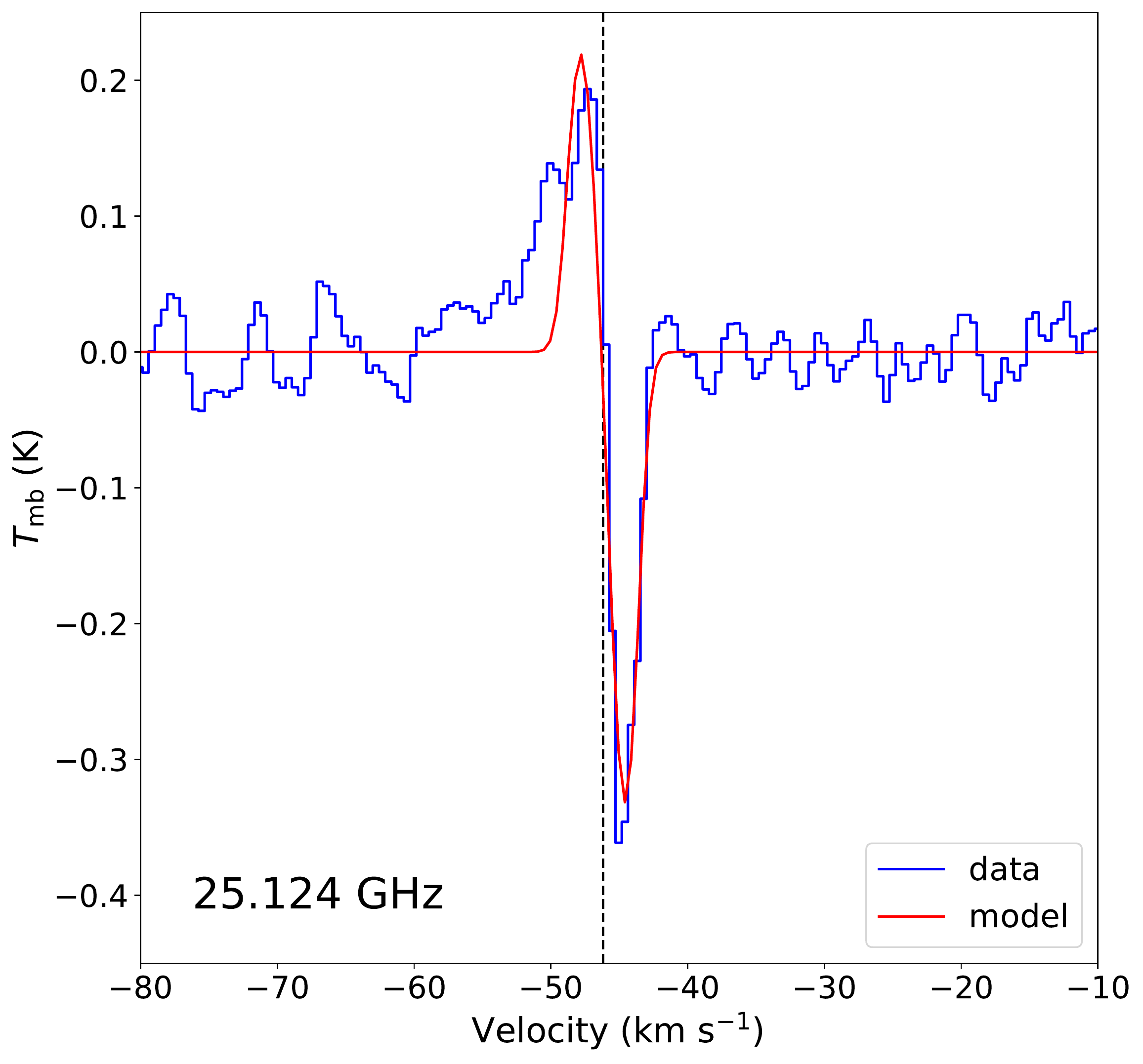}
\includegraphics[width=0.24\textwidth]{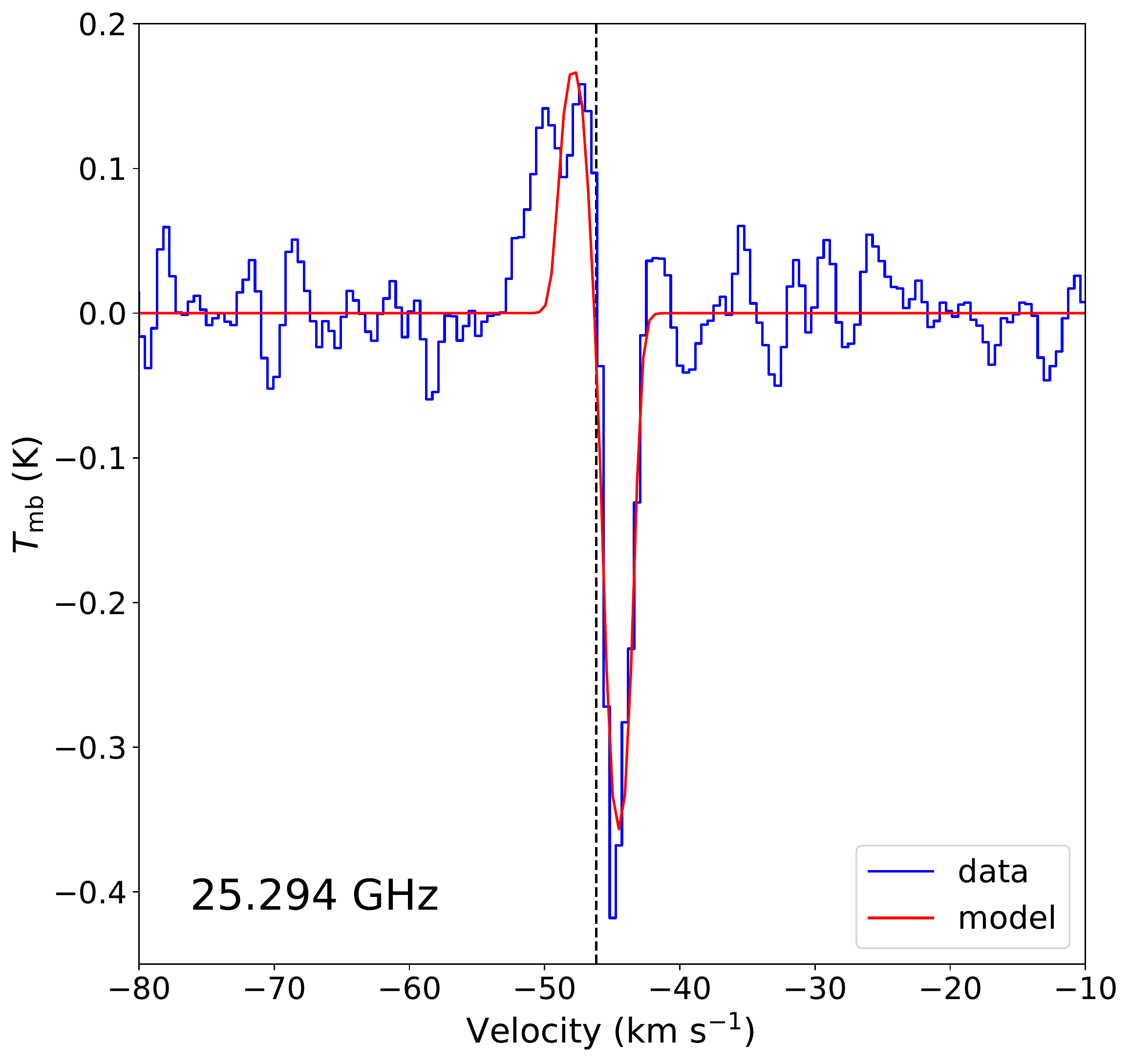}
\includegraphics[width=0.24\textwidth]{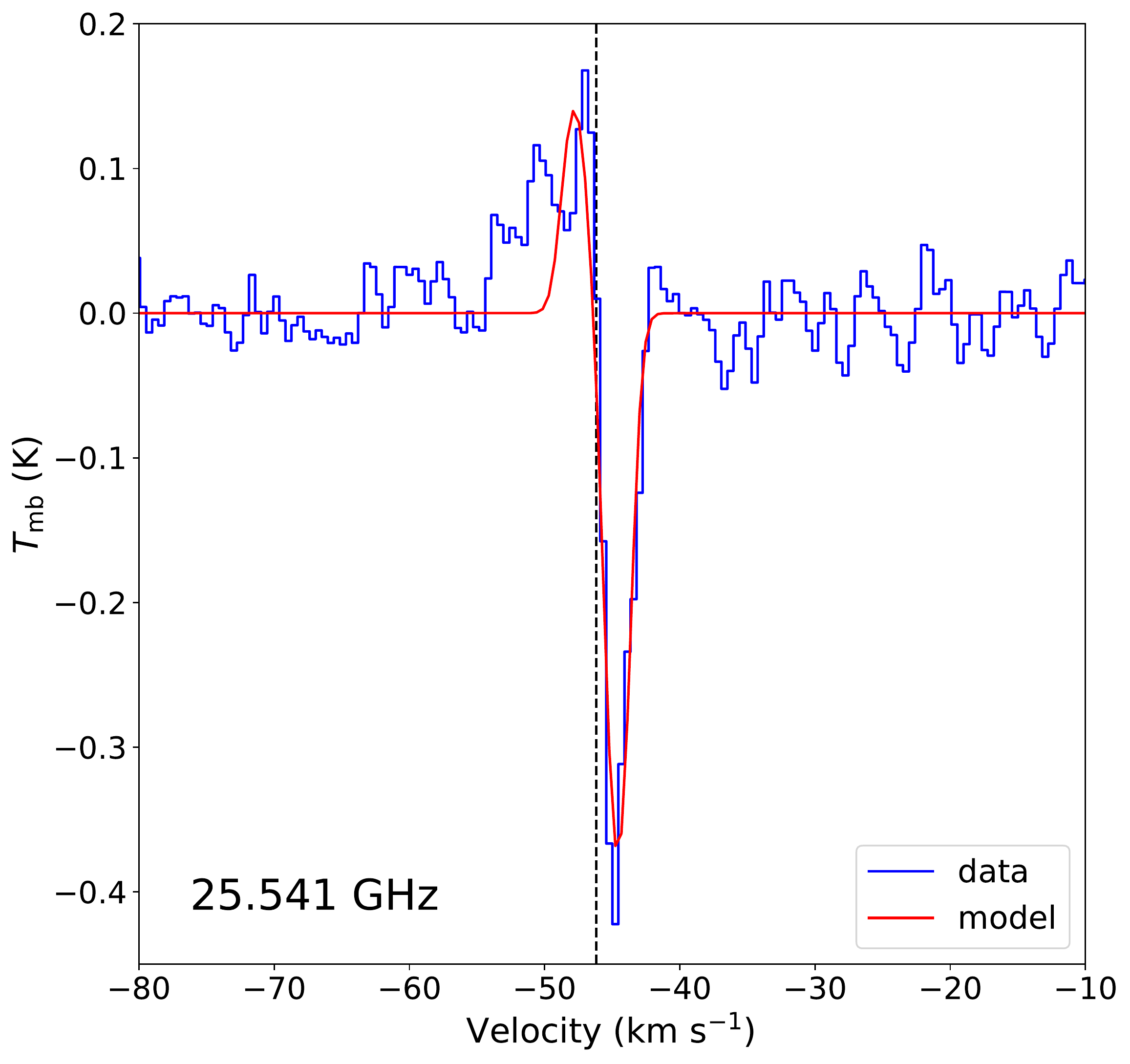}
\includegraphics[width=0.24\textwidth]{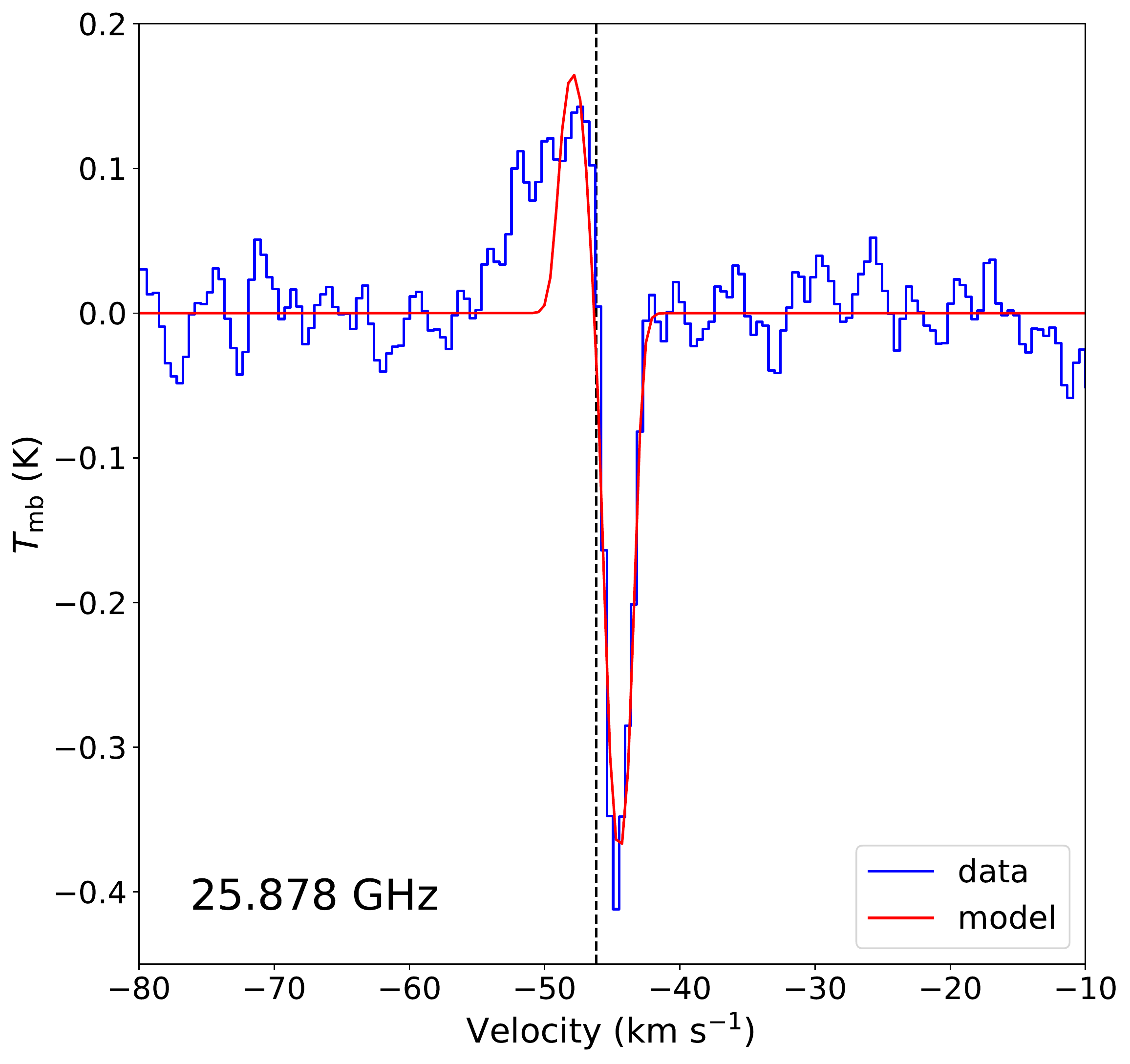}
\caption{Observed and modelled spectra for the methanol transitions with absorption features towards W3(OH). The short line name of each transition is denoted in the lower left corner of each panel. The blue spectra represent the observed data, and the red curves indicate the fitted line from the two-layer model with adopted parameters listed in Table~\ref{Tab:model}. The vertical black dashed lines represent the systemic velocity of $V_{\rm LSR}$=$-$46.19~\kms.
\label{fig:model-w3oh}}
\end{figure*}

In order to derive the free parameters in the model together with their uncertainties, we used the \textit{emcee} code \citep{2013PASP..125..306F} to perform Monte Carlo Markov chain (MCMC) calculations with the affine-invariant ensemble sampler \citep{2010CAMCS...5...65G}. Uniform priors are assumed for the five free parameters. The likelihood function is assumed to be ${\rm e}^{-\chi^{2}/2}$, and $\chi^{2}$ is defined as:
\begin{equation}
 \chi^{2} = \Sigma_{i}(T_{\rm{obs},i}-T_{\rm{mod},i})^{2}/\sigma_{\rm{obs},i}^{2}   
\end{equation}
where $T_{\rm{obs},i}$ and $T_{\rm{mod},i}$ are the observed and modelled brightness temperatures for each channel, and $\sigma_{\rm{obs},i}$ is the standard deviation of $T_{\rm{obs},i}$. The posterior distribution of these parameters is estimated by the product of the prior and likelihood functions. For the MCMC calculations, we run 20 walkers and 10000 steps after the burn-in phase. The 1$\sigma$ errors in the fitted parameters are estimated by the 16th and 84th percentiles of the posterior distribution. In order to help the convergence, the parameter spaces are set to be 0.5--10~\kms\/ for $V_{\rm in}$, 0.05--30 for $\tau_0$, 0.5--3~\kms\/ for $\sigma$, 0--150~K for $T_{\rm f}$ and $T_{\rm r}$.
The MCMC fitting to the 37.7~GHz CH$_{3}$OH line in W31C is shown in Fig.~\ref{fig:model-37}. The same method is applied to all the methanol transitions showing redshifted absorption, and the fitted spectra are presented in Figs.~\ref{fig:model}--\ref{fig:model-w3oh} for W31C and W3(OH), respectively, and Table~\ref{Tab:model} shows the fitting parameters.

We note that the infall velocity is the key parameter in deriving the infall rates, which are of central interest for understanding mass accretion in star formation. We tested the model with different fixed parameters to study how $V_{\rm in}$ is affected.
We find that the derived infall velocity is nearly independent of the assumed parameters. This is generally in line with previous studies \citep{2012A&A...544L...7P} in which such behaviour was verified using a different method (i.e. minimised-$\chi^{2}$ fitting). However, we also note that the presence of maser emission components in the redshifted velocity range might lead to underestimation of $V_{\rm in}$. Furthermore, the uncertainties in the assumed systemic velocity will propagate to $V_{\rm in}$, so there may be additional uncertainties of 0.44~\kms\,and 0.02~\kms\,in $V_{\rm in}$ for W31C and W3(OH) (see Secs.~\ref{Sec:thermal} and \ref{Sec:w3oh}). On the other hand, $\tau_0$, $T_{\rm f}$ and $T_{\rm r}$ are coupled (see Eqs.~\ref{eq:twolayer1}--\ref{eq:twolayer4}), so their values can be significantly affected by varying the fixed parameters. %In addition, the poor signal-to-noise ratios give rise to the large uncertainties in the fitted parameters (see the 86.9 GHz line in Fig.~\ref{fig:model} for example). 

The $V_{\rm in}$ value of 1.07$^{+0.23}_{-0.21}$~\kms\/ in the 25.541~GHz transition may be underestimated, because the observed spectrum is not well fitted by the model (see Fig.~\ref{fig:model}). The cause of the anomaly may be due to an inaccurate $V_{\rm LSR}$ for the 25.541 GHz line. If we use a velocity of $-$2~\kms\/, around which velocity the emission transits to absorption, we can reproduce a better fit giving $V_{\rm in} = 2.32^{+0.40}_{-0.94}$~\kms. $V_{\rm in}$ derived from the 86.6 and 86.9 GHz lines has a large uncertainty due to poor S/N ratios. 

Except for the 25.541, 86.6 and 86.9 GHz transitions, the derived $V_{\rm in}$ in W31C ranges from 2.0 to 3.5~\kms\,at corresponding linear scales of 0.3--1.0 pc, which is generally consistent with the infall velocities derived by other tracers at a linear scale of 0.2--0.3 pc \citep[e.g.][]{2013MNRAS.436.1335L,2017A&A...597A..70L}. Good agreement with other estimates of the infall velocity is also seen in W3(OH). Our derived $V_{\rm in}$ in W3(OH) lies in the range 1.6--1.8~\kms\,at a linear scale of $\sim$0.4~pc, consistent with previously reported infall velocities \citep{1987ApJ...323L.117K,1991A&A...251..220W}. These results suggest that redshifted methanol absorption can reliably estimate the infall velocities. 
Further discussion of the relationship between the infall velocity and the upper level, critical density and velocity dispersion can be found in Appendix~\ref{sec:appendix}.

The $V_{\rm in}$ values we obtain are lower than the infall velocities (4.5--6.5~\kms) measured on smaller scales of 0.02--0.05~pc in W31C \citep{1987ApJ...318..712K,1988ApJ...324..920K,2002ApJ...568..754K,2005ApJ...624L..49S}. As already mentioned, our beams cover both W3(H$_2$O) and W3(OH). High angular HCO$^+$ (3--2) resolution observations suggest that the infall velocity of W3(H$_2$O) is 2.7$\pm$0.3 \kms\,at a linear scale of $\sim$0.02~pc \citep{2016MNRAS.456.2681Q}, higher than our values of 1.5--1.8~\kms. Both cases suggest that the infall velocities are lower at larger scales. This is readily explained by gravity-dominated velocity fields that are commonly assumed to follow a power-law distribution \citep{1988ApJ...324..920K,2003cdsf.conf..157E,2018MNRAS.477.4951L}, for instance, the free-fall velocity follows $V_{\rm in}\sim r^{-0.5}$. Therefore, our observations are likely to trace the large-scale inward motions of the two sources. 
Our spectra of W31C indicate that inward motions may be present on scales larger than previously reported. However, our observations can only loosely constrain the linear scale ($\lesssim$1~pc) with the beam size, so further mapping observations are needed to better constrain the scale of inward motions. 
%Especially, our spectra of W31C provide evidence for the presence of inward motions at a scale of $\lesssim$1~pc which is the to-date largest scale to show inward motions in W31C. 

Assuming a uniform density sphere undergoing spherically symmetric free fall for both sources \citep[e.g.][]{2012A&A...544L...7P}, we estimate the mass infall rates to be (3--15)$\times 10^{-3}$~$M_{\odot}$~yr$^{-1}$ for W31C and (1--2)$\times 10^{-3}$~$M_{\odot}$~yr$^{-1}$ for W3(OH), similar to the values estimated for massive clumps in other studies \citep[e.g,][]{2010A&A...520A..49S}. Mass infall rates of this magnitude are able to overcome the radiation pressure due to the luminosity of the central star during the high-mass star formation process \citep{2003ApJ...585..850M}. The large-scale inward motions and the high mass infall rates favour a global collapse for both sources, consistent with the global hierarchical collapse and clump-fed scenario \citep[e.g.][]{2018ARA&A..56...41M,2019MNRAS.490.3061V} that is different from the low-mass star formation picture (see also Appendix~\ref{sec:appendix}).

%\begin{figure*}[!htbp]
%\centering
%\includegraphics[width=0.33\textwidth]{mcmc-w3oh-20.1.pdf}
%\includegraphics[width=0.33\textwidth]{mcmc-w3oh-23.4.pdf}
%\includegraphics[width=0.33\textwidth]{mcmc-w3oh-24.928.pdf}
%\includegraphics[width=0.33\textwidth]{mcmc-w3oh-24.933.pdf}
%\includegraphics[width=0.33\textwidth]{mcmc-w3oh-24.934.pdf}
%\includegraphics[width=0.33\textwidth]{mcmc-w3oh-24.959.pdf}
%\includegraphics[width=0.33\textwidth]{mcmc-w3oh-25.018.pdf}
%\includegraphics[width=0.33\textwidth]{mcmc-w3oh-25.124.pdf}
%\includegraphics[width=0.33\textwidth]{mcmc-w3oh-25.294.pdf}
%\includegraphics[width=0.33\textwidth]{mcmc-w3oh-25.541.pdf}
%\includegraphics[width=0.33\textwidth]{mcmc-w3oh-25.878.pdf}
%\caption{
%\label{fig:mcmc-w3oh-all}}
%\end{figure*}

\subsection{Methanol absorption in different environments}\label{sec:unique}
Our observations have revealed redshifted absorption in  multiple methanol
transitions towards both W31C and W3(OH). However, absorption features are not always observed in the same transitions for different sources. For example, 
redshifted absorption features are detected in the $2_1-3_0E$ line at 
19.9~GHz and the $9_2-10_1A^+$ line at 23.1~GHz towards W31C, but these two 
transitions are found to show (class II) maser emission in W3(OH), and thus may also affect absorption of W31C, towards which also the  6.7 and 12.1 GHz transitions show absorption as well as maser features.

Furthermore, towards W3(OH), the nine $J_2-J_1E$ lines all show (apparent) 
inverse P-Cygni profiles towards W3(OH). These resemble very 
closely the profiles of the 24/25 GHz $J,K$ inversion 
transitions of NH$_3$ observed towards this region (like us) with the Effelsberg 100 m telescope by \citet{Wilson1978, Mauersberger1986}. Mapping of the $J,K = (5,5)$ line by 
\citet{Mauersberger1986} (with the $38''$ Effelsberg beam) showed that the emission component arises from a position $\approx 5''$ east of the absorption, which arises from the 
UCH{\scriptsize II} region. Consequently, \citet{Mauersberger1986} 
persuasively argued that the emission arises from the hot core(s) discussed in Sec.~\ref{sec:intro}, which actually was one of the very first hot molecular cores ever identified  \citep[by][]{TurnerWelch1984} at the site of H$_2$O maser emission in the region \citep{Forster1977}. 

%% Inverse P-Cygni profiles become more prominent in W31C only for the $J_2-J_1E$ transitions with $J>$8. 
%This is likely caused by different excitation conditions and continuum levels in the two sources.
Because of this source-by-source variation, it might not be straightforward to use a single methanol transition to systematically search for infall motions towards high-mass star formation regions. Abundant methanol transitions that are accessible by ground-based telescopes cover a broad range of energy levels and critical densities (see Table~\ref{Tab:freq}). This allows us to investigate inward motions at different scales and for different environments. Therefore, observations of multiple methanol transitions provide a suitable choice for such studies.

%A possible reason for the difference is that the excitation conditions for methanol in W31C are different from those in W3(OH). The maser emission at 19.9 and 23.1 GHz belong to class II methanol maser, which are pumped by infrared radiation, while the $J_2-J_1E$ lines are classified as class I methanol maser that arise in the collision-dominated regions.

Based on K-band single-dish observations, NGC7538 IRS1 shows methanol maser emission and absorption features similar to W3(OH), but the absorption velocities with respect to their respective systemic velocities are different. In both sources, absorption features have been detected in the $J_2-J_1E$ ($J$=2,3,4,5,6,9) lines near 25~GHz \citep{1986A&A...157..318M}, the $10_1-9_2A^-$ line at 23.4~GHz, and the $11_1-10_2A^+$ line at 20.1~GHz \citep{1986A&A...169..271M}, while maser emission has been detected in the $2_1-3_0E$ line at 19.9~GHz \citep{1985A&A...147L..19W} and the $9_2-10_1A^+$ line at 23.1~GHz \citep{1984A&A...134L...7W}. The systemic velocity of NGC7538 IRS1 is about $-$57.4~\kms\,derived from the $2_K-1_K$ thermal methanol quartet lines near 96 GHz \citep{2002A&A...387..179M}. The methanol transitions at 20.1, 23.4~GHz and six $J_2-J_1E$ lines near 25~GHz show absorption and the velocities of these absorption features range from $-$59.9 to $-$59.1~\kms, revealing that the absorption features are actually blueshifted with respect to the systemic velocity. This result is also in line with previous NH$_3$ (1,1), (2,2), (3,3) measurements \citep[e.g.][]{1983A&A...127L..19W,1984ApJ...282L..93H,1991ApJ...371..163K}. These blueshifted absorption lines are indicative of expansion in NGC7538 IRS1 which is likely caused by its associated powerful outflows \citep[e.g.][]{2011ApJ...728....6Q}. This is in stark contrast to our observed redshifted absorption in W3(OH). This indicates that these two sources have very different dominant large-scale motions on the clump-scales probed by single-dish observations. Therefore, our single-dish methanol observations support that both W31C and W3(OH) host ongoing large-scale inward motions, although both sources also show outflows on smaller scales \citep{2010ApJ...725.2190L,2011ApJ...740L..19Z,2016MNRAS.456.2681Q}. On the other hand, higher angular resolution ($\sim$0$\rlap{.}\arcsec$2) observations of the high-energy line HCN (4$-$3) $v_2$=1 ($E_{\rm up}$ = 1050~K) shows redshifted absorption towards NGC7538 IRS1, indicating infall motions in its innermost regions \citep{2013A&A...558A..81B}. Therefore, one might need high energy and high critical density methanol transitions to disentangle the infall motions and outflows in sources exhibiting complex velocity structures. Nevertheless, the single-dish redshifted methanol absorption measurements have been demonstrated to be a good tracer of the large-scale inward motions for high-mass star-formation regions, providing motivation for follow-up studies with higher angular resolution observations.

It is important to note that CH$_3$OH absorption in the 6.7 and 12.1 GHz lines has also been detected in low-mass star formation regions including TMC1, L183, and NGC 1333 \citep{1988A&A...197..271W,2008A&A...489.1175P}. These generally do not show centimetre continuum radiation. Instead, the absorption is produced by overcooling against the cosmic microwave background (see Sec.~\ref{sec:intro}). In these sources the absorption dips are at the systemic velocity, that is, the absorption is neither redshifted nor blueshifted, although they are known to show infall motions from other studies \citep[e.g.][]{2001ApJ...562..770D,2007ApJ...671.1839S,2009ApJ...701.1044K}. There are two possible explanations for this. One is that the relatively compact size will cause a significant dilution in single-dish observations. Unlike the case of high-mass star formation regions, which can show large-scale collapse, low-mass star formation regions usually exhibit local collapse, that is, collapse only becomes prominent towards compact dense cores. For the NGC1333 IRAS4A observations, the beam size was about 40$\arcsec$\,for the 6.7 GHz methanol absorption observations \citep{2008A&A...489.1175P}, while the collapsing region is known to have an angular scale of $\sim$5--7$\arcsec$ from interferometric observations of H$_2$CO $3_{12}-2_{11}$ and N$_2$H$^+$ (1$-$0) \citep{2001ApJ...562..770D}. This implies that the filling factor was only 2\%--3\%. On the other hand, anti-inversion of the 6.7 GHz methanol transition is expected in a more extended region with densities of $<10^{6}$~cm$^{-3}$ \citep{2008A&A...489.1175P}. Therefore, the redshifted methanol absorption might be largely diluted by the large beam of the single-dish observations. The second possible explanation is that the infall velocities of these low-mass star formation regions are $\lesssim$0.5~\kms\,\citep[e.g.][]{2001ApJ...562..770D,2007ApJ...671.1839S}, generally lower than those of their high-mass counterparts. The low infall velocities will result in small velocity shifts in the spectra which are difficult to disentangle. For example, the velocity shift is as small as $\lesssim$0.15~\kms\,in TMC1 \citep{2007ApJ...671.1839S}. 
Interferometric observations of CH$_3$OH absorption with high spectral resolution are therefore required to determine whether redshifted methanol absorption can trace infall motions towards low-mass star formation. 

\subsection{Redshifted methanol absorption is common or not in high-mass star formation regions}
As discussed in Sec.~\ref{sec:model}, the two studied objects contain UCH{\scriptsize II} regions with high free-free continuum flux densities. If bright continuum sources are required to be able to detect the absorption, then redshifted methanol absorption can only be observed in a limited number of sources at a late evolutionary stage where they host a bright UCH{\scriptsize II} region. In order to roughly estimate the number of potential sources, we have used the results from the Coordinated Radio ``N" Infrared Survey for High-mass star formation (CORNISH) project that covers 110 deg$^2$ of the northern sky (10$^\circ<l<65^\circ$, $|b|<$1$^\circ$) using the VLA in B and BnA configurations at 5~GHz \citep{2012PASP..124..939H,2013ApJS..205....1P}. Based on the CORNISH survey, \cite{2018A&A...615A.103K} compiled a catalogue of 239 UCH{\sc ii} regions. In this catalogue, the flux density of W31C(G10.6234$-$0.3837) at 5~GHz is measured to be 1952.22~mJy. We find only 10 out of 239 sources (including W31C) have flux densities of $\geq$1952.22~mJy and if we adopt that as the threshold criterion, it implies that redshifted methanol absorption may be detected in only 4\% of the UCH{\sc ii} regions, if we neglect the collisional pumping effects. 
It is worth noting that the CORNISH survey is only sensitive to compact UCH{\sc ii} regions due to the missing short spacing problem. 
%For instance, G30.783$-$0.027, an extended H{\sc ii} region with a radius of 89$\arcsec$, shows an integrated flux density of 9979.49~mJy at 5.8~GHz from the VLA observations in the D configuration \citep{2019A&A...627A.175M}, which is significantly higher than the CORNISH value of 87.53~mJy at 5~GHz. 
Given the single-dish measurements that could be pointed against extended H{\sc ii} regions, the estimated percentage of $\sim$4\% based on the CORNISH survey should be underestimated.
%To examine this, we check the radio source catalogue (28$^\circ<l<36^\circ$, $|b|<$1$^\circ$) of the Global View of Star Formation in the Milky Way (GLOSTAR) survey \citep{2019A&A...627A.175M} and find that G30.783$-$0.027, an extended H{\sc ii} region with a radius of 89$\arcsec$, shows an integrated flux density of 9979.49~mJy at 5.8~GHz, compared to 87.53~mJy reported from the CORNISH survey.
On the other hand, the W31C flux threshold appears to contradict the observations of \citet{1992MNRAS.254..301P} who reported a detection rate of 66\% (38/58) for methanol absorption at 12.2 GHz towards bright southern H{\sc ii} regions and some selected dark clouds. 
This suggests that the criteria of using the flux density of W31C is an unnecessarily high threshold, however, the detection of methanol absorption towards sources without UCH{\scriptsize II} regions suggests that collisional pumping effects cannot be neglected in at least a handful of cases \citep[e.g.][]{1988A&A...197..271W,2008A&A...489.1175P}. %This is also supported by our observational results that different absorption intensities for different transitions which are close in frequency.
This is also supported by our observations that the absorption intensities differ for different transitions with similar rest frequencies.
The overcooling of methanol transitions due to collisional pumping can greatly aid in the detection of methanol absorption. 
%It is worth mentioning that \citet{1992MNRAS.254..301P} found there is a velocity offset in statistics of 2.7~\kms\, between absorption features at 12.2~GHz and H{\sc ii} regions traced by H109$\alpha$. Their data also shows that 16 (nearly half) sources with redshifted methanol absorption compared to H109$\alpha$ velocity, which may indicate the detection of redshifted methanol absorption is not that rare than we predict.

Based on our current work, we can only give a lower limit of $\sim$4\% for the fraction of potential UCH{\scriptsize II} regions showing methanol absorption. 
For many regions the overcooling of methanol transitions may also enhance absorption, as is observed towards low-mass sources and dark clouds, but this requires specific physical conditions \citep[e.g.][]{1988A&A...197..271W,2008A&A...489.1175P,2016A&A...592A..31L}.
The work undertaken to date suggests that redshifted methanol absorption may be most suitable for studying infall motions towards high-mass star formation regions hosting bright H{\scriptsize II} regions.
%On the other hand, the overcooling requires specific physical conditions \citep[e.g.][]{2008A&A...489.1175P,2016A&A...592A..31L}.
%This may indicate that redshifted methanol absorption is not suitable to study infall motions of high-mass star formation regions at different evolutionary stages. 

\section{Summary}\label{Sec:sum}
We have performed observations of multiple methanol transitions towards two well-known collapsing dense clumps, W31C(G10.6$-$0.4) and W3(OH), with the Effelsberg-100 m, IRAM-30 m, and APEX-12 m telescopes. We detected 14 and 11 redshifted methanol absorption towards W31C and W3(OH), respectively. Redshifted methanol absorption in the 20.1, 25.541, 25.878, 37.7, 38.3, 38.5, 85.6, 86.6 and 86.9 GHz transitions are reported towards W31C for the first time. %The nine $J_{2}$--$J_{1}$ lines of E-type methanol near 25 GHz show inverse P-Cygni profiles in W3(OH). 
The infall velocities estimated using a two-layer model agree with previously reported values derived from other tracers, suggesting that redshifted methanol absorption is a reliable tracer of infall motions within high-mass star formation regions hosting bright H{\scriptsize II} regions. Furthermore, our observations indicate the presence of large-scale ($\lesssim$1 pc) inward motions, and the mass infall rates are estimated to be $\gtrsim$10$^{-3}$~$M_{\odot}$~yr$^{-1}$, which supports the global hierarchical collapse and clump-fed scenario.

%We detected 14 methanol transitions with absorption feature were detected towards W31C, at the frequencies of 6.7, 12.2, 19.9, 20.1, 23.1, 23.4, 25.541, 25.878, 37.7, 38.3, 38.5, 85.6, 86.6 and 86.9 GHz.
%Nine of them were discovered for the first time, the frequencies of these transitions are 20.1, 25.541, 25.878, 37.7, 38.3, 38.5, 85.6, 86.6 and 86.9 GHz.
%All 14 detected absorption features are redshifted to the systemic velocity, and the CH$_{3}$OH transitions at 19.9, 25.541, 25.878 and 85.6 GHz show an inverse P-Cygni profile, while the class I CH$_{3}$OH maser are blueshifted to the systemic velocity.
%Towards W3(OH), a series of nine $J_2-J_1E$ CH$_{3}$OH lines ($J$ from 2 to 10) near 25 GHz show an inverse P-Cygni profile.

%The redshifted absorption features and inverse P-Cygni profiles of these transitions indicate that these two sources are harbouring infall motions in clump scale.

\section*{ACKNOWLEDGMENTS}\label{sec:ack}
We greatly thank Dr. Yuxin Lin for her help on the critical density calculations and discussions about the multi-layer view of global collapse.
We acknowledge the IRAM-30 m, APEX-12 m, and Effelsberg-100 m staff for their assistance with our observations. 
%This work is based on observations carried out under project numbers 181-10, 043-19, and 045-19 with the IRAM 30m telescope. 
IRAM is supported by INSU/CNRS (France), MPG (Germany) and IGN (Spain). The research leading to these results has received funding from the European Union’s Horizon 2020 research and innovation programme under grant agreement No 730562 [RadioNet]. 
%This publication is based on data acquired with the Atacama Pathfinder Experiment (APEX) under programme 0103.F-9516A. 
APEX is a collaboration between the Max-Planck-Institut f{\"u}r Radioastronomie, the European Southern Observatory, and the Onsala Space Observatory. This work is also based on observations with the 100-m telescope of the MPIfR (Max-Planck-Institut für Radioastronomie) at Effelsberg. This work made use of Python libraries including Astropy\footnote{\url{https://www.astropy.org/}} \citep{2013A&A...558A..33A},  NumPy\footnote{\url{https://www.numpy.org/}} \citep{5725236}, SciPy\footnote{\url{https://www.scipy.org/}} \citep{jones2001scipy}, Matplotlib\footnote{\url{https://matplotlib.org/}} \citep{Hunter:2007}, APLpy \citep{2012ascl.soft08017R}, emcee\footnote{\url{https://emcee.readthedocs.io/en/stable/}} \citep{2013PASP..125..306F}, and corner.py \citep{Foreman-Mackey2016}. This paper was prepared using the Overleaf\footnote{\url{https://www.overleaf.com}} web application.

% seaborn\footnote{\url{https://seaborn.pydata.org/}}, 

\bibliographystyle{aa}
\bibliography{references}

\begin{thebibliography}{139}
\expandafter\ifx\csname natexlab\endcsname\relax\def\natexlab#1{#1}\fi

\bibitem[{{Ahmadi} {et~al.}(2018){Ahmadi}, {Beuther}, {Mottram}, {Bosco},
  {Linz}, {Henning}, {Winters}, {Kuiper}, {Pudritz}, {S{\'a}nchez-Monge},
  {Keto}, {Beltran}, {Bontemps}, {Cesaroni}, {Csengeri}, {Feng},
  {Galvan-Madrid}, {Johnston}, {Klaassen}, {Leurini}, {Longmore}, {Lumsden},
  {Maud}, {Menten}, {Moscadelli}, {Motte}, {Palau}, {Peters}, {Ragan},
  {Schilke}, {Urquhart}, {Wyrowski}, \& {Zinnecker}}]{Ahmadi2018}
{Ahmadi}, A., {Beuther}, H., {Mottram}, J.~C., {et~al.} 2018, \aap, 618, A46

\bibitem[{{Astropy Collaboration} {et~al.}(2013){Astropy Collaboration},
  {Robitaille}, {Tollerud}, {Greenfield}, {Droettboom}, {Bray}, {Aldcroft},
  {Davis}, {Ginsburg}, {Price-Whelan}, {Kerzendorf}, {Conley}, {Crighton},
  {Barbary}, {Muna}, {Ferguson}, {Grollier}, {Parikh}, {Nair}, {Unther},
  {Deil}, {Woillez}, {Conseil}, {Kramer}, {Turner}, {Singer}, {Fox}, {Weaver},
  {Zabalza}, {Edwards}, {Azalee Bostroem}, {Burke}, {Casey}, {Crawford},
  {Dencheva}, {Ely}, {Jenness}, {Labrie}, {Lim}, {Pierfederici}, {Pontzen},
  {Ptak}, {Refsdal}, {Servillat}, \& {Streicher}}]{2013A&A...558A..33A}
{Astropy Collaboration}, {Robitaille}, T.~P., {Tollerud}, E.~J., {et~al.} 2013,
  \aap, 558, A33

\bibitem[{{Batrla} {et~al.}(1987){Batrla}, {Matthews}, {Menten}, \&
  {Walmsley}}]{1987Natur.326...49B}
{Batrla}, W., {Matthews}, H.~E., {Menten}, K.~M., \& {Walmsley}, C.~M. 1987,
  \nat, 326, 49

\bibitem[{{Beuther} {et~al.}(2013){Beuther}, {Linz}, \&
  {Henning}}]{2013A&A...558A..81B}
{Beuther}, H., {Linz}, H., \& {Henning}, T. 2013, \aap, 558, A81

\bibitem[{{Breen} {et~al.}(2019){Breen}, {Contreras}, {Dawson}, {Ellingsen},
  {Voronkov}, \& {McCarthy}}]{2019MNRAS.484.5072B}
{Breen}, S.~L., {Contreras}, Y., {Dawson}, J.~R., {et~al.} 2019, \mnras, 484,
  5072

\bibitem[{{Breen} {et~al.}(2014){Breen}, {Ellingsen}, {Caswell}, {Green},
  {Voronkov}, {Avison}, {Fuller}, {Quinn}, \& {Titmarsh}}]{2014MNRAS.438.3368B}
{Breen}, S.~L., {Ellingsen}, S.~P., {Caswell}, J.~L., {et~al.} 2014, \mnras,
  438, 3368

\bibitem[{{Breen} {et~al.}(2010){Breen}, {Ellingsen}, {Caswell}, \&
  {Lewis}}]{2010MNRAS.401.2219B}
{Breen}, S.~L., {Ellingsen}, S.~P., {Caswell}, J.~L., \& {Lewis}, B.~E. 2010,
  \mnras, 401, 2219

\bibitem[{{Carter} {et~al.}(2012){Carter}, {Lazareff}, {Maier}, {Chenu},
  {Fontana}, {Bortolotti}, {Boucher}, {Navarrini}, {Blanchet}, {Greve}, {John},
  {Kramer}, {Morel}, {Navarro}, {Pe{\~n}alver}, {Schuster}, \&
  {Thum}}]{2012A&A...538A..89C}
{Carter}, M., {Lazareff}, B., {Maier}, D., {et~al.} 2012, \aap, 538, A89

\bibitem[{{Caswell}(2009)}]{2009PASA...26..454C}
{Caswell}, J.~L. 2009, \pasa, 26, 454

\bibitem[{{Caswell} {et~al.}(1995{\natexlab{a}}){Caswell}, {Vaile},
  {Ellingsen}, \& {Norris}}]{1995MNRAS.274.1126C}
{Caswell}, J.~L., {Vaile}, R.~A., {Ellingsen}, S.~P., \& {Norris}, R.~P.
  1995{\natexlab{a}}, \mnras, 274, 1126

\bibitem[{{Caswell} {et~al.}(1995{\natexlab{b}}){Caswell}, {Vaile},
  {Ellingsen}, {Whiteoak}, \& {Norris}}]{1995MNRAS.272...96C}
{Caswell}, J.~L., {Vaile}, R.~A., {Ellingsen}, S.~P., {Whiteoak}, J.~B., \&
  {Norris}, R.~P. 1995{\natexlab{b}}, \mnras, 272, 96

\bibitem[{{Chen} {et~al.}(2012){Chen}, {Ellingsen}, {He}, {Xu}, {Gan}, {Shen},
  {An}, {Sun}, \& {Ju}}]{2012ApJS..200....5C}
{Chen}, X., {Ellingsen}, S.~P., {He}, J.-H., {et~al.} 2012, \apjs, 200, 5

\bibitem[{{Coudert} {et~al.}(2015){Coudert}, {Gutl{\'e}}, {Huet}, {Grabow}, \&
  {Levshakov}}]{2015JChPh.143d4304C}
{Coudert}, L.~H., {Gutl{\'e}}, C., {Huet}, T.~R., {Grabow}, J.~U., \&
  {Levshakov}, S.~A. 2015, \jcp, 143, 044304

\bibitem[{{Cragg} {et~al.}(2005){Cragg}, {Sobolev}, \&
  {Godfrey}}]{2005MNRAS.360..533C}
{Cragg}, D.~M., {Sobolev}, A.~M., \& {Godfrey}, P.~D. 2005, \mnras, 360, 533

\bibitem[{{Csengeri} {et~al.}(2016){Csengeri}, {Leurini}, {Wyrowski},
  {Urquhart}, {Menten}, {Walmsley}, {Bontemps}, {Wienen}, {Beuther}, {Motte},
  {Nguyen-Luong}, {Schilke}, {Schuller}, {Zavagno}, \&
  {Sanna}}]{2016A&A...586A.149C}
{Csengeri}, T., {Leurini}, S., {Wyrowski}, F., {et~al.} 2016, \aap, 586, A149

\bibitem[{{Di Francesco} {et~al.}(2001){Di Francesco}, {Myers}, {Wilner},
  {Ohashi}, \& {Mardones}}]{2001ApJ...562..770D}
{Di Francesco}, J., {Myers}, P.~C., {Wilner}, D.~J., {Ohashi}, N., \&
  {Mardones}, D. 2001, \apj, 562, 770

\bibitem[{{Dreher} \& {Welch}(1981)}]{1981ApJ...245..857D}
{Dreher}, J.~W. \& {Welch}, W.~J. 1981, \apj, 245, 857

\bibitem[{{Ellingsen} {et~al.}(2011){Ellingsen}, {Breen}, {Sobolev},
  {Voronkov}, {Caswell}, \& {Lo}}]{2011ApJ...742..109E}
{Ellingsen}, S.~P., {Breen}, S.~L., {Sobolev}, A.~M., {et~al.} 2011, \apj, 742,
  109

\bibitem[{{Ellingsen} {et~al.}(2012){Ellingsen}, {Sobolev}, {Cragg}, \&
  {Godfrey}}]{2012ApJ...759L...5E}
{Ellingsen}, S.~P., {Sobolev}, A.~M., {Cragg}, D.~M., \& {Godfrey}, P.~D. 2012,
  \apjl, 759, L5

\bibitem[{{Ellingsen} {et~al.}(2018){Ellingsen}, {Voronkov}, {Breen},
  {Caswell}, \& {Sobolev}}]{2018MNRAS.480.4851E}
{Ellingsen}, S.~P., {Voronkov}, M.~A., {Breen}, S.~L., {Caswell}, J.~L., \&
  {Sobolev}, A.~M. 2018, \mnras, 480, 4851

\bibitem[{{Elmegreen} \& {Scalo}(2004)}]{2004ARA&A..42..211E}
{Elmegreen}, B.~G. \& {Scalo}, J. 2004, \araa, 42, 211

\bibitem[{{Evans}(2003)}]{2003cdsf.conf..157E}
{Evans}, Neal, I. 2003, in SFChem 2002: Chemistry as a Diagnostic of Star
  Formation, ed. C.~L. {Curry} \& M.~{Fich}, 157

\bibitem[{Foreman-Mackey(2016)}]{Foreman-Mackey2016}
Foreman-Mackey, D. 2016, Journal of Open Source Software, 1, 24

\bibitem[{{Foreman-Mackey} {et~al.}(2013){Foreman-Mackey}, {Hogg}, {Lang}, \&
  {Goodman}}]{2013PASP..125..306F}
{Foreman-Mackey}, D., {Hogg}, D.~W., {Lang}, D., \& {Goodman}, J. 2013, \pasp,
  125, 306

\bibitem[{{Forster} {et~al.}(1990){Forster}, {Caswell}, {Okumura}, {Hasegawa},
  \& {Ishiguro}}]{1990A&A...231..473F}
{Forster}, J.~R., {Caswell}, J.~L., {Okumura}, S.~K., {Hasegawa}, T., \&
  {Ishiguro}, M. 1990, \aap, 231, 473

\bibitem[{{Forster} {et~al.}(1977){Forster}, {Welch}, \&
  {Wright}}]{Forster1977}
{Forster}, J.~R., {Welch}, W.~J., \& {Wright}, M.~C.~H. 1977, \apjl, 215, L121

\bibitem[{{Gildas Team}(2013)}]{2013ascl.soft05010G}
{Gildas Team}. 2013, {GILDAS: Grenoble Image and Line Data Analysis Software}

\bibitem[{{Goodman} \& {Weare}(2010)}]{2010CAMCS...5...65G}
{Goodman}, J. \& {Weare}, J. 2010, Communications in Applied Mathematics and
  Computational Science, 5, 65

\bibitem[{{Green} {et~al.}(2010){Green}, {Caswell}, {Fuller}, {Avison},
  {Breen}, {Ellingsen}, {Gray}, {Pestalozzi}, {Quinn}, {Thompson}, \&
  {Voronkov}}]{2010MNRAS.409..913G}
{Green}, J.~A., {Caswell}, J.~L., {Fuller}, G.~A., {et~al.} 2010, \mnras, 409,
  913

\bibitem[{{G{\"u}sten} {et~al.}(2006){G{\"u}sten}, {Nyman}, {Schilke},
  {Menten}, {Cesarsky}, \& {Booth}}]{2006A&A...454L..13G}
{G{\"u}sten}, R., {Nyman}, L.~{\r{A}}., {Schilke}, P., {et~al.} 2006, \aap,
  454, L13

\bibitem[{{Haschick} \& {Baan}(1989)}]{1989ApJ...339..949H}
{Haschick}, A.~D. \& {Baan}, W.~A. 1989, \apj, 339, 949

\bibitem[{{Henkel} {et~al.}(1984){Henkel}, {Wilson}, \&
  {Johnston}}]{1984ApJ...282L..93H}
{Henkel}, C., {Wilson}, T.~L., \& {Johnston}, K.~J. 1984, \apjl, 282, L93

\bibitem[{{Hills} {et~al.}(1975){Hills}, {Pankonin}, \&
  {Landecker}}]{Hills1975}
{Hills}, R., {Pankonin}, V., \& {Landecker}, T.~L. 1975, \aap, 39, 149

\bibitem[{{Ho} \& {Haschick}(1986)}]{1986ApJ...304..501H}
{Ho}, P.~T.~P. \& {Haschick}, A.~D. 1986, \apj, 304, 501

\bibitem[{{Ho} \& {Young}(1996)}]{1996ApJ...472..742H}
{Ho}, P. T.~P. \& {Young}, L.~M. 1996, \apj, 472, 742

\bibitem[{{Hoare} {et~al.}(2012){Hoare}, {Purcell}, {Churchwell}, {Diamond},
  {Cotton}, {Chandler}, {Smethurst}, {Kurtz}, {Mundy}, {Dougherty}, {Fender},
  {Fuller}, {Jackson}, {Garrington}, {Gledhill}, {Goldsmith}, {Lumsden},
  {Mart{\'\i}}, {Moore}, {Muxlow}, {Oudmaijer}, {Pandian}, {Paredes},
  {Shepherd}, {Spencer}, {Thompson}, {Umana}, {Urquhart}, \&
  {Zijlstra}}]{2012PASP..124..939H}
{Hoare}, M.~G., {Purcell}, C.~R., {Churchwell}, E.~B., {et~al.} 2012, \pasp,
  124, 939

\bibitem[{{Houghton} \& {Whiteoak}(1995)}]{1995MNRAS.273.1033H}
{Houghton}, S. \& {Whiteoak}, J.~B. 1995, \mnras, 273, 1033

\bibitem[{Hunter(2007)}]{Hunter:2007}
Hunter, J.~D. 2007, Computing in Science \& Engineering, 9, 90

\bibitem[{{Hunter} {et~al.}(2014){Hunter}, {Brogan}, {Cyganowski}, \&
  {Young}}]{2014ApJ...788..187H}
{Hunter}, T.~R., {Brogan}, C.~L., {Cyganowski}, C.~J., \& {Young}, K.~H. 2014,
  \apj, 788, 187

\bibitem[{Jones {et~al.}(2001)Jones, Oliphant, Peterson,
  {et~al.}}]{jones2001scipy}
Jones, E., Oliphant, T., Peterson, P., {et~al.} 2001, {SciPy}: Open source
  scientific tools for {Python}

\bibitem[{{Kalcheva} {et~al.}(2018){Kalcheva}, {Hoare}, {Urquhart}, {Kurtz},
  {Lumsden}, {Purcell}, \& {Zijlstra}}]{2018A&A...615A.103K}
{Kalcheva}, I.~E., {Hoare}, M.~G., {Urquhart}, J.~S., {et~al.} 2018, \aap, 615,
  A103

\bibitem[{{Kang} {et~al.}(2016){Kang}, {Byun}, {Kim}, {Kim}, {Lyo}, \&
  {Vlemmings}}]{2016ApJS..227...17K}
{Kang}, J.-h., {Byun}, D.-Y., {Kim}, K.-T., {et~al.} 2016, \apjs, 227, 17

\bibitem[{{Keto}(2002)}]{2002ApJ...568..754K}
{Keto}, E. 2002, \apj, 568, 754

\bibitem[{{Keto}(1990)}]{1990ApJ...355..190K}
{Keto}, E.~R. 1990, \apj, 355, 190

\bibitem[{{Keto}(1991)}]{1991ApJ...371..163K}
{Keto}, E.~R. 1991, \apj, 371, 163

\bibitem[{{Keto} {et~al.}(1987{\natexlab{a}}){Keto}, {Ho}, \&
  {Haschick}}]{1987ApJ...318..712K}
{Keto}, E.~R., {Ho}, P. T.~P., \& {Haschick}, A.~D. 1987{\natexlab{a}}, \apj,
  318, 712

\bibitem[{{Keto} {et~al.}(1988){Keto}, {Ho}, \&
  {Haschick}}]{1988ApJ...324..920K}
{Keto}, E.~R., {Ho}, P. T.~P., \& {Haschick}, A.~D. 1988, \apj, 324, 920

\bibitem[{{Keto} {et~al.}(1987{\natexlab{b}}){Keto}, {Ho}, \&
  {Reid}}]{1987ApJ...323L.117K}
{Keto}, E.~R., {Ho}, P. T.~P., \& {Reid}, M.~J. 1987{\natexlab{b}}, \apjl, 323,
  L117

\bibitem[{{Keto} {et~al.}(1995){Keto}, {Welch}, {Reid}, \&
  {Ho}}]{1995ApJ...444..765K}
{Keto}, E.~R., {Welch}, W.~J., {Reid}, M.~J., \& {Ho}, P.~T.~P. 1995, \apj,
  444, 765

\bibitem[{{Kim} {et~al.}(2019){Kim}, {Kim}, \& {Kim}}]{2019ApJS..244....2K}
{Kim}, W.-J., {Kim}, K.-T., \& {Kim}, K.-T. 2019, \apjs, 244, 2

\bibitem[{{Kim} {et~al.}(2020){Kim}, {Wyrowski}, {Urquhart},
  {P{\'e}rez-Beaupuits}, {Pillai}, {Tiwari}, \& {Menten}}]{2020A&A...644A.160K}
{Kim}, W.~J., {Wyrowski}, F., {Urquhart}, J.~S., {et~al.} 2020, \aap, 644, A160

\bibitem[{{Kirk} {et~al.}(2009){Kirk}, {Crutcher}, \&
  {Ward-Thompson}}]{2009ApJ...701.1044K}
{Kirk}, J.~M., {Crutcher}, R.~M., \& {Ward-Thompson}, D. 2009, \apj, 701, 1044

\bibitem[{{Klaassen} \& {Wilson}(2008)}]{2008ApJ...684.1273K}
{Klaassen}, P.~D. \& {Wilson}, C.~D. 2008, \apj, 684, 1273

\bibitem[{{Klein} {et~al.}(2012){Klein}, {Hochg{\"u}rtel}, {Kr{\"a}mer},
  {Bell}, {Meyer}, \& {G{\"u}sten}}]{2012A&A...542L...3K}
{Klein}, B., {Hochg{\"u}rtel}, S., {Kr{\"a}mer}, I., {et~al.} 2012, \aap, 542,
  L3

\bibitem[{{Kurtz}(2005)}]{2005IAUS..227..111K}
{Kurtz}, S. 2005, in Massive Star Birth: A Crossroads of Astrophysics, ed.
  R.~{Cesaroni}, M.~{Felli}, E.~{Churchwell}, \& M.~{Walmsley}, Vol. 227,
  111--119

\bibitem[{{Kurtz} {et~al.}(2004){Kurtz}, {Hofner}, \&
  {{\'A}lvarez}}]{2004ApJS..155..149K}
{Kurtz}, S., {Hofner}, P., \& {{\'A}lvarez}, C.~V. 2004, \apjs, 155, 149

\bibitem[{{Larson}(1981)}]{1981MNRAS.194..809L}
{Larson}, R.~B. 1981, \mnras, 194, 809

\bibitem[{{Lees} \& {Haque}(1974)}]{1974CaJPh..52.2250L}
{Lees}, R.~M. \& {Haque}, S.~S. 1974, Canadian Journal of Physics, 52, 2250

\bibitem[{{Leung} \& {Brown}(1977)}]{1977ApJ...214L..73L}
{Leung}, C.~M. \& {Brown}, R.~L. 1977, \apjl, 214, L73

\bibitem[{{Leurini} {et~al.}(2016){Leurini}, {Menten}, \&
  {Walmsley}}]{2016A&A...592A..31L}
{Leurini}, S., {Menten}, K.~M., \& {Walmsley}, C.~M. 2016, \aap, 592, A31

\bibitem[{{Li}(2018)}]{2018MNRAS.477.4951L}
{Li}, G.-X. 2018, \mnras, 477, 4951

\bibitem[{{Lin} {et~al.}(2016){Lin}, {Liu}, {Li}, {Zhang}, {Ginsburg},
  {Pineda}, {Qian}, {Galv{\'a}n-Madrid}, {McLeod}, {Rosolowsky}, {Dale},
  {Immer}, {Koch}, {Longmore}, {Walker}, \& {Testi}}]{2016ApJ...828...32L}
{Lin}, Y., {Liu}, H.~B., {Li}, D., {et~al.} 2016, \apj, 828, 32

\bibitem[{{Liu}(2017)}]{2017A&A...597A..70L}
{Liu}, H.~B. 2017, \aap, 597, A70

\bibitem[{{Liu} {et~al.}(2010{\natexlab{a}}){Liu}, {Ho}, \&
  {Zhang}}]{2010ApJ...725.2190L}
{Liu}, H.~B., {Ho}, P. T.~P., \& {Zhang}, Q. 2010{\natexlab{a}}, \apj, 725,
  2190

\bibitem[{{Liu} {et~al.}(2010{\natexlab{b}}){Liu}, {Ho}, {Zhang}, {Keto}, {Wu},
  \& {Li}}]{2010ApJ...722..262L}
{Liu}, H.~B., {Ho}, P. T.~P., {Zhang}, Q., {et~al.} 2010{\natexlab{b}}, \apj,
  722, 262

\bibitem[{{Liu} {et~al.}(2012){Liu}, {Quintana-Lacaci}, {Wang}, {Ho}, {Li},
  {Zhang}, \& {Zhang}}]{2012ApJ...745...61L}
{Liu}, H.~B., {Quintana-Lacaci}, G., {Wang}, K., {et~al.} 2012, \apj, 745, 61

\bibitem[{{Liu} {et~al.}(2011){Liu}, {Zhang}, \& {Ho}}]{2011ApJ...729..100L}
{Liu}, H.~B., {Zhang}, Q., \& {Ho}, P. T.~P. 2011, \apj, 729, 100

\bibitem[{{Liu} {et~al.}(2013){Liu}, {Wu}, {Wu}, {Qin}, \&
  {Zhang}}]{2013MNRAS.436.1335L}
{Liu}, T., {Wu}, Y., {Wu}, J., {Qin}, S.-L., \& {Zhang}, H. 2013, \mnras, 436,
  1335

\bibitem[{{Mangum} \& {Wootten}(1993)}]{1993ApJS...89..123M}
{Mangum}, J.~G. \& {Wootten}, A. 1993, \apjs, 89, 123

\bibitem[{{Mauersberger} {et~al.}(1986){Mauersberger}, {Wilson}, \&
  {Walmsley}}]{Mauersberger1986}
{Mauersberger}, R., {Wilson}, T.~L., \& {Walmsley}, C.~M. 1986, \aap, 166, L26

\bibitem[{{McKee} \& {Tan}(2003)}]{2003ApJ...585..850M}
{McKee}, C.~F. \& {Tan}, J.~C. 2003, \apj, 585, 850

\bibitem[{{Mehrotra} {et~al.}(1985){Mehrotra}, {Dreizler}, \&
  {M{\"a}der}}]{1985ZNatA..40..683M}
{Mehrotra}, S.~C., {Dreizler}, H., \& {M{\"a}der}, H. 1985, Zeitschrift
  Naturforschung Teil A, 40, 683

\bibitem[{{Menten}(1991{\natexlab{a}})}]{1991ASPC...16..119M}
{Menten}, K.~M. 1991{\natexlab{a}}, in Astronomical Society of the Pacific
  Conference Series, Vol.~16, Atoms, Ions and Molecules: New Results in
  Spectral Line Astrophysics, ed. A.~D. {Haschick} \& P.~T.~P. {Ho}, 119--136

\bibitem[{{Menten}(1991{\natexlab{b}})}]{1991ApJ...380L..75M}
{Menten}, K.~M. 1991{\natexlab{b}}, \apjl, 380, L75

\bibitem[{{Menten} {et~al.}(1986{\natexlab{a}}){Menten}, {Walmsley}, {Henkel},
  \& {Wilson}}]{1986A&A...157..318M}
{Menten}, K.~M., {Walmsley}, C.~M., {Henkel}, C., \& {Wilson}, T.~L.
  1986{\natexlab{a}}, \aap, 157, 318

\bibitem[{{Menten} {et~al.}(1988){Menten}, {Walmsley}, {Henkel}, \&
  {Wilson}}]{Menten1988}
{Menten}, K.~M., {Walmsley}, C.~M., {Henkel}, C., \& {Wilson}, T.~L. 1988,
  \aap, 198, 267

\bibitem[{{Menten} {et~al.}(1986{\natexlab{b}}){Menten}, {Walmsley}, {Henkel},
  {Wilson}, {Snyder}, {Hollis}, \& {Lovas}}]{1986A&A...169..271M}
{Menten}, K.~M., {Walmsley}, C.~M., {Henkel}, C., {et~al.} 1986{\natexlab{b}},
  \aap, 169, 271

\bibitem[{{Minier} \& {Booth}(2002)}]{2002A&A...387..179M}
{Minier}, V. \& {Booth}, R.~S. 2002, \aap, 387, 179

\bibitem[{{Motte} {et~al.}(2018){Motte}, {Bontemps}, \&
  {Louvet}}]{2018ARA&A..56...41M}
{Motte}, F., {Bontemps}, S., \& {Louvet}, F. 2018, \araa, 56, 41

\bibitem[{{M{\"u}ller} {et~al.}(2004){M{\"u}ller}, {Menten}, \&
  {M{\"a}der}}]{2004A&A...428.1019M}
{M{\"u}ller}, H.~S.~P., {Menten}, K.~M., \& {M{\"a}der}, H. 2004, \aap, 428,
  1019

\bibitem[{{M{\"u}ller} {et~al.}(2005){M{\"u}ller}, {Schl{\"o}der}, {Stutzki},
  \& {Winnewisser}}]{2005JMoSt.742..215M}
{M{\"u}ller}, H. S.~P., {Schl{\"o}der}, F., {Stutzki}, J., \& {Winnewisser}, G.
  2005, Journal of Molecular Structure, 742, 215

\bibitem[{{Myers} {et~al.}(1996){Myers}, {Mardones}, {Tafalla}, {Williams}, \&
  {Wilner}}]{1996ApJ...465L.133M}
{Myers}, P.~C., {Mardones}, D., {Tafalla}, M., {Williams}, J.~P., \& {Wilner},
  D.~J. 1996, \apjl, 465, L133

\bibitem[{{Ortiz-Le{\'o}n} {et~al.}(2021){Ortiz-Le{\'o}n}, {Menten},
  {Brunthaler}, {Csengeri}, {Urquhart}, {Wyrowski}, {Gong}, {Rugel}, {Dzib},
  {Yang}, {Nguyen}, {Cotton}, {Medina}, {Dokara}, {K{\"o}nig}, {Beuther},
  {Pandian}, {Reich}, \& {Roy}}]{2021A&A...651A..87O}
{Ortiz-Le{\'o}n}, G.~N., {Menten}, K.~M., {Brunthaler}, A., {et~al.} 2021,
  \aap, 651, A87

\bibitem[{{Pandian} {et~al.}(2008){Pandian}, {Leurini}, {Menten}, {Belloche},
  \& {Goldsmith}}]{2008A&A...489.1175P}
{Pandian}, J.~D., {Leurini}, S., {Menten}, K.~M., {Belloche}, A., \&
  {Goldsmith}, P.~F. 2008, \aap, 489, 1175

\bibitem[{{Peng} \& {Whiteoak}(1989)}]{1989PASA....8..204P}
{Peng}, R.~S. \& {Whiteoak}, J.~B. 1989, \pasa, 8, 204

\bibitem[{{Peng} \& {Whiteoak}(1992)}]{1992MNRAS.254..301P}
{Peng}, R.~S. \& {Whiteoak}, J.~B. 1992, \mnras, 254, 301

\bibitem[{{Pety}(2005)}]{2005sf2a.conf..721P}
{Pety}, J. 2005, in SF2A-2005: Semaine de l'Astrophysique Francaise, ed.
  F.~{Casoli}, T.~{Contini}, J.~M. {Hameury}, \& L.~{Pagani}, 721

\bibitem[{{Pineda} {et~al.}(2012){Pineda}, {Maury}, {Fuller}, {Testi},
  {Garc{\'\i}a-Appadoo}, {Peck}, {Villard}, {Corder}, {van Kempen}, {Turner},
  {Tachihara}, \& {Dent}}]{2012A&A...544L...7P}
{Pineda}, J.~E., {Maury}, A.~J., {Fuller}, G.~A., {et~al.} 2012, \aap, 544, L7

\bibitem[{{Purcell} {et~al.}(2013){Purcell}, {Hoare}, {Cotton}, {Lumsden},
  {Urquhart}, {Chandler}, {Churchwell}, {Diamond}, {Dougherty}, {Fender},
  {Fuller}, {Garrington}, {Gledhill}, {Goldsmith}, {Hindson}, {Jackson},
  {Kurtz}, {Mart{\'\i}}, {Moore}, {Mundy}, {Muxlow}, {Oudmaijer}, {Pandian},
  {Paredes}, {Shepherd}, {Smethurst}, {Spencer}, {Thompson}, {Umana}, \&
  {Zijlstra}}]{2013ApJS..205....1P}
{Purcell}, C.~R., {Hoare}, M.~G., {Cotton}, W.~D., {et~al.} 2013, \apjs, 205, 1

\bibitem[{{Purcell} {et~al.}(2012){Purcell}, {Longmore}, {Walsh}, {Whiting},
  {Breen}, {Britton}, {Brooks}, {Burton}, {Cunningham}, {Green},
  {Harvey-Smith}, {Hindson}, {Hoare}, {Indermuehle}, {Jones}, {Lo}, {Lowe},
  {Phillips}, {Thompson}, {Urquhart}, {Voronkov}, \&
  {White}}]{2012MNRAS.426.1972P}
{Purcell}, C.~R., {Longmore}, S.~N., {Walsh}, A.~J., {et~al.} 2012, \mnras,
  426, 1972

\bibitem[{{Qin} {et~al.}(2016){Qin}, {Schilke}, {Wu}, {Liu}, {Wu},
  {S{\'a}nchez-Monge}, \& {Liu}}]{2016MNRAS.456.2681Q}
{Qin}, S.-L., {Schilke}, P., {Wu}, J., {et~al.} 2016, \mnras, 456, 2681

\bibitem[{{Qin} {et~al.}(2015){Qin}, {Schilke}, {Wu}, {Wu}, {Liu}, {Liu}, \&
  {S{\'a}nchez-Monge}}]{2015ApJ...803...39Q}
{Qin}, S.-L., {Schilke}, P., {Wu}, J., {et~al.} 2015, \apj, 803, 39

\bibitem[{{Qiu} {et~al.}(2011){Qiu}, {Zhang}, \&
  {Menten}}]{2011ApJ...728....6Q}
{Qiu}, K., {Zhang}, Q., \& {Menten}, K.~M. 2011, \apj, 728, 6

\bibitem[{{Rabli} \& {Flower}(2010)}]{2010MNRAS.406...95R}
{Rabli}, D. \& {Flower}, D.~R. 2010, \mnras, 406, 95

\bibitem[{{Robitaille} \& {Bressert}(2012)}]{2012ascl.soft08017R}
{Robitaille}, T. \& {Bressert}, E. 2012, {APLpy: Astronomical Plotting Library
  in Python}

\bibitem[{{Salii} \& {Sobolev}(2006)}]{2006ARep...50..965S}
{Salii}, S.~V. \& {Sobolev}, A.~M. 2006, Astronomy Reports, 50, 965

\bibitem[{{Sanna} {et~al.}(2014){Sanna}, {Reid}, {Menten}, {Dame}, {Zhang},
  {Sato}, {Brunthaler}, {Moscadelli}, \& {Immer}}]{2014ApJ...781..108S}
{Sanna}, A., {Reid}, M.~J., {Menten}, K.~M., {et~al.} 2014, \apj, 781, 108

\bibitem[{{Schnee} {et~al.}(2007){Schnee}, {Caselli}, {Goodman}, {Arce},
  {Ballesteros-Paredes}, \& {Kuchibhotla}}]{2007ApJ...671.1839S}
{Schnee}, S., {Caselli}, P., {Goodman}, A., {et~al.} 2007, \apj, 671, 1839

\bibitem[{{Schneider} {et~al.}(2010){Schneider}, {Csengeri}, {Bontemps},
  {Motte}, {Simon}, {Hennebelle}, {Federrath}, \&
  {Klessen}}]{2010A&A...520A..49S}
{Schneider}, N., {Csengeri}, T., {Bontemps}, S., {et~al.} 2010, \aap, 520, A49

\bibitem[{{Shirley}(2015)}]{2015PASP..127..299S}
{Shirley}, Y.~L. 2015, \pasp, 127, 299

\bibitem[{{Slysh} {et~al.}(1992){Slysh}, {Kalenskii}, \&
  {Val'tts}}]{1992ApJ...397L..43S}
{Slysh}, V.~I., {Kalenskii}, S.~V., \& {Val'tts}, I.~E. 1992, \apjl, 397, L43

\bibitem[{{Slysh} {et~al.}(1999){Slysh}, {Kalenskii}, {Val'TTS}, {Golubev}, \&
  {Mead}}]{1999ApJS..123..515S}
{Slysh}, V.~I., {Kalenskii}, S.~V., {Val'TTS}, I.~E., {Golubev}, V.~V., \&
  {Mead}, K. 1999, \apjs, 123, 515

\bibitem[{{Slysh} {et~al.}(2002){Slysh}, {Kalenski{\u{i}}}, \&
  {Val'tts}}]{2002ARep...46...49S}
{Slysh}, V.~I., {Kalenski{\u{i}}}, S.~V., \& {Val'tts}, I.~E. 2002, Astronomy
  Reports, 46, 49

\bibitem[{{Sobolev} {et~al.}(1997){Sobolev}, {Cragg}, \&
  {Godfrey}}]{sobolev1997}
{Sobolev}, A.~M., {Cragg}, D.~M., \& {Godfrey}, P.~D. 1997, \aap, 324, 211

\bibitem[{{Sobolev} \& {Deguchi}(1994)}]{Sobolev1994}
{Sobolev}, A.~M. \& {Deguchi}, S. 1994, \aap, 291, 569

\bibitem[{{Sollins} \& {Ho}(2005)}]{2005ApJ...630..987S}
{Sollins}, P.~K. \& {Ho}, P. T.~P. 2005, \apj, 630, 987

\bibitem[{{Sollins} {et~al.}(2005){Sollins}, {Zhang}, {Keto}, \&
  {Ho}}]{2005ApJ...624L..49S}
{Sollins}, P.~K., {Zhang}, Q., {Keto}, E., \& {Ho}, P. T.~P. 2005, \apjl, 624,
  L49

\bibitem[{{Sun} \& {Gao}(2009)}]{2009MNRAS.392..170S}
{Sun}, Y. \& {Gao}, Y. 2009, \mnras, 392, 170

\bibitem[{{Tang} {et~al.}(2018){Tang}, {Henkel}, {Wyrowski}, {Giannetti},
  {Menten}, {Csengeri}, {Leurini}, {Urquhart}, {K{\"o}nig}, {G{\"u}sten},
  {Lin}, {Zheng}, {Esimbek}, \& {Zhou}}]{2018A&A...611A...6T}
{Tang}, X.~D., {Henkel}, C., {Wyrowski}, F., {et~al.} 2018, \aap, 611, A6

\bibitem[{{Turner} \& {Welch}(1984)}]{TurnerWelch1984}
{Turner}, J.~L. \& {Welch}, W.~J. 1984, \apjl, 287, L81

\bibitem[{{Val'tts} {et~al.}(1999){Val'tts}, {Ellingsen}, {Slysh}, {Kalenskii},
  {Otrupcek}, \& {Voronkov}}]{1999MNRAS.310.1077V}
{Val'tts}, I.~E., {Ellingsen}, S.~P., {Slysh}, V.~I., {et~al.} 1999, \mnras,
  310, 1077

\bibitem[{{van der Tak} {et~al.}(2019){van der Tak}, {Shipman}, {Jacq},
  {Herpin}, {Braine}, \& {Wyrowski}}]{2019A&A...625A.103V}
{van der Tak}, F.~F.~S., {Shipman}, R.~F., {Jacq}, T., {et~al.} 2019, \aap,
  625, A103

\bibitem[{{van der Walt} {et~al.}(2011){van der Walt}, {Colbert}, \&
  {Varoquaux}}]{5725236}
{van der Walt}, S., {Colbert}, S.~C., \& {Varoquaux}, G. 2011, Computing in
  Science Engineering, 13, 22

\bibitem[{{V{\'a}zquez-Semadeni} {et~al.}(2019){V{\'a}zquez-Semadeni}, {Palau},
  {Ballesteros-Paredes}, {G{\'o}mez}, \&
  {Zamora-Avil{\'e}s}}]{2019MNRAS.490.3061V}
{V{\'a}zquez-Semadeni}, E., {Palau}, A., {Ballesteros-Paredes}, J.,
  {G{\'o}mez}, G.~C., \& {Zamora-Avil{\'e}s}, M. 2019, \mnras, 490, 3061

\bibitem[{{Voronkov} {et~al.}(2006){Voronkov}, {Brooks}, {Sobolev},
  {Ellingsen}, {Ostrovskii}, \& {Caswell}}]{2006MNRAS.373..411V}
{Voronkov}, M.~A., {Brooks}, K.~J., {Sobolev}, A.~M., {et~al.} 2006, \mnras,
  373, 411

\bibitem[{{Voronkov} {et~al.}(2011){Voronkov}, {Walsh}, {Caswell}, {Ellingsen},
  {Breen}, {Longmore}, {Purcell}, \& {Urquhart}}]{2011MNRAS.413.2339V}
{Voronkov}, M.~A., {Walsh}, A.~J., {Caswell}, J.~L., {et~al.} 2011, \mnras,
  413, 2339

\bibitem[{{Walmsley} {et~al.}(1988){Walmsley}, {Batrla}, {Matthews}, \&
  {Menten}}]{1988A&A...197..271W}
{Walmsley}, C.~M., {Batrla}, W., {Matthews}, H.~E., \& {Menten}, K.~M. 1988,
  \aap, 197, 271

\bibitem[{{Welch} {et~al.}(1987){Welch}, {Dreher}, {Jackson}, {Terebey}, \&
  {Vogel}}]{1987Sci...238.1550W}
{Welch}, W.~J., {Dreher}, J.~W., {Jackson}, J.~M., {Terebey}, S., \& {Vogel},
  S.~N. 1987, Science, 238, 1550

\bibitem[{{Whiteoak} {et~al.}(1988){Whiteoak}, {Gardner}, {Caswell}, {Norris},
  {Wellington}, \& {Peng}}]{1988MNRAS.235..655W}
{Whiteoak}, J.~B., {Gardner}, F.~F., {Caswell}, J.~L., {et~al.} 1988, \mnras,
  235, 655

\bibitem[{{Whiteoak} \& {Peng}(1989)}]{1989MNRAS.239..677W}
{Whiteoak}, J.~B. \& {Peng}, R.~S. 1989, \mnras, 239, 677

\bibitem[{{Wilson} {et~al.}(1978){Wilson}, {Bieging}, \& {Downes}}]{Wilson1978}
{Wilson}, T.~L., {Bieging}, J., \& {Downes}, D. 1978, \aap, 63, 1

\bibitem[{{Wilson} {et~al.}(1993){Wilson}, {Huettemeister}, {Dahmen}, \&
  {Henkel}}]{1993A&A...268..249W}
{Wilson}, T.~L., {Huettemeister}, S., {Dahmen}, G., \& {Henkel}, C. 1993, \aap,
  268, 249

\bibitem[{{Wilson} {et~al.}(1991){Wilson}, {Johnston}, \&
  {Mauersberger}}]{1991A&A...251..220W}
{Wilson}, T.~L., {Johnston}, K.~J., \& {Mauersberger}, R. 1991, \aap, 251, 220

\bibitem[{{Wilson} {et~al.}(1983){Wilson}, {Mauersberger}, {Walmsley}, \&
  {Batrla}}]{1983A&A...127L..19W}
{Wilson}, T.~L., {Mauersberger}, R., {Walmsley}, C.~M., \& {Batrla}, W. 1983,
  \aap, 127, L19

\bibitem[{{Wilson} {et~al.}(1985){Wilson}, {Walmsley}, {Menten}, \&
  {Hermsen}}]{1985A&A...147L..19W}
{Wilson}, T.~L., {Walmsley}, C.~M., {Menten}, K.~M., \& {Hermsen}, W. 1985,
  \aap, 147, L19

\bibitem[{{Wilson} {et~al.}(1984){Wilson}, {Walmsley}, {Snyder}, \&
  {Jewell}}]{1984A&A...134L...7W}
{Wilson}, T.~L., {Walmsley}, C.~M., {Snyder}, L.~E., \& {Jewell}, P.~R. 1984,
  \aap, 134, L7

\bibitem[{{Wu} \& {Evans}(2003)}]{2003ApJ...592L..79W}
{Wu}, J. \& {Evans}, Neal~J., I. 2003, \apjl, 592, L79

\bibitem[{{Wu} {et~al.}(2010){Wu}, {Evans}, {Shirley}, \&
  {Knez}}]{2010ApJS..188..313W}
{Wu}, J., {Evans}, Neal~J., I., {Shirley}, Y.~L., \& {Knez}, C. 2010, \apjs,
  188, 313

\bibitem[{{Wyrowski} {et~al.}(2016){Wyrowski}, {G{\"u}sten}, {Menten},
  {Wiesemeyer}, {Csengeri}, {Heyminck}, {Klein}, {K{\"o}nig}, \&
  {Urquhart}}]{2016A&A...585A.149W}
{Wyrowski}, F., {G{\"u}sten}, R., {Menten}, K.~M., {et~al.} 2016, \aap, 585,
  A149

\bibitem[{{Wyrowski} {et~al.}(2012){Wyrowski}, {G{\"u}sten}, {Menten},
  {Wiesemeyer}, \& {Klein}}]{2012A&A...542L..15W}
{Wyrowski}, F., {G{\"u}sten}, R., {Menten}, K.~M., {Wiesemeyer}, H., \&
  {Klein}, B. 2012, \aap, 542, L15

\bibitem[{{Wyrowski} {et~al.}(2006){Wyrowski}, {Heyminck}, {G{\"u}sten}, \&
  {Menten}}]{2006A&A...454L..95W}
{Wyrowski}, F., {Heyminck}, S., {G{\"u}sten}, R., \& {Menten}, K.~M. 2006,
  \aap, 454, L95

\bibitem[{{Wyrowski} {et~al.}(1999){Wyrowski}, {Schilke}, {Walmsley}, \&
  {Menten}}]{Wyrowski1999}
{Wyrowski}, F., {Schilke}, P., {Walmsley}, C.~M., \& {Menten}, K.~M. 1999,
  \apjl, 514, L43

\bibitem[{{Xu} {et~al.}(2006){Xu}, {Reid}, {Zheng}, \&
  {Menten}}]{2006Sci...311...54X}
{Xu}, Y., {Reid}, M.~J., {Zheng}, X.~W., \& {Menten}, K.~M. 2006, Science, 311,
  54

\bibitem[{{Yang} {et~al.}(2019){Yang}, {Thompson}, {Tian}, {Bihr}, {Beuther},
  \& {Hindson}}]{2019MNRAS.482.2681Y}
{Yang}, A.~Y., {Thompson}, M.~A., {Tian}, W.~W., {et~al.} 2019, \mnras, 482,
  2681

\bibitem[{{Yang} {et~al.}(2021){Yang}, {Urquhart}, {Thompson}, {Menten},
  {Wyrowski}, {Brunthaler}, {Tian}, {Rugel}, {Yang}, {Yao}, \&
  {Mutale}}]{2021A&A...645A.110Y}
{Yang}, A.~Y., {Urquhart}, J.~S., {Thompson}, M.~A., {et~al.} 2021, \aap, 645,
  A110

\bibitem[{{Zapata} {et~al.}(2011){Zapata}, {Rodr{\'\i}guez-Garza},
  {Rodr{\'\i}guez}, {Girart}, \& {Chen}}]{2011ApJ...740L..19Z}
{Zapata}, L.~A., {Rodr{\'\i}guez-Garza}, C., {Rodr{\'\i}guez}, L.~F., {Girart},
  J.~M., \& {Chen}, H.-R. 2011, \apjl, 740, L19

\bibitem[{{Zhang} \& {Ho}(1997)}]{1997ApJ...488..241Z}
{Zhang}, Q. \& {Ho}, P. T.~P. 1997, \apj, 488, 241

\bibitem[{{Zhang} {et~al.}(1998){Zhang}, {Ho}, \&
  {Ohashi}}]{1998ApJ...494..636Z}
{Zhang}, Q., {Ho}, P. T.~P., \& {Ohashi}, N. 1998, \apj, 494, 636

\bibitem[{{Zhou} {et~al.}(1993){Zhou}, {Evans}, {Koempe}, \&
  {Walmsley}}]{1993ApJ...404..232Z}
{Zhou}, S., {Evans}, Neal~J., I., {Koempe}, C., \& {Walmsley}, C.~M. 1993,
  \apj, 404, 232

\end{thebibliography}

\begin{appendix}

\begin{table*}[!hbt]
\section{Tables}\label{sec:table}

\caption{Methanol transitions. }\label{Tab:freq}
%\normalsize
\small
%\tiny
\renewcommand\arraystretch{0.97}
\centering
\begin{tabular}{clrrcrccc}
\hline \hline 
No. & CH$_3$OH  & Rest frequency &  Line  & Class & $E_{\rm up}$ & $n_{\rm crit}$ & W31C & Note \\ 
& Transition    & (MHz)      & (GHz) & &   (K) & (cm$^{-3}$) &Reference   &    \\   
\hline
& $J_0-(J+1)_{-1}E$  series\\
\hline
1& $0_{0}-1_{-1}E$&108 893.963(7)$^*$ & 108 & II &13.1 & 2.28$\times$10$^4$ &this work & maser?/thermal\\
2& $2_0-3_{-1}E$& 12 178.597(4)$^*$ & 12.2 & II &20.1 & 7.58$\times$10$^4$ & this work,(1)$-$(3) &maser \& absorption\\
3& $4_{-1}-3_0E$&36 169.265(30)$^*$ & 36 & I &28.8 & 5.01$\times$10$^4$ &  this work,(4),(5)  &maser\\
4& $5_{-1}-4_0E$ & 84 521.169(10)$^*$ & 84 & I &40.4 & 1.04$\times$10$^5$ &this work,(5)  &maser\\
5& $6_{-1}-5_0E$ & 132 890.692(10)$^*$ & 132.9  & I &54.3 & 1.88$\times$10$^5$ & (6) &maser?/thermal\\
6& $8_{-1}-7_{0}E$&229 758.760(50)$^*$ & 229 & I &89.1 & 4.92$\times$10$^5$ &this work & maser?/thermal\\
\hline
& $J_{-2}-(J+1)_{-1}E$ series\\
\hline
7& $3_{-2}-4_{-1}E$ &230 027.047(11)$^\alpha$ & 230 & - &39.8 & 2.93$\times$10$^6$ &this work & thermal\\
8& $5_{-2}-6_{-1}E$ &133 605.439(11)$^\alpha$ &133.6 & II &60.7 & 3.19$\times$10$^6$ & (6) & maser?/thermal\\
9& $6_{-2}-7_{-1}E$ &85 568.084(10)$^*$ &85.6 & II &74.7 & 3.54$\times$10$^6$ & this work &inverse P-Cygni\\
10& $7_{-2}-8_{-1}E$ &37 703.700(30)$^*$ &37.7 & II &90.9 & 3.98$\times$10$^6$ & this work &absorption\\
11& $11_{-1}-10_{-2}E$&104 300.414(7)$^*$ &104 & I &158.6 & 1.49$\times$10$^6$ &this work & maser?/thermal\\
\hline
& $J_1-(J+1)_{0}E$ series\\
\hline
12& $2_1-3_0E$& 19 967.3961(2)$^*$  & 19.9 & II &28.0 & 1.53$\times$10$^5$ & this work,(7) &inverse P-Cygni$^{(a)}$\\
13& $8_0-7_1E$& 220 078.561(8)$^\alpha$ & 220 & - &96.6 & 1.25$\times$10$^6$ & this work &thermal\\
\hline
& $J_2-J_1E$ @25 GHz series\\
\hline
14& $2_2-2_1E$ & 24 934.3801(16)$^\beta$ & 24.934 & I & 29.2 & 2.85$\times$10$^4$ & this work,(8)&maser?/thermal$^{(b)}$ \\
15& $3_2-3_1E$ & 24 928.6994(8)$^\beta$ & 24.928 & I & 36.3 & 6.76$\times$10$^4$ & this work,(8) &maser?/thermal$^{(b)}$\\
16& $4_2-4_1E$ & 24 933.4693(8)$^\beta$ & 24.933 & I & 45.5 & 1.31$\times$10$^5$ & this work,(8) &maser?/thermal$^{(b)}$\\
17& $5_2-5_1E$ & 24 959.0789(4)$^*$ & 24.959 & I & 57.1 & 2.29$\times$10$^5$ & this work,(8) &maser$^{(b)}$\\
18& $6_2-6_1E$ & 25 018.1225(4)$^*$ & 25.018 & I & 71.0 & 3.71$\times$10$^5$ & this work,(8) &maser$^{(b)}$\\
19& $7_2-7_1E$ & 25 124.8719(4)$^*$ & 25.124 & I & 87.3 & 5.69$\times$10$^5$ & this work &maser$^{(b)}$\\
20& $8_2-8_1E$ & 25 294.4165(2)$^*$ & 25.294 & I & 105.8 & 8.14$\times$10$^5$ & this work &maser$^{(b)}$\\
21& $9_2-9_1E$ & 25 541.3979(4)$^*$ & 25.541 & I & 126.7 & 1.12$\times$10$^6$ & this work &inverse P-Cygni$^{(b)}$\\
22& $10_2-10_1E$ & 25 878.2661(4)$^*$ & 25.878 & I & 150.0 & 1.53$\times$10$^6$ & this work &inverse P-Cygni$^{(b)}$\\
\hline
23& $5_{1}-4_{2}E$ &216 945.521(12)$^\alpha$ & 216 & II &55.9 & 4.84$\times$10$^5$ &this work & maser?/thermal\\
24& $4_{2}-3_{1}E$ &218 440.063(13)$^\alpha$ & 218 & I &45.5 & 1.31$\times$10$^5$ &this work & maser?/thermal\\
\hline
& $J_1-(J+1)_0 A^{+}$ series\\
\hline
25& $3_1-4_0A^{+}$& 107 013.803(5)$^*$ & 107 & II &28.3 & 9.17$\times$10$^5$ &this work & maser?/thermal\\
26& $5_1-6_0A^{+}$& 6 668.5188(4)$^\beta$ & 6.7 & II &49.1 & 1.66$\times$10$^6$ &this work,(9)$-$(12) & maser \& absorption\\
27& $7_0-6_1A^{+}$& 44 069.410(10)$^*$ & 44 & I &65.0 & 2.80$\times$10$^5$ &this work,(13)$-$(15) & maser\\
28& $8_0-7_1A^{+}$& 95 169.463(10)$^*$ & 95 & I &83.5 & 4.29$\times$10$^5$ &this work,(16) & maser\\
\hline
& $(J+1)_2-J_3A^{\mp}$ series\\
\hline
29& $6_2-5_3A^{-}$ & 38 293.268(50)$^*$ & 38.3 & II  &86.5 & 3.76$\times$10$^6$ &this work &absorption\\
30& $6_2-5_3A^{+}$ & 38 452.677(50)$^*$ & 38.5 & II &86.5 & 3.81$\times$10$^6$ &this work  &absorption\\
31& $7_2-6_3A^{-}$ & 86 615.600(5)$^*$ & 86.6 & II &102.7 & 4.30$\times$10$^6$ &this work  &absorption\\
32& $7_2-6_3A^{+}$ & 86 902.949(5)$^*$ & 86.9 & II &102.7 & 4.26$\times$10$^6$ &this work   &absorption\\
33& $10_2-9_3A^{-}$ & 231 281.100(12)$^\alpha$ & 231 & II &165.3 & 6.22$\times$10$^6$ &this work  &maser?/thermal\\
34& $10_2-9_3A^{+}$ & 232 418.521(12)$^\alpha$ & 232 & II &165.4 & 6.36$\times$10$^6$ &this work   &maser?/thermal\\
\hline
& $J_2-(J+1)_1A^{-}$ series\\
\hline
35& $4_2-5_1A^{-}$ & 234 683.370(12)$^\alpha$ & 234 & -  &60.9 & 2.99$\times$10$^6$ &this work   &thermal\\
36& $6_2-7_1A^{-}$ & 132 621.824(12)$^\alpha$ & 132.6 & II  &86.5 & 3.76$\times$10$^6$ & (6)  &maser?/thermal\\
37& $10_1-9_2A^{-}$& 23 444.778(2)$^\gamma$ &23.4 & I & 143.3 & 1.44$\times$10$^6$ & this work,(17)  &absorption$^{(c)}$\\
\hline
& $J_2-(J+1)_1A^{+}$ series\\
\hline
38& $7_2-8_1A^{+}$ & 111 289.550(10)$^*$ & 111 & - &102.7 & 4.26$\times$10$^6$ & this work  & thermal\\
39& $9_2-10_1A^{+}$ & 23 121.0242(5)$^*$ & 23.1 & II &142.2 & 5.68$\times$10$^6$ & this work,(18)  &absorption$^{(a)}$\\
40& $11_1-10_2A^{+}$ & 20 171.089(2)$^\gamma$ & 20.1 & - & 166.4 & 6.50$\times$10$^6$ & this work & absorption$^{(c)}$\\
\hline
& $(J+1)_1-J_1A^{\pm}$ series\\
\hline
41& $2_1-1_1A^{+}$ & 95 914.309(5)$^*$ & 95.914 & - & 21.4 & 7.04$\times$10$^5$ & -  & -$^{(d)}$\\
42& $2_1-1_1A^{-}$ & 97 582.804(7)$^*$ & 97.582 & - & 21.6 & 5.14$\times$10$^5$ & -  & -$^{(d)}$\\
43& $3_1-2_1A^{+}$ & 143 865.801(10)$^*$ & 143.865 & - & 28.3 & 9.17$\times$10$^5$ & -  & -$^{(d)}$\\
\hline
\end{tabular}
\tablefoot{Column 1 gives the serial number of each transition (for reference in the text).
Columns 2--4 give the transition, rest frequency and short line name of each methanol transition, respectively. 
The frequencies of methanol transitions are adopted from \citet{2004A&A...428.1019M}, marked by an asterisk, and references for the other methanol transitions are: $^\alpha$~the CDMS database \citep{2005JMoSt.742..215M}; $^\beta$~\cite{2015JChPh.143d4304C}; $^\gamma$~\cite{1985ZNatA..40..683M}.
The frequency uncertainties are given in parentheses in units of the least significant figure.
Column 5 shows the classification of methanol maser transitions.
Column 6 gives the upper level energy of each transition. We note that the $E$-type ground state (1$_{-1}$) level is at 7.9 K.
Column 7 gives the critical density of each transition which is calculated by assuming optically thin transitions and a gas kinetic temperature of 100 K. The corresponding code is available from \url{https://github.com/yxlinaqua/molecule_basic}.
Column 8 lists the reference for W31C:
(1) \citet{1995MNRAS.274.1126C}; (2) \citet{2010MNRAS.401.2219B}; (3) \citet{2014MNRAS.438.3368B}; (4) \citet{1989ApJ...339..949H}; (5) \citet{2019MNRAS.484.5072B}; (6) \citet{1999ApJS..123..515S}; (7) \citet{1985A&A...147L..19W}; (8) \citet{1986A&A...157..318M}; (9) \citet{1991ApJ...380L..75M}; (10) \citet{1995MNRAS.272...96C}; (11) \citet{2009PASA...26..454C}; (12) \citet{2010MNRAS.409..913G}; (13) \citet{1990A&A...231..473F}; (14) \citet{2004ApJS..155..149K}; (15) \citet{2019ApJS..244....2K}; (16) \citet{2012ApJS..200....5C}; (17) \citet{1986A&A...169..271M}; (18) \citet{1984A&A...134L...7W}.
Column 9 gives the note for W31C, and the note for the corresponding transitions towards W3(OH) can be found here:
(a) maser;
(b) all observed $J_2-J_1E$ lines near 25 GHz are shown inverse P-Cygni profile;
(c) absorption;
(d) thermal emission.\\
}
\normalsize
\end{table*}

\begin{table*}[!hbt]
\caption{Observational parameters for W31C and W3(OH).}\label{Tab:obs}
\normalsize
\centering
\begin{tabular}{lcccccc}
\hline \hline 
Line  &Obs. date & beam size &$V_{\rm res}$  & rms &  $\eta_{\rm mb}$  & Scaling factor \\ 
(GHz) &(yyyymmdd) & ($\arcsec$) &(km s$^{-1}$)  & (Jy)  & & (Jy/K)\\ 
\hline
\multicolumn{7}{c}{W31C} \\
\hline
\multicolumn{2}{l}{Effelsberg-100 m observations} \\
\hline
6.7 &20210105 &109 &0.16 &0.02  &0.69  &0.66\\%1.51
12.2 &20210105 &62 &0.13 &0.02  &0.67  &0.70\\%1.43
19.9 &20210514 &40 &0.66 &0.01  &0.62  &0.83\\%1.21
20.1 &20210514 &40 &0.66 &0.01  &0.62  &0.83\\%1.21
23.1 &20210514 &36 &0.57 &0.02  &0.60  &0.97\\%1.03
23.4 &20210514 &36 &0.57 &0.02  &0.60  &0.97\\%1.03
25 series &20210514 &34 &0.53  &0.02  &0.58  &1.02\\%0.98
36 &20200914 &24 &0.37 &0.04  &0.47 &1.19\\%0.84
37.7 &20200914 &23 &0.35 &0.04  &0.44 &1.27\\%0.79
38.3 &20200914 &23 &0.35 &0.04  &0.44 &1.27\\%0.79
38.5 &20200914 &23 &0.34 &0.05  &0.44 &1.27\\%0.79
44 &20200530 &20 &0.30 &0.04  &0.39 &1.47\\%0.68
\hline
\multicolumn{2}{l}{IRAM-30 m observations} \\
\hline
84 &20110408 &29 &0.80 &0.26  &0.81  &5.82\\
85.6 &20200629 &29 &0.79 &0.08  &0.81  &5.82\\
86.6 &20200629 &28 &0.78 &0.09  &0.81  &5.82\\ 
86.9 &20200629 &28 &0.78 &0.08  &0.81  &5.82\\
95 &20110409 &26 &0.71 &0.16  &0.81  &5.82\\
104 &20200629 &24 &0.65 &0.10  &0.78  &5.98\\
107 &20200629 &23 &0.61 &0.13  &0.78  &5.98\\
108 &20110411 &23 &0.62 &0.20  &0.78  &5.98\\
111 &20110411 &22 &0.61 &0.29  &0.78  &5.98\\
\hline
\multicolumn{2}{l}{APEX-12 m observations} \\
\hline
216 &20190612 &29 &0.17 &0.60  &0.72  &40\\
218 &20190612 &29 &0.17 &0.71 &0.72  &40\\
220 &20190612 &28 &0.17 &0.64  &0.72  &40\\
229 &20190601 &27 &0.16 &0.74  &0.72  &40\\
230 &20190601 &27 &0.16 &0.72  &0.72  &40\\
231 &20190601 &27 &0.16 &1.14  &0.72  &40\\
232 &20190601 &27 &0.16 &0.73  &0.72  &40\\
234 &20190601 &27 &0.16 &0.88  &0.72  &40\\
\hline
\multicolumn{7}{c}{W3(OH)} \\
\hline
\multicolumn{2}{l}{Effelsberg-100 m observations} \\
\hline
19.9 &20210513 &40 &0.66 &0.01  &0.62  &0.83\\%1.21
20.1 &20210513 &40 &0.66 &0.01  &0.62  &0.83\\%1.21
23.1 &20210513 &36 &0.57 &0.01  &0.60  &0.97\\%1.03
23.4 &20210513 &36 &0.57 &0.01  &0.60  &0.97\\%1.03
25 series &20210513 &34 &0.53  &0.01  &0.58  &1.02\\%0.98
\hline
\multicolumn{2}{l}{IRAM-30 m observations} \\
\hline
95.914 &20190629 &26 &0.71 &0.03  &0.81  &5.82\\
97.582 &20190629 &25 &0.70 &0.02  &0.81  &5.82\\
143.865 &20190629 &17 &0.47 &0.03  &0.73  &6.32\\
\hline
\end{tabular}
\tablefoot{
The scaling factor is used for the conversion between antenna temperature and flux density. For APEX observations, the velocity resolution is given after once Hanning smoothing. For information on the Effelsberg-100 m: \url{https://eff100mwiki.mpifr-bonn.mpg.de/doku.php?id=information_for_astronomers:rx_list}, for the IRAM-30 m: \url{https://publicwiki.iram.es/Iram30mEfficiencies}, for the APEX telescope: \url{http://www.apex-telescope.org/telescope/efficiency/index.php?orderBy=mJyK&planetBy=all&yearBy=2019&sortAs=DESC}
}
\normalsize
\end{table*}
%For IRAM 30m observations, the observed HPBWs can be well fitted by HPBW/arcsec=2460/Freq/GHz.

\begin{table}[!hbtp]
\caption{Observational results of CH$_3$OH transitions in W31C.}\label{Tab:result}
\small
%\tiny
\renewcommand\arraystretch{0.97}
\centering
\begin{tabular}{lrrrr}
\hline 
\hline
Line &$V_{\rm lsr}$ & $\Delta V$ &$S_{\rm pk}$  & $\int S{\rm d}V$\\ 
(GHz)  &(km s$^{-1}$) &(km s$^{-1}$)  & (Jy)  & (Jy km s$^{-1}$)\\ 
\hline
6.7 & $-$8.19 (0.01) & 0.59 (0.03) & 1.42 (0.21) & 0.89 (0.04)\\
6.7 & $-$1.93 (0.07) & 3.34 (0.16) & $-$0.65 (0.21) & $-$2.31 (0.09)\\
6.7 & 4.78 (0.01) & 0.88 (0.02) & 3.70 (0.21) & 3.45 (0.05)\\
12.2 & $-$1.23 (0.03) & 5.40 (0.05) & $-$0.42 (0.05) & $-$2.39 (0.02)\\
12.2 &  4.84 (0.01) & 0.96 (0.02) & 0.53 (0.05) & 0.55 (0.01)\\
19.9 & $-$0.90 (0.11) & 3.99 (0.21) & $-$0.11 (0.01) & $-$0.46 (0.03)\\
20.1 & $-$0.40 (0.31) & 4.18 (0.63) & $-$0.06 (0.02) & $-$0.27 (0.03)\\
23.1 & $-$0.74 (0.01) & 4.69 (0.16) & $-$0.23 (0.02) & $-$1.15 (0.04)\\
23.4 & $-$0.51 (0.22) & 3.87 (0.49) & $-$0.13 (0.03) & $-$0.54 (0.05)\\
24.928 & $-$9.25 (0.13) & 1.34 (0.36) & 0.04 (0.01) & 0.06 (0.02) \\
24.928 & $-$5.18 (0.03) & 4.88 (0.31) & 0.24 (0.01) & 1.26 (0.07) \\
24.928 & $-$1.90 (0.07) & 2.37 (0.15) & 0.18 (0.01) & 0.46 (0.04) \\
24.928 & 2.08 (0.13) & 3.31 (0.32) & 0.11 (0.01) & 0.37 (0.03) \\
24.933 & $-$8.04 (0.01) & 2.79 (0.29) & 0.15 (0.01) & 0.44 (0.03) \\
24.933 & $-$5.40 (0.04) & 2.71 (0.18) & 0.23 (0.01) & 0.66 (0.04) \\
24.933 & $-$2.90 (0.01) & 2.19 (0.13) & 0.25 (0.01) & 0.59 (0.04) \\
24.933 & $-$0.52 (0.05) & 2.74 (0.48) & 0.15 (0.01) & 0.45 (0.07) \\
24.933 & 2.64 (0.26) & 1.85 (0.40) & 0.07 (0.01) & 0.14 (0.04) \\
24.934 & $-$4.77 (0.14) & 5.85 (0.31) & 0.16 (0.01) & 0.98 (0.05) \\
24.934 & $-$1.99 (0.12) & 1.03 (0.46) & 0.06 (0.01) & 0.06 (0.04) \\
24.959 & $-$8.02 (0.11) & 1.57 (0.18) & 0.13 (0.01) & 0.22 (0.01) \\
24.959 & $-$5.98 (0.04) & 1.89 (0.20) & 0.28 (0.01) & 0.56 (0.07) \\
24.959 & $-$2.80 (0.15) & 3.74 (0.17) & 0.28 (0.01) & 1.11 (0.03) \\
24.959 & 2.09 (0.14) & 5.36 (1.01) & 0.08 (0.01) & 0.48 (0.07) \\
25.018 & $-$8.00 (0.13) & 2.20 (0.32) & 0.12 (0.01) & 0.28 (0.03) \\
25.018 & $-$5.97 (0.06) & 1.12 (0.15) & 0.30 (0.01) & 0.35 (0.07) \\
25.018 & $-$4.44 (0.11) & 1.67 (0.42) & 0.19 (0.01) & 0.33 (0.12) \\
25.018 & $-$2.18 (0.14) & 2.49 (0.32) & 0.21 (0.01) & 0.56 (0.09) \\
25.018 & 2.21 (0.24) & 3.04 (0.62) & 0.07 (0.01) & 0.22 (0.04) \\
25.124 & $-$7.78 (0.12) & 1.69 (0.75) & 0.08 (0.01) & 0.15 (0.04) \\
25.124 & $-$5.98 (0.05) & 1.08 (0.17) & 0.20 (0.01) & 0.23 (0.05) \\
25.124 & $-$2.92 (0.11) & 4.14 (0.33) & 0.19 (0.01) & 0.82 (0.05) \\
25.124 & 3.10 (0.17) & 2.59 (0.46) & 0.08 (0.01) & 0.22 (0.03) \\
25.294 & $-$4.78 (0.14) & 5.30 (0.27) & 0.12 (0.02) & 0.67 (0.03) \\
25.541 & $-$4.93 (0.18) & 4.83 (0.60) & 0.10 (0.01) & 0.49 (0.05) \\
25.541 & 0.20 (0.18) & 2.21 (0.32) & $-$0.07 (0.01) & $-$0.17 (0.04) \\
25.878 & $-$5.97 (0.17) & 3.10 (0.34) & 0.07 (0.01) & 0.24 (0.03) \\
25.878 & 0.10 (0.14) & 2.73 (0.31) & $-$0.09 (0.01) & $-$0.26 (0.03) \\
36 & $-$7.35 (0.04) & 2.64 (0.03) & 3.52 (0.41) & 9.91 (0.25)\\
36 & $-$6.19 (0.03) & 1.43 (0.01) & 12.70 (0.41) & 19.31 (0.31)\\
36 & $-$2.84 (0.03) & 7.59 (0.15) & 1.69 (0.41) & 13.67 (0.21)\\
37.7 & $-$0.68 (0.05) & 5.35 (0.11) & $-$0.89 (0.05) & $-$5.05 (0.09)\\
38.3 & $-$0.74 (0.10) & 4.77 (0.22) & $-$0.38 (0.04)  &$-$1.94 (0.08)\\
38.5 & $-$0.65 (0.14) & 4.81 (0.28) & $-$0.30 (0.06) & $-$1.54 (0.08)\\
44 & $-$7.26 (0.03) & 2.86 (0.05) &7.58 (0.69) & 23.11 (0.03)\\
44 & $-$6.65 (0.00) & 0.76 (0.01) & 44.87 (0.69) & 36.51 (0.33)\\
84 & $-$6.94 (0.03) & 2.36 (0.09) & 9.05 (0.39) & 22.70 (1.16)\\
84 & $-$3.82 (0.08) & 8.41 (0.13) & 9.77 (0.39) & 87.38 (1.64)\\
85.6 & 0.22 (0.19) & 3.80 (0.35) & $-$0.55 (0.08) & $-$2.22 (0.21) \\ 
86.6 & 0.20 (0.56) & 3.79 (1.12) & $-$0.17 (0.10) & $-$0.70 (0.20)\\
86.9 & 0.43 (0.31) & 3.72 (0.88) & $-$0.29 (0.07) & $-$1.17 (0.21)\\
95 & $-$7.00 (0.01) & 2.33 (0.03) & 13.87 (0.70) & 34.32 (0.55) \\
95 & $-$3.65 (0.07) & 9.29 (0.09) & 6.13 (0.70) & 60.54 (0.80) \\
104 & $-$2.88 (0.18) & 5.81 (0.51) & 0.75 (0.12) & 4.65 (0.33)\\
107 & $-$3.95 (0.04) & 7.00 (0.11) & 6.03 (0.32) & 44.98 (0.58)\\
108 & $-$4.00 (0.05) & 7.83 (0.12) & 6.41 (0.28) & 53.43 (0.64)\\
111 & $-$4.13 (0.26) & 4.23 (0.68) & 1.32 (0.24) & 5.96 (0.75)\\
216 & $-$3.26 (0.05) & 7.26 (0.12) & 14.10 (0.79) & 108.72 (1.47)\\
218 & $-$3.33 (0.02) & 7.54 (0.02) & 82.92 (1.70) & 665.33 (1.36)\\
220 & $-$2.84 (0.04) & 7.56 (0.09) & 14.30 (0.86) & 114.85 (1.15)\\
229 & $-$3.47 (0.01) & 8.18 (0.03) & 43.30 (1.09) & 376.80 (1.24)\\
230 &  $-$3.21 (0.05) & 6.96 (0.12) & 10.00 (0.78) & 74.41 (1.14)\\
231 & $-$2.35 (0.23) & 6.43 (0.52) & 3.37 (0.91) &23.12 (1.72)\\
232 & $-$2.83 (0.17) & 6.87 (0.47) & 3.45 (0.74) & 25.19 (1.35)\\
234 & $-$2.92 (0.06) & 7.03 (0.15) & 11.41 (0.99) & 84.74 (1.38)\\
\hline
%\end{longtable}
\end{tabular}
\tablefoot{Focusing on the absorption, we  fitted single Gaussian profile to only to the absorption component of 19.9, 20.1, 23.4, 37.7, 38.3, 38.5, 85.6, 86.6 and 86.9 GHz lines.}%, to better depict the absorption feature.}
\end{table}
\normalsize

\begin{table}[!hbtp]
\caption{Observational results of CH$_3$OH transitions in W3(OH).}\label{Tab:result-w3oh}
\small
\centering
\begin{tabular}{lrrrr}
\hline 
\hline
Line &$V_{\rm lsr}$ & $\Delta V$ &$S_{\rm pk}$  & $\int S{\rm d}V$\\  
(GHz)  &(km s$^{-1}$) &(km s$^{-1}$)  & (Jy)  & (Jy km s$^{-1}$)\\ 
\hline
19.9 & $-$43.53 (0.00) & 1.61 (0.01) & 30.16 (0.96) & 51.65 (0.10)\\
20.1 & $-$44.40 (0.03) & 2.31 (0.07) & $-$0.21 (0.01) & $-$0.51 (0.01)\\
23.1 & $-$43.36 (0.00) & 1.45 (0.01) & 2.86 (0.15) & 4.42 (0.03)\\
23.4 & $-$46.41 (0.47) & 19.84 (1.02) & 0.07 (0.01) & 1.53 (0.07)\\
23.4 & $-$44.47 (0.03) & 2.16 (0.08) & $-$0.36 (0.01) & $-$0.83 (0.03)\\
24.928 & $-$46.43 (0.24) & 4.56 (0.34) & 0.19 (0.02) & 0.93 (0.11)\\
24.928 & $-$44.62 (0.04) & 2.44 (0.09) & $-$0.49 (0.02) & $-$1.27 (0.11)\\
24.933 & $-$46.78 (0.19) & 4.87 (0.28) & 0.16 (0.03) & 0.84 (0.07)\\
24.933 & $-$44.62 (0.02) & 2.26 (0.07) & $-$0.46 (0.03) & $-$1.10 (0.06)\\
24.934 & $-$45.20 (0.18) & 6.26 (0.50) & 0.14 (0.02) & 0.95 (0.09)\\
24.934 & $-$44.51 (0.03) & 2.51 (0.08) & $-$0.53 (0.02) & $-$1.42 (0.08)\\
24.959 & $-$47.50 (0.26) & 6.51 (0.41) & 0.14 (0.02) & 0.94 (0.07)\\
24.959 & $-$44.59 (0.03) & 2.12 (0.09) & $-$0.40 (0.02) & $-$0.90 (0.06)\\
25.018 & $-$47.24 (0.21) & 6.58 (0.34) & 0.14 (0.02) & 1.00 (0.06)\\
25.018 & $-$44.65 (0.03) & 2.21 (0.09) & $-$0.37 (0.02) & $-$0.87 (0.05)\\
25.124 & $-$47.66 (0.30) & 7.45 (0.57) & 0.10 (0.01) & 0.82 (0.07)\\
25.124 & $-$44.63 (0.04) & 2.14 (0.10) & $-$0.29 (0.01) & $-$0.66 (0.04)\\
25.294 & $-$47.36 (0.42) & 6.66 (0.53) & 0.09 (0.02) & 0.66 (0.08)\\
25.294 & $-$44.64 (0.04) & 2.14 (0.13) & $-$0.30 (0.02) & $-$0.69 (0.06)\\
25.541 & $-$47.90 (0.36) & 7.94 (0.56) & 0.07 (0.02) & 0.58 (0.05)\\
25.541 & $-$44.62 (0.03) & 2.17 (0.09) & $-$0.28 (0.02) & $-$0.65 (0.03)\\
25.878 & $-$48.12 (0.39) & 7.39 (0.59) & 0.08 (0.01) & 0.64 (0.06)\\
25.878 & $-$44.57 (0.04) & 2.15 (0.10) & $-$0.29 (0.01) & $-$0.66 (0.05)\\
95.914 & $-$46.19 (0.01) & 4.81 (0.03) & 4.72 (0.39) & 24.13 (0.13)\\
97.865 & $-$46.23 (0.01) & 4.76 (0.03) & 5.01 (0.42) & 25.39 (0.11)\\
143.865 & $-$46.15 (0.01) & 4.81 (0.02) & 11.13 (0.73) & 56.91 (0.15)\\
\hline
%\end{longtable}
\end{tabular}
%\tablefoot{}
\end{table}
\normalsize

\begin{table*}[!hbt]
\caption{Two-layer modelling results for the methanol transitions with redshifted absorption feature towards W31C and W3(OH).}\label{Tab:model}
\normalsize
\renewcommand\arraystretch{1.5}
\centering
\begin{tabular}{lcccccc}
\hline \hline 
Line & \multicolumn{1}{c}{Fixed parameters} & \multicolumn{5}{c}{Output parameters}\\
\cmidrule(lr){2-2} \cmidrule(lr){3-7} 
     & $T_{\rm c}$ & $V_{\rm in}$ & $\tau_0$ & $\sigma$ & $T_{\rm f}$ & $T_{\rm r}$ \\
(GHz)  & (K)           & (\kms) &  &(\kms)  & (K) & (K)\\
\hline
\multicolumn{7}{c}{W31C}\\
\hline
6.7  & 6.83 & 1.93$^{+0.04}_{-0.03}$ & 0.15$^{+0.07}_{-0.04}$ & 1.88$^{+0.04}_{-0.05}$ & 1.41$^{+0.58}_{-0.30}$ & 57.36$^{+31.46}_{-33.22}$\\
12.2 & 6.39 & 2.17$^{+0.03}_{-0.03}$ & 0.24$^{+0.05}_{-0.03}$ & 2.04$^{+0.02}_{-0.02}$ & 1.85$^{+0.53}_{-0.51}$ & 2.05$^{+1.83}_{-0.79}$\\ 
19.9 & 5.53 & 2.04$^{+0.63}_{-0.68}$ & 0.25$^{+0.26}_{-0.10}$ & 2.06$^{+0.41}_{-0.40}$ & 7.35$^{+3.01}_{-2.17}$ & 39.41$^{+36.41}_{-27.49}$\\
20.1 & 5.42 & 2.72$^{+0.13}_{-0.13}$ & 0.11$^{+0.01}_{-0.00}$ & 2.09$^{+0.10}_{-0.10}$ & 3.82$^{+0.31}_{-0.14}$ & 2.50$^{+2.27}_{-1.12}$\\
23.1 & 5.03 & 2.68$^{+0.09}_{-0.09}$ & 0.14$^{+0.06}_{-0.03}$ & 1.95$^{+0.08}_{-0.07}$ & 2.79$^{+0.72}_{-0.77}$ & 9.88$^{+11.43}_{-6.47}$\\
23.4 & 4.90 & 2.95$^{+0.16}_{-0.16}$ & 0.11$^{+0.02}_{-0.01}$ & 1.58$^{+0.15}_{-0.15}$ & 2.60$^{+0.35}_{-0.29}$ & 3.08$^{+3.33}_{-1.56}$\\
25.541 & 4.62 & 1.07$^{+0.23}_{-0.21}$ & 0.31$^{+0.08}_{-0.05}$ & 1.93$^{+0.14}_{-0.14}$ & 11.83$^{+0.74}_{-1.29}$ & 88.35$^{+8.52}_{-14.81}$\\
25.878 & 4.53 & 3.07$^{+0.15}_{-0.18}$ & 0.39$^{+0.99}_{-0.22}$ & 1.22$^{+0.16}_{-0.15}$ & 6.28$^{+3.68}_{-2.03}$ & 29.61$^{+44.67}_{-24.72}$\\
37.7 & 4.52 & 2.73$^{+0.05}_{-0.05}$ & 1.40$^{+0.47}_{-0.43}$ & 1.91$^{+0.09}_{-0.08}$ & 2.46$^{+0.27}_{-0.30}$ & 4.12$^{+4.73}_{-2.27}$\\
38.3 & 4.38 & 2.52$^{+0.13}_{-0.13}$ & 0.39$^{+1.13}_{-0.20}$ & 2.00$^{+0.14}_{-0.23}$ & 5.16$^{+2.21}_{-1.45}$ & 37.05$^{+41.65}_{-29.37}$\\
38.5 & 4.35 & 2.60$^{+0.17}_{-0.14}$ & 3.57$^{+4.84}_{-3.40}$ & 1.53$^{+0.53}_{-0.28}$ & 3.54$^{+0.66}_{-0.15}$ & 2.84$^{+31.72}_{-1.48}$\\
85.6 & 0.48 & 3.47$^{+0.20}_{-0.19}$ & 2.21$^{+4.75}_{-2.02}$ & 1.42$^{+0.37}_{-0.29}$ & 3.27$^{+1.82}_{-0.17}$ & 19.93$^{+23.69}_{-1.81}$\\
86.6 & 0.48  & 1.63$^{+1.24}_{-0.73}$ & 0.17$^{+0.17}_{-0.06}$ & 2.03$^{+0.49}_{-0.52}$ & 6.24$^{+2.25}_{-1.90}$ & 51.23$^{+25.50}_{-20.75}$\\
86.9 & 0.48 & 3.89$^{+0.28}_{-0.31}$ & 0.25$^{+1.22}_{-0.12}$ & 1.26$^{+0.40}_{-0.32}$ & 4.32$^{+1.62}_{-1.11}$ & 32.12$^{+19.18}_{-13.20}$\\ 
\hline
\multicolumn{7}{c}{W3(OH)}\\
\hline
20.1 & 5.82 & 1.78$^{+0.02}_{-0.02}$ & 0.13$^{+0.03}_{-0.02}$ & 0.95$^{+0.02}_{-0.02}$ & 2.90$^{+0.42}_{-0.46}$ & 9.41$^{+8.03}_{-5.79}$\\
23.4 & 5.32 & 1.75$^{+0.04}_{-0.04}$ & 1.26$^{+0.79}_{-0.72}$ & 0.65$^{+0.05}_{-0.05}$ & 4.62$^{+1.41}_{-0.32}$ & 7.05$^{+20.61}_{-4.78}$\\
24.928 & 5.28 & 1.77$^{+0.02}_{-0.02}$ & 3.18$^{+0.74}_{-0.66}$ & 0.68$^{+0.03}_{-0.03}$ & 5.23$^{+0.19}_{-0.15}$ & 13.00$^{+2.41}_{-1.86}$\\
24.933 & 5.27 & 1.74$^{+0.02}_{-0.02}$ & 4.37$^{+1.08}_{-0.86}$ & 0.62$^{+0.03}_{-0.03}$ & 5.07$^{+0.14}_{-0.12}$ & 11.15$^{+1.65}_{-1.41}$\\
24.934 & 5.27 & 1.70$^{+0.02}_{-0.02}$ & 1.51$^{+0.56}_{-0.50}$ & 0.73$^{+0.04}_{-0.04}$ & 4.54$^{+0.49}_{-0.27}$ & 8.05$^{+7.29}_{-3.93}$\\
24.959 & 5.26 & 1.74$^{+0.02}_{-0.02}$ & 3.70$^{+1.85}_{-1.50}$ & 0.65$^{+0.07}_{-0.05}$ & 5.25$^{+0.40}_{-0.22}$ & 12.24$^{+4.94}_{-2.63}$\\
25.018 & 5.24 & 1.72$^{+0.03}_{-0.03}$ & 4.53$^{+1.67}_{-1.34}$ & 0.65$^{+0.05}_{-0.04}$ & 5.28$^{+0.20}_{-0.14}$ & 11.93$^{+2.63}_{-1.61}$\\
25.124 & 5.20 & 1.62$^{+0.05}_{-0.06}$ & 0.68$^{+2.53}_{-0.37}$ & 0.83$^{+0.09}_{-0.18}$ & 7.20$^{+3.97}_{-2.38}$ & 36.31$^{+49.27}_{-29.52}$\\
25.294 & 5.12 & 1.68$^{+0.06}_{-0.06}$ & 2.14$^{+2.40}_{-1.66}$ & 0.71$^{+0.13}_{-0.10}$ & 4.65$^{+2.80}_{-0.33}$ & 6.44$^{+36.47}_{-4.19}$\\
25.541 & 5.03 & 1.60$^{+0.05}_{-0.05}$ & 0.76$^{+1.27}_{-0.52}$ & 0.81$^{+0.07}_{-0.09}$ & 5.40$^{+4.19}_{-1.14}$ & 18.47$^{+57.08}_{-15.48}$\\
25.878 & 4.89 & 1.72$^{+0.04}_{-0.04}$ & 1.90$^{+1.32}_{-1.34}$ & 0.72$^{+0.11}_{-0.07}$ & 4.69$^{+2.14}_{-0.33}$ & 9.23$^{+28.17}_{-4.23}$\\
\hline
\end{tabular}
\tablefoot{$V_{\rm in}$ is the infall velocity, $\tau_0$ is the peak optical depth of each layer, $\sigma$ is the velocity dispersion, $T_{\rm f}$ and $T_{\rm r}$ are the excitation temperatures of the ``front" and ``rear" layers, respectively. $\Phi$= 0.9 and $T_{\rm b}$=2.73~K are fixed for modelling the lines in both W31C and W3(OH). $V_{\rm LSR}$ are fixed at $-$3.43~\kms\/ and $-$46.19~\kms\/ for modelling the lines in W31C and W3(OH), respectively. }
\normalsize
\end{table*}

\clearpage
\section{Infall velocity as a function of the upper energy level, critical density, and velocity dispersion}\label{sec:appendix}

In this Appendix, we compare the derived infall velocity with the upper energy level, critical density as well as velocity dispersion.

\begin{figure}[!htbp]
\centering
\includegraphics[width=0.471\textwidth]{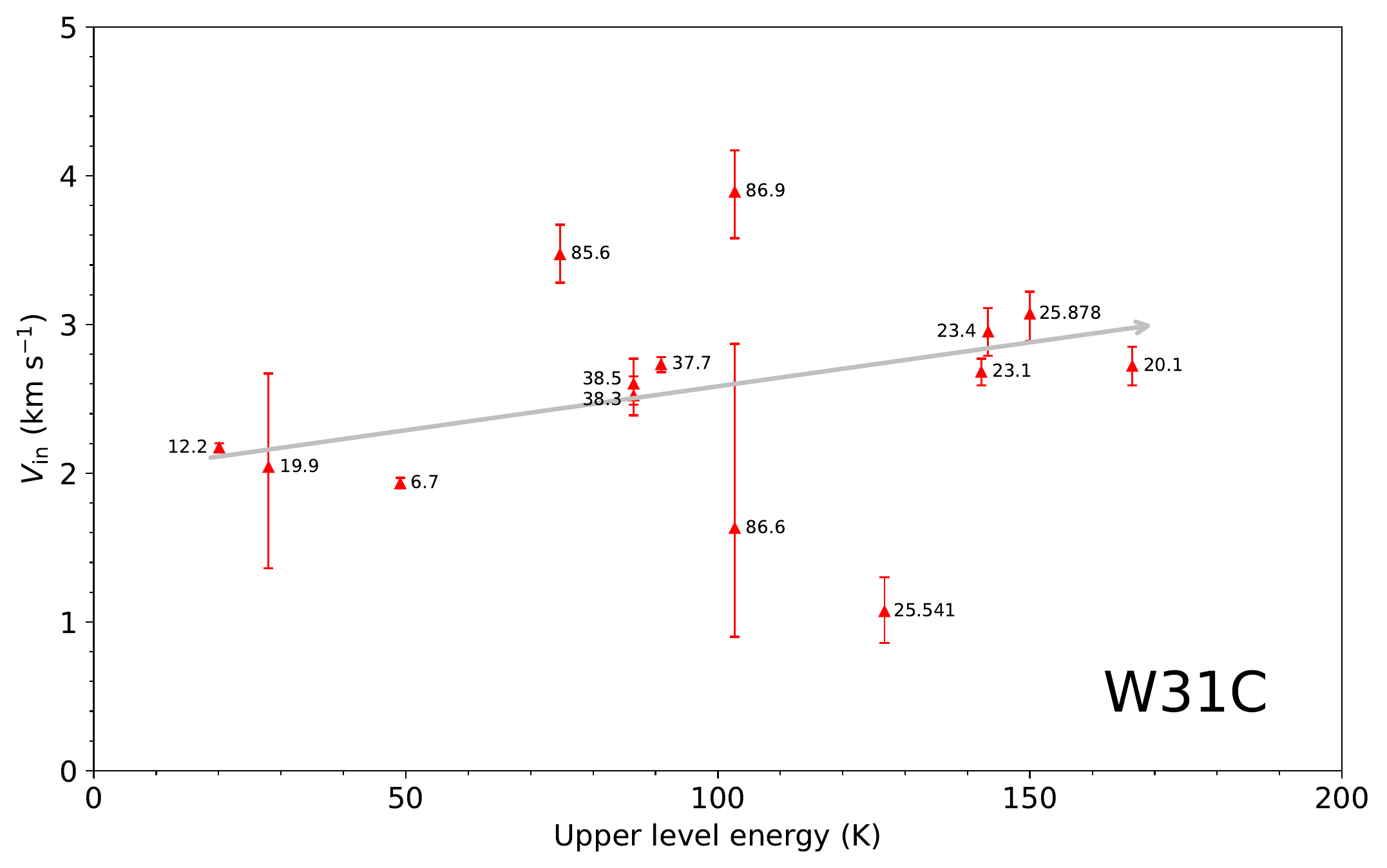}
\includegraphics[width=0.471\textwidth]{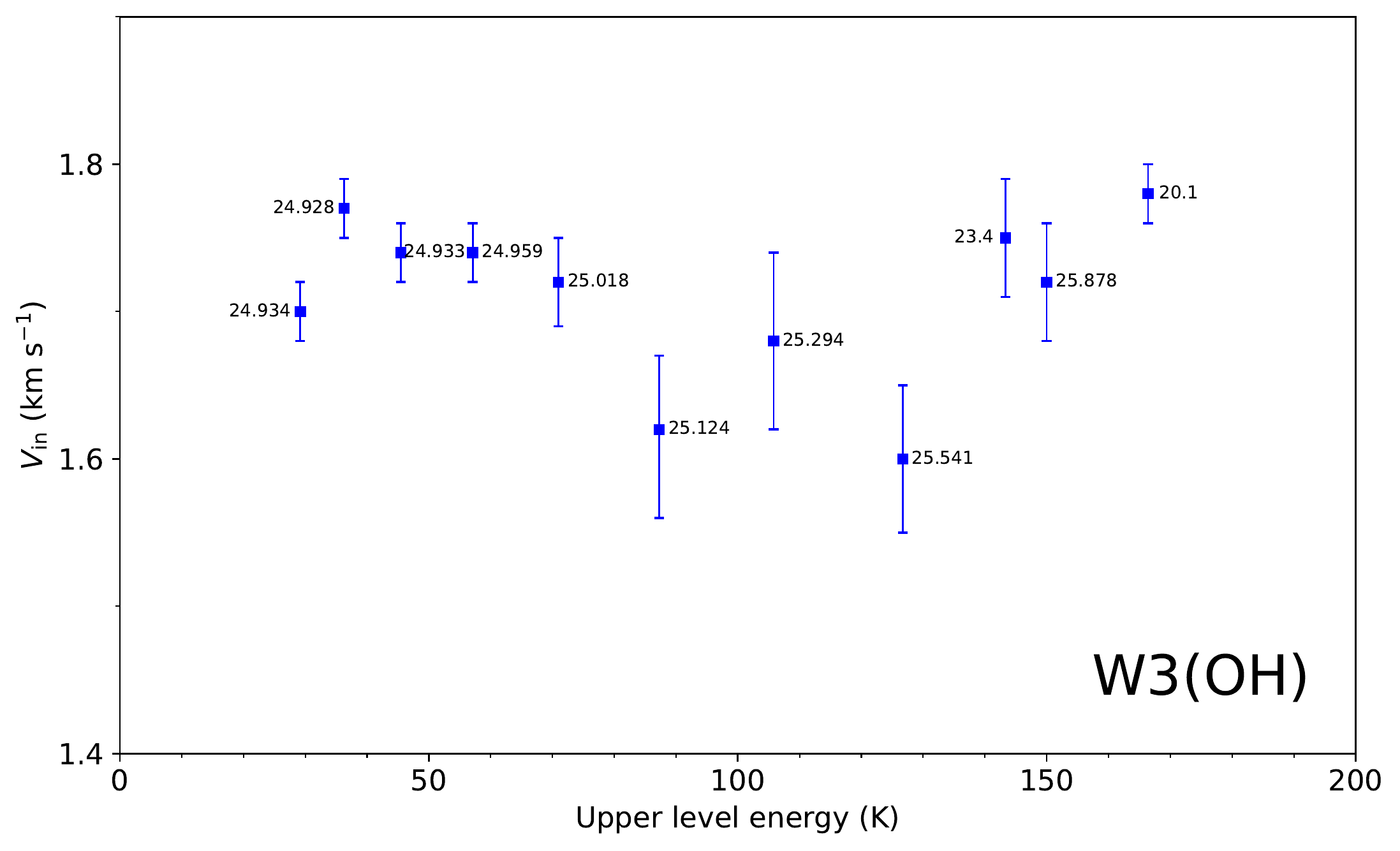}
\caption{Infall velocity as a function of the upper energy level for both W31C and W3(OH), respectively. The grey arrow roughly depicts a slight upward trend in W31C.
\label{fig:Vin-e}}
\end{figure}

\begin{figure}[!htbp]
\centering
\includegraphics[width=0.471\textwidth]{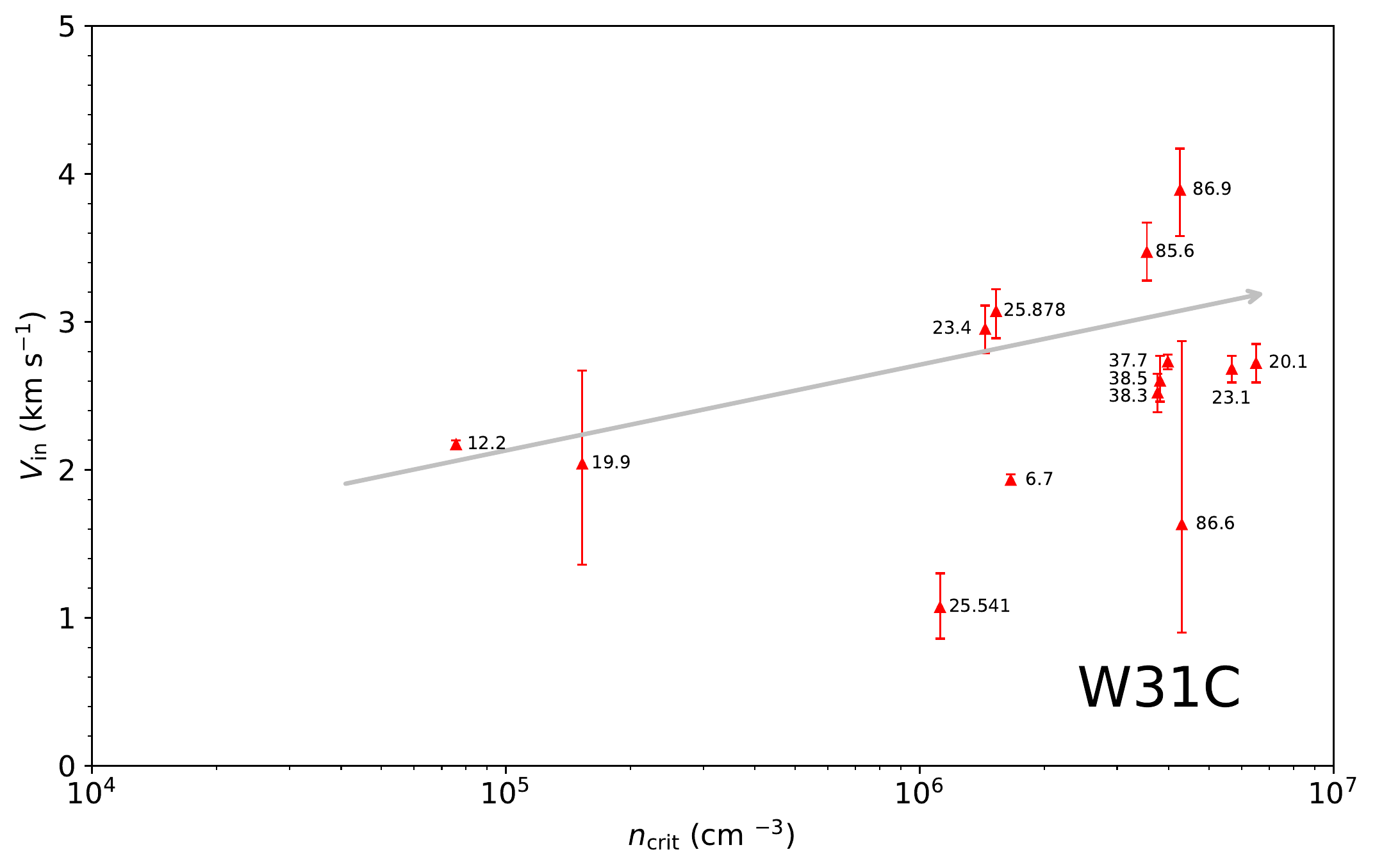}
\includegraphics[width=0.471\textwidth]{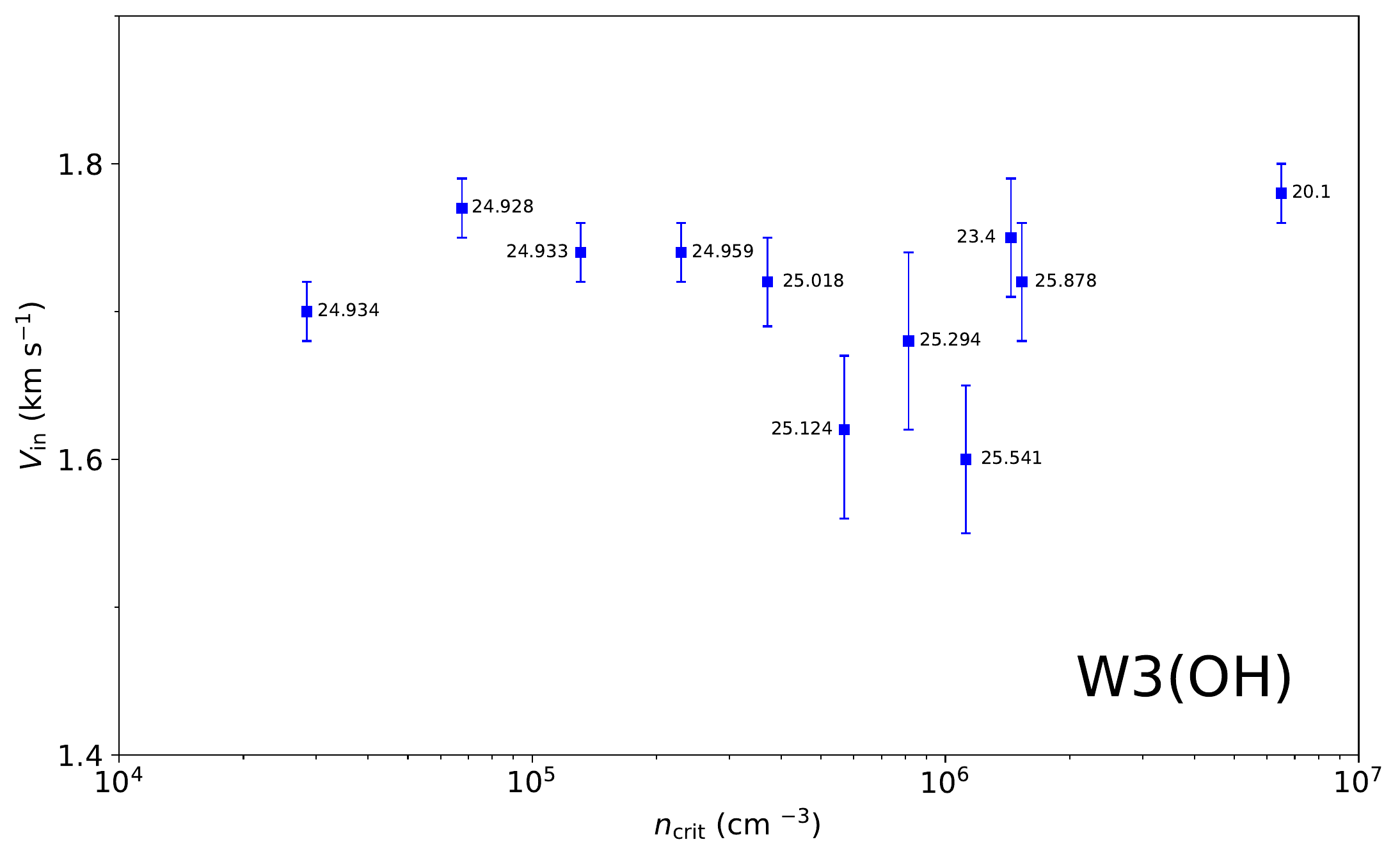}
\caption{Infall velocity as a function of the critical density for both W31C and W3(OH), respectively. The grey arrow roughly depicts a slight upward trend in W31C.
\label{fig:Vin-n}}
\end{figure}

Figures~\ref{fig:Vin-e} and \ref{fig:Vin-n} show the infall velocity in W31C and W3(OH) as a function of the upper energy level and critical density, respectively.
In W31C, apart from the CH$_3$OH transitions at 25.541, 86.6 and 86.9~GHz (which may have inaccurate $V_{\rm LSR}$ or poor S/N ratio), the derived $V_{\rm in}$ shows an increasing trend with increasing upper energy level and critical density. %implying that a multi-layer view of global collapse with hierarchical inflows in W31C.
While in W3(OH), $V_{\rm in}$ distributes within a narrow velocity range and shows no clear trend with the upper energy level and critical density.

\begin{figure}[!htbp]
\centering
\includegraphics[width=0.5\textwidth]{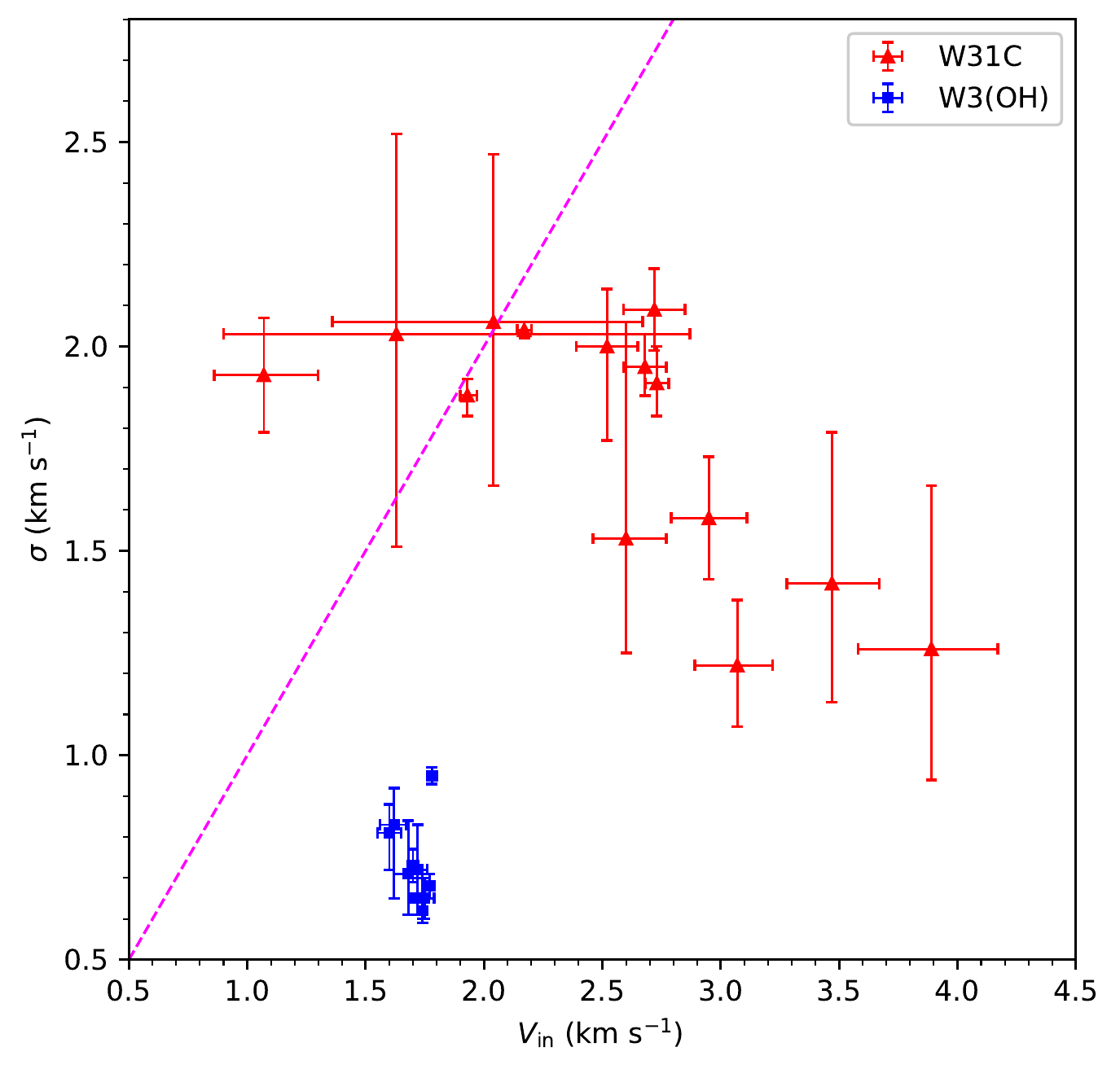}
\caption{Velocity dispersion as a function of infall velocity for both W31C and W3(OH) are marked with red triangles and blue squares, respectively. The magenta dashed line represents where velocity dispersion equals infall velocity.
\label{fig:sigma-Vin}}
\end{figure}

Figure~\ref{fig:sigma-Vin} shows that there is a different behaviour between W31C and W3(OH) when comparing $\sigma$ with $V_{\rm in}$ that are derived from the MCMC fitting. 
We find that $V_{\rm in}$ is comparable to $\sigma$ in W31C for $V_{\rm in}\lesssim$2.5~\kms, indicating that the collapse can sustain the turbulence in W31C at a large scale. For $V_{\rm in}\gtrsim$2.5~\kms, $\sigma$ becomes lower than $V_{\rm in}$. Because the higher $V_{\rm in}$ is derived from lines with high energies and critical densities, the behaviour is likely to occur at small scales. This is also the case for W3(OH). At the small scales, the dissipation rate should be higher because of higher densities \citep[e.g.][]{2004ARA&A..42..211E}. The more efficient dissipation rate may explain the observed behaviour that $\sigma$ is lower than $V_{\rm in}$.  If the observed $\sigma$ follows the typical line width-size scaling relation \citep[e.g.][]{1981MNRAS.194..809L}, the lower $\sigma$ corresponds to the smaller scale. Hence, the trend that $V_{\rm in}$ increases with decreasing $\sigma$ might indicate a multi-layer view of global collapse with hierarchical inflows in W31C.

%while $V_{\rm in}$ is larger than $\sigma$ in W3(OH), indicating that the collapse sustains the turbulence in W31C, while the dissipation rate is higher than the kinetic energy injection of collapse in W3(OH). 
%The velocity dispersion in W3(OH) appears to be similar, while the velocity dispersion in W31C span a wide range, which may be evidence that W31C has a more extensive inflow zone given the line width-size scaling relation, while W3(OH) has relatively confined infall region.

\end{appendix}

\end{document}